%% file: cwp-arxiv.tex
\documentclass[12pt,a4paper]{article}
\usepackage{jheppub}

\usepackage{graphicx}
\usepackage{subcaption}
\usepackage[titletoc]{appendix}
\usepackage{xspace}

\usepackage[utf8]{inputenc}

\usepackage[style=numeric-comp,sorting=none]{biblatex}
\include{cwp-abstract}

\begin{document}

\noindent
\begin{tabular*}{\linewidth}{lc@{\extracolsep{\fill}}r@{\extracolsep{0pt}}}
 & & HSF-CWP-2017-01 \\
 & & December 15, 2017 \\ 
 & & \\
\end{tabular*}
\vspace{2.0cm}

\title{A Roadmap for\\HEP Software and Computing R\&D\\for the 2020s}

\author{HEP Software Foundation\footnote{Authors are listed at the end of this report.}}

\maketitle

\newpage

\input{cwp-main}

\newpage
\input{appendices}

\sloppy
\raggedright
\clearpage
\printbibliography[title={References},heading=bibintoc]

\clearpage
\section*{The HEP Software Foundation}
\input{authors}

\end{document}

%% file: cwp-abstract.tex
\abstract{Particle physics has an ambitious and broad experimental 
programme for the coming decades. This programme requires large 
investments in detector hardware, either to build new facilities 
and experiments, or to upgrade existing ones.
Similarly, it requires commensurate investment in the R\&D of software to acquire, manage, process, and analyse the shear amounts of data to be recorded.
In planning for the HL-LHC in particular, it is critical that all of the
collaborating stakeholders agree on the software goals and priorities,
and that the efforts complement each other. In this spirit, this white paper
describes the R\&D activities required to prepare for this software upgrade.}

%% file: cwp-main.tex


\hypertarget{introduction}{%
\section{Introduction}\label{introduction}}

Particle physics has an ambitious experimental programme for the
coming decades. The programme supports the strategic goals of the
particle physics community that have been laid out by the European
Strategy for Particle Physics~\cite{ESPP2013} and by
the Particle Physics Project Prioritization Panel (P5)~\cite{DOE-P5} in the United
States~\cite{2014pwa}. Broadly speaking, the
scientific goals are:

\begin{itemize}
\item Exploit the discovery of the Higgs boson as a precision tool
  for investigating Standard Model (SM) and Beyond the Standard Model
  (BSM) physics.
\item
  Study the decays of $b$- and $c$-hadrons, and tau leptons, in the search
  for manifestations of BSM physics, and investigate
  matter-antimatter differences.
\item
  Search for signatures of dark matter.
\item
  Probe neutrino oscillations and masses.
\item
  Study the Quark Gluon Plasma state of matter in heavy-ion collisions.
\item
  Explore the unknown.
\end{itemize}

\begin{figure*}[bthp]
    \centering
    \includegraphics[width=0.94\textwidth]{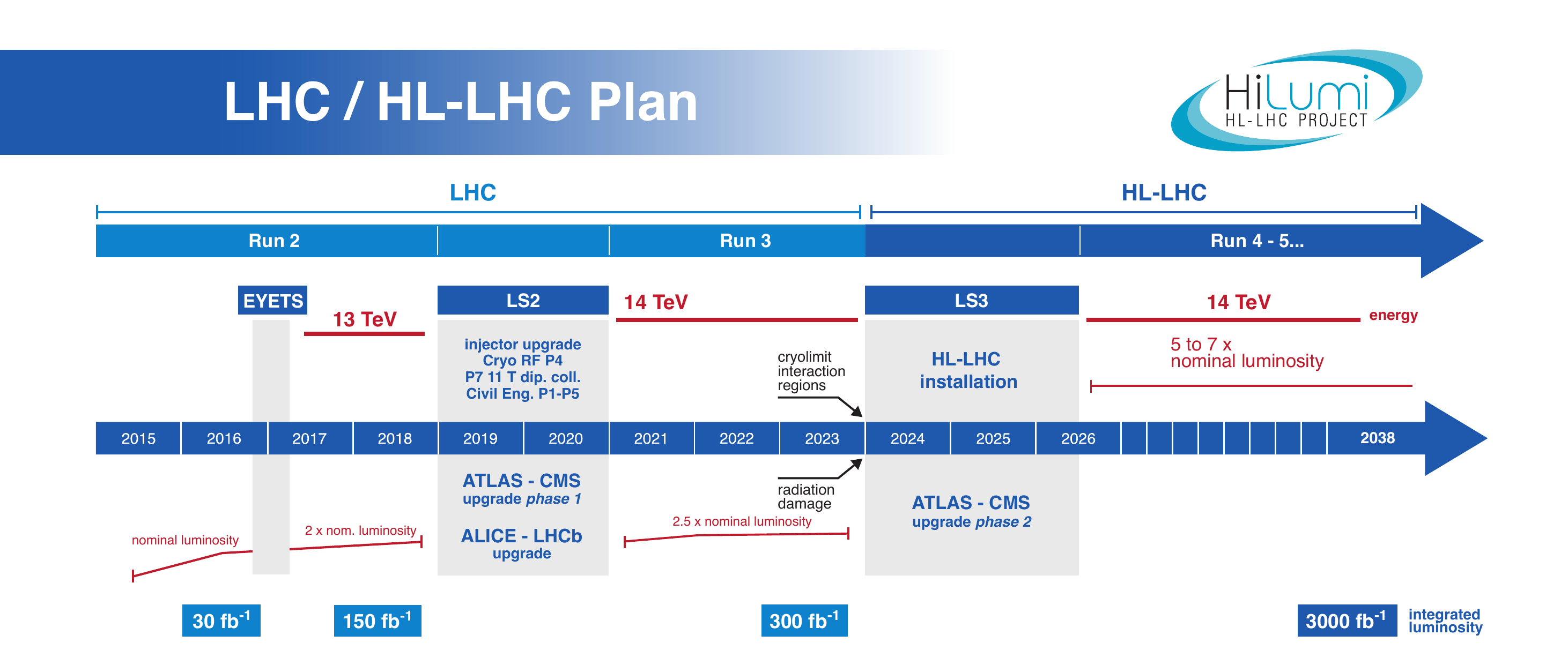}
    \caption{The current schedule for the LHC and HL-LHC upgrade
and run~\cite{HL-LHC}. Currently, the start of the HL-LHC run is foreseen for mid
2026. The long shutdowns, LS2 and LS3, will be used to upgrade both the
accelerator and the detector hardware.}
    \label{fig:hl-lhc}
\end{figure*}

The High-Luminosity Large Hadron Collider
(HL-LHC)~\cite{HL-LHC,1742-6596-515-1-012012,Apollinari:2284929}
will be a major
upgrade of the current LHC~\cite{LHC} supporting the aim of
an in-depth investigation of the properties of the Higgs boson and its
couplings to other particles (Figure~\ref{fig:hl-lhc}). The ATLAS~\cite{ATLAS} and CMS~\cite{CMS}
collaborations will continue to make measurements in the Higgs sector,
while searching for new physics Beyond the Standard Model (BSM). Should
a BSM discovery be made, a full exploration of that physics will be
pursued. Such BSM physics may help shed light on the nature of dark
matter, which we know makes up the majority of gravitational matter in
the universe, but which does not interact via the electromagnetic or
strong nuclear forces~\cite{Mangano2016}.

The LHCb experiment at the LHC~\cite{LHCb} and the Belle II experiment at
KEK~\cite{BelleII} study various aspects of heavy
flavour physics (b- and c-quark, and tau-lepton physics), where quantum
influences of very high mass particles manifest themselves in lower
energy phenomena. Their primary goal is to look for BSM physics, either
by studying CP violation (that is, asymmetries in the behaviour of
particles and their corresponding antiparticles) or modifications in
rate or angular distributions in rare heavy-flavour decays. Current
manifestations of such asymmetries do not explain why our universe is so
matter dominated. These flavour physics programmes are related to BSM
searches through effective field theory, and powerful constraints on new
physics keep coming from such studies.

\begin{figure*}
\vspace*{0.3cm}
    \centering
    \includegraphics[width=0.94\textwidth]{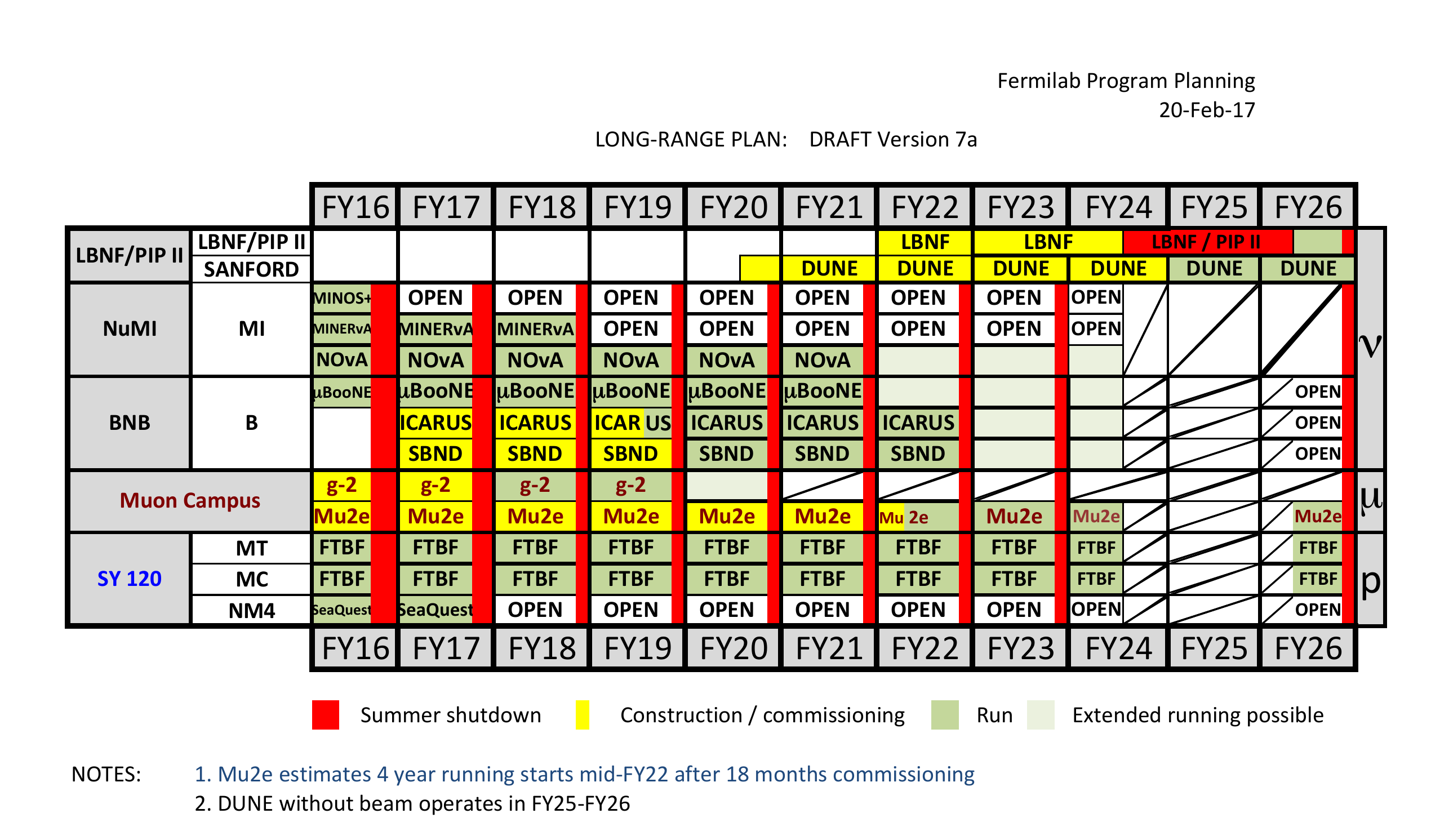}
    \caption{Run schedule for the Fermilab facility until 2026~\cite{fnal-schedule}.}
    \label{fig:fnal-if}
\end{figure*}

The study of neutrinos, their mass and oscillations, can also shed light
on matter-antimatter asymmetry. The DUNE experiment will provide a huge
improvement in our ability to probe neutrino physics, detecting
neutrinos from the Long Baseline Neutrino Facility at Fermilab, as well
as linking to astro-particle physics programmes, in particular through
the potential detection of supernovas and relic neutrinos. An overview
of the experimental programme scheduled at the Fermilab facility is
given in Figure~\ref{fig:fnal-if}.

In the study of the early universe immediately after the Big Bang, it is
critical to understand the phase transition between the highly
compressed quark-gluon plasma and the nuclear matter in the universe
today. The ALICE experiment at the LHC~\cite{ALICE} and
the CBM~\cite{CBM} and PANDA~\cite{PANDA} experiments at the Facility for Antiproton and Ion
Research (FAIR) are specifically designed to probe this aspect of
nuclear and particle physics. In addition ATLAS, CMS and LHCb all
contribute to the LHC heavy-ion programme.

These experimental programmes require large investments in detector
hardware, either to build new facilities and experiments (e.g., FAIR and DUNE) or to upgrade existing ones (HL-LHC, Belle II). Similarly, they
require commensurate investment in the research and development
necessary to deploy software to acquire, manage, process, and analyse
the data recorded.

For the HL-LHC, which is scheduled to begin taking data in 2026 (Figure~\ref{fig:hl-lhc}) and to run into the 2030s, some 30 times more data than
the LHC has currently produced will be collected by ATLAS and CMS. As
the total amount of LHC data already collected is close to an exabyte,
it is clear that the problems to be solved require approaches beyond
simply scaling current solutions, assuming Moore's Law and more or less
constant operational budgets. The nature of computing hardware
(processors, storage, networks) is evolving with radically new
paradigms, the quantity of data to be processed is increasing
dramatically, its complexity is increasing, and more sophisticated
analyses will be required to maximise physics yield. Developing and
deploying sustainable software for future and upgraded experiments,
given these constraints, is both a technical and a social challenge, as
detailed in this paper. An important message of this report is that a
``software upgrade'' is needed to run in parallel with the hardware
upgrades planned for the HL-LHC in order to take full advantage of these hardware upgrades and to complete the HL-LHC physics programme.

In planning for the HL-LHC in particular, it is critical that all of the
collaborating stakeholders agree on the software goals and priorities,
and that the efforts complement each other. In this spirit, the HEP
Software Foundation (HSF) began a planning exercise in late 2016 to
prepare a Community White Paper (CWP)~\cite{HSF2017} at the behest of the Worldwide LHC
Computing Grid (WLCG) project~\cite{WLCG2016}. The role of the HSF is to
facilitate coordination and common efforts in HEP software and computing
internationally and to provide a structure for the community to set
goals and priorities for future work. The objective of the CWP is to
provide a roadmap for software R\&D in preparation for the HL-LHC and
for other HEP experiments on a similar timescale, which would identify
and prioritise the software research and development investments
required:

\begin{itemize}
\item
  to achieve improvements in software efficiency, scalability and
  performance, and to make use of advances in CPU, storage and network
  technologies in order to cope with the challenges ahead;
\item
  to enable new approaches to computing and software that can radically
  extend the physics reach of the detectors;
\item
  to ensure the long-term sustainability of the software through the
  lifetime of the HL- LHC;
\item
  to ensure data and knowledge preservation beyond the lifetime of
  individual experiments;
\item
  to attract the required new expertise by offering appropriate career
  recognition to physicists specialising in software development and by
  an effective training effort to target all software contributors in the
  community.
\end{itemize}

The CWP process, organised by the HSF with the participation of the LHC
experiments and the wider HEP software and computing community, began
with a kick-off workshop at the San Diego Supercomputer Centre (SDSC),
USA, in January 2017 and concluded after a final workshop in June 2017 at
the Laboratoire d'Annecy de Physique des Particules (LAPP), France, with
a large number of intermediate topical workshops and meetings
(Appendix \ref{appendix-a---list-of-workshops}). The
entire CWP process involved an estimated 250 participants.

To reach more widely than the LHC experiments, specific contact was made
with individuals with software and computing responsibilities in the
Fermilab muon and neutrino experiments, Belle II, the Linear Collider
community, as well as various national computing organisations. The CWP
process was able to build on all the links established since the
inception of the HSF in 2014.

Working groups were established on various topics which were expected to
be important parts of the HL-LHC roadmap: \emph{Careers, Staffing and
Training; Conditions Database; Data Organisation, Management and Access;
Data Analysis and Interpretation; Data and Software Preservation;
Detector Simulation; Data-Flow Processing Frameworks; Facilities and
Distributed Computing; Machine Learning; Physics Generators; Security;
Software Development, Deployment and Validation/Verification; Software
Trigger and Event Reconstruction;} and \emph{Visualisation.} The work of
each working group is summarised in this document.

This document is the result of the CWP process. Investing in the roadmap
outlined here will be fruitful for the whole of the HEP programme and
may also benefit other projects with similar technical challenges,
particularly in astrophysics, e.g., the Square Kilometre Array (SKA)~\cite{SKA},
the Cherenkov Telescope Array (CTA)~\cite{CTA} and the Large Synoptic Survey
Telescope (LSST)~\cite{LSST}.

\hypertarget{software-and-computing-challenges}{%
\section{Software and Computing
Challenges}\label{software-and-computing-challenges}}

Run 2 for the LHC started in 2015 and delivered a proton-proton
collision energy of
\href{http://home.cern/about/engineering/restarting-lhc-why-13-tev}{13
TeV}. By the end of LHC Run 2 in 2018, it is expected that about 150
fb\textsuperscript{-1} of physics data will have been collected by both
ATLAS and CMS. Together with ALICE and LHCb, the total size of LHC data
storage pledged by sites for the year 2017 is around 1 exabyte, as shown
in Table~\ref{tab:crsg2017} from the LHC's Computing Resource Scrutiny Group
(CRSG)~\cite{Lucchesi:2284575}. The CPU allocation from the CRSG for 2017 to each
experiment is also shown.

\begin{table*}
    \centering
\begin{tabular}{lp{0.15\textwidth}p{0.15\textwidth}p{0.15\textwidth}p{0.15\textwidth}}
\hline
\textbf{Experiment} & \textbf{2017 Disk Pledges (PB)} & \textbf{2017 Tape Pledges (PB)} & \textbf{Total Disk and Tape Pledges (PB)} & \textbf{2017 CPU Pledges (kHS06)}\\
\hline
ALICE & 67 & 68 & 138 & 807\\
ATLAS & 172 & 251 & 423 & 2194\\
CMS & 123 & 204 & 327 & 1729\\
LHCb & 35 & 67 & 102 & 413\\
\textbf{Total} & \textbf{400} & \textbf{591} & \textbf{990} &
\textbf{5143}\\
\hline
\end{tabular}
    \caption{Resources pledged by WLCG sites to the 4 LHC experiments for
the year 2017 as described at the September 2017 session of the
Computing Resources Scrutiny Group (CRSG).}
    \label{tab:crsg2017}
\end{table*}

Using an approximate conversion from HS06~\cite{HS06} to CPU cores of 10
means that LHC computing in 2017 is supported by about 500k CPU cores.
These resources are deployed ubiquitously, from close to the experiments
themselves at CERN to a worldwide distributed computing infrastructure,
the WLCG~\cite{WLCG}. Each experiment has developed its own workflow management and
data management software to manage its share of WLCG resources.

In order to process the data, the 4 largest LHC experiments have written
tens of millions of lines of program code over the last 15 years
\cite{1742-6596-898-7-072013,Elmer2014,OHAliRoot,OHAliPhysics}. This
has involved contributions from thousands of physicists and many
computing professionals, encompassing a wide range of skills and
abilities. The majority of this code was written for a single
architecture (x86\_64) and with a serial processing model in mind. There
is considerable anxiety in the experiments that much of this software is
not sustainable, with the original authors no longer in the field and
much of the code itself in a poorly maintained state, ill-documented, and
lacking tests. This code, which is largely experiment-specific, manages
the entire experiment data flow, including data acquisition, high-level
triggering, calibration and alignment, simulation, reconstruction (of
both real and simulated data), visualisation, and final data analysis.

HEP experiments are typically served with a large set of integrated and
configured common software components, which have been developed either
in-house or externally. Well-known examples include ROOT~\cite{Brun1996}, which is a data analysis toolkit that
also plays a critical role in the implementation of experiments' data
storage systems, and Geant4~\cite{Agostinelli2003}, a
simulation framework through which most detector simulation is achieved.
Other packages provide tools for supporting the development process;
they include compilers and scripting languages, as well as tools for
integrating, building, testing, and generating documentation. Physics
simulation is supported by a wide range of event generators provided by
the theory community (PYTHIA~\cite{PYTHIA}, SHERPA~\cite{Gleisberg:2008ta}, ALPGEN~\cite{Mangano:565290}, MADGRAPH~\cite{MADGRAPH}, HERWIG~\cite{HERWIG},
amongst many others). There is also code developed to support the
computing infrastructure itself, such as the CVMFS distributed caching
filesystem~\cite{1742-6596-331-4-042003}, the Frontier database caching
mechanism~\cite{Frontier}, the XRootD file access software~\cite{XRootD} and a
number of storage systems (dCache, DPM, EOS). This list of packages is 
by no means exhaustive, but illustrates the range of software employed
and its critical role in almost every aspect of the programme.

Already in Run 3 LHCb will process more than 40 times the number of
collisions that it does today, and ALICE will read out Pb-Pb collisions
continuously at 50 kHz. The upgrade to the HL-LHC for Run 4 then
produces a step change for ATLAS and CMS. The beam intensity will rise
substantially, giving bunch crossings where the number of discrete
proton-proton interactions (pileup) will rise to about 200, from about 60
today. This has important consequences for the operation of the
detectors and for the performance of the reconstruction software. The
two experiments will upgrade their trigger systems to record 5-10 times
as many events as they do today. It is anticipated that HL-LHC will
deliver about 300 fb\textsuperscript{-1} of data each year.

\begin{figure*}
    \centering
    \begin{subfigure}[b]{0.47\textwidth}
        \includegraphics[width=\textwidth]{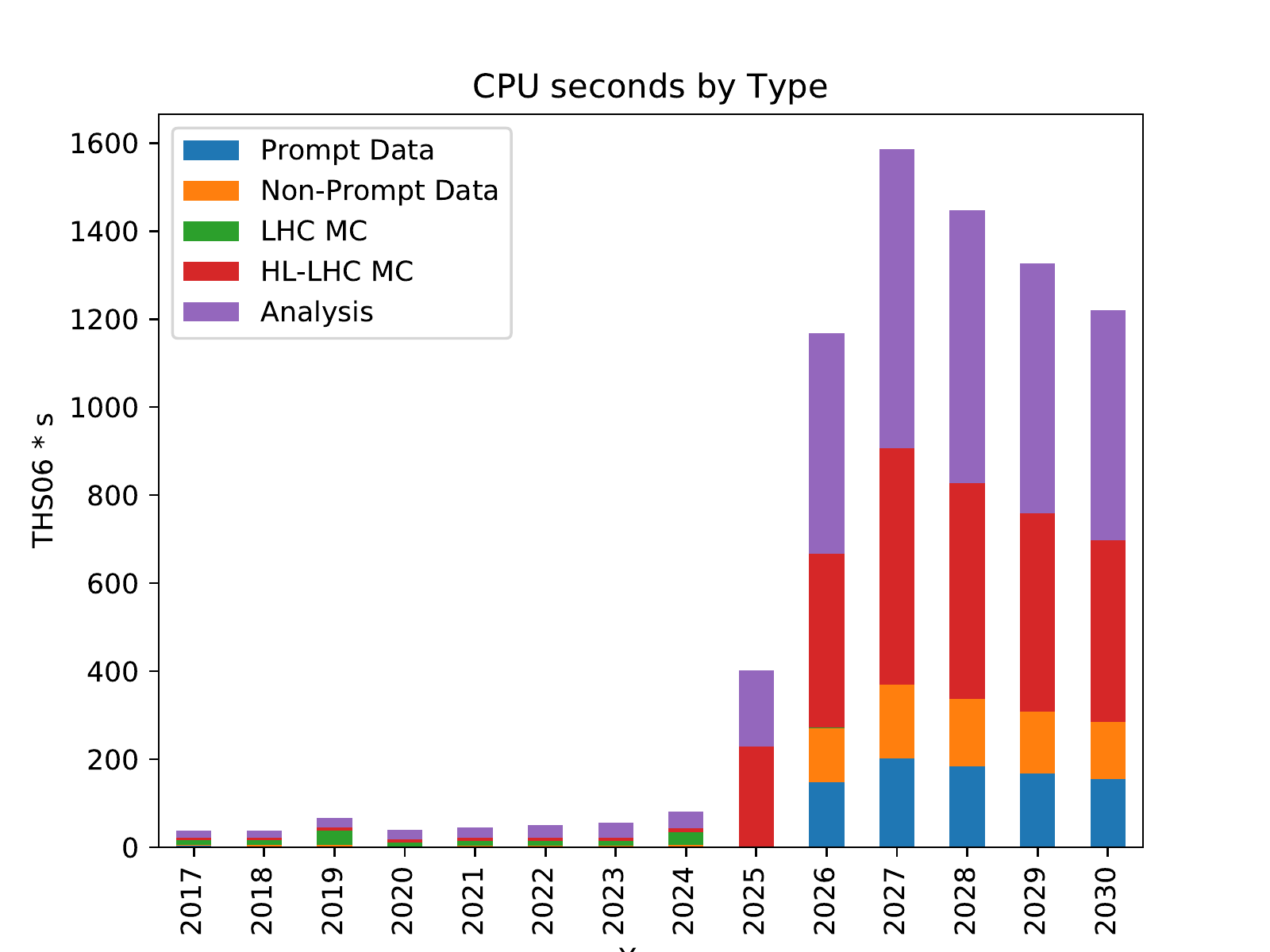}
        \caption{}
        \label{fig:cms-cpu}
    \end{subfigure}
    \quad
    \begin{subfigure}[b]{0.47\textwidth}
        \includegraphics[width=\textwidth]{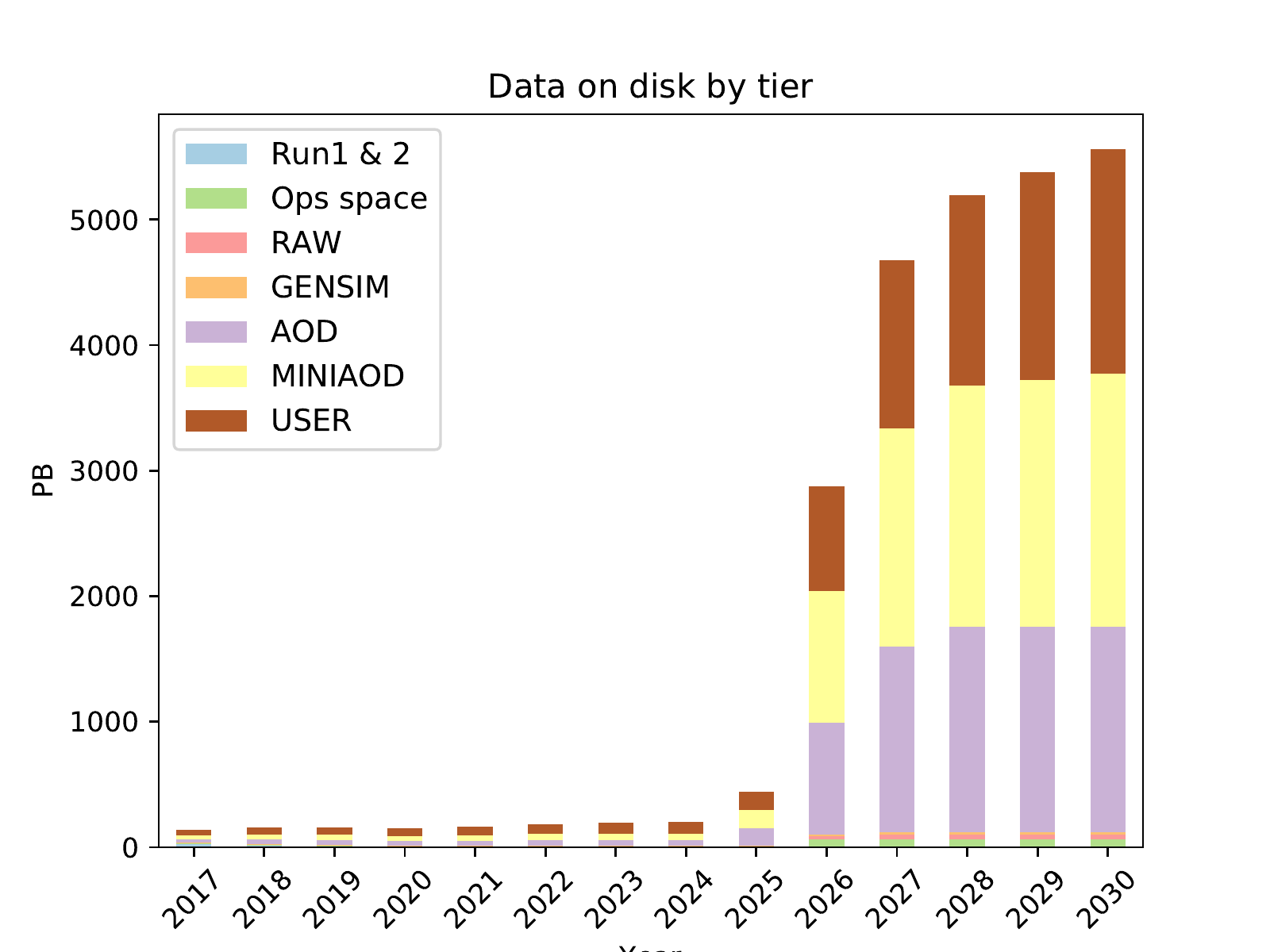}
        \caption{}
        \label{fig:cms-disk}
    \end{subfigure}
    \caption{CMS estimated CPU (\ref{fig:cms-cpu}) and disk space (\ref{fig:cms-disk})
    resources required into the HL-LHC era, using the current computing model
    with parameters projected out for the next 12 years~\cite{1742-6596-1085-2-022006}.}
    \label{fig:cms-resources}
\end{figure*}

\begin{figure*}
    \centering
    \begin{subfigure}[b]{0.9\textwidth}
    	    \centering
        \includegraphics[width=0.6\textwidth]{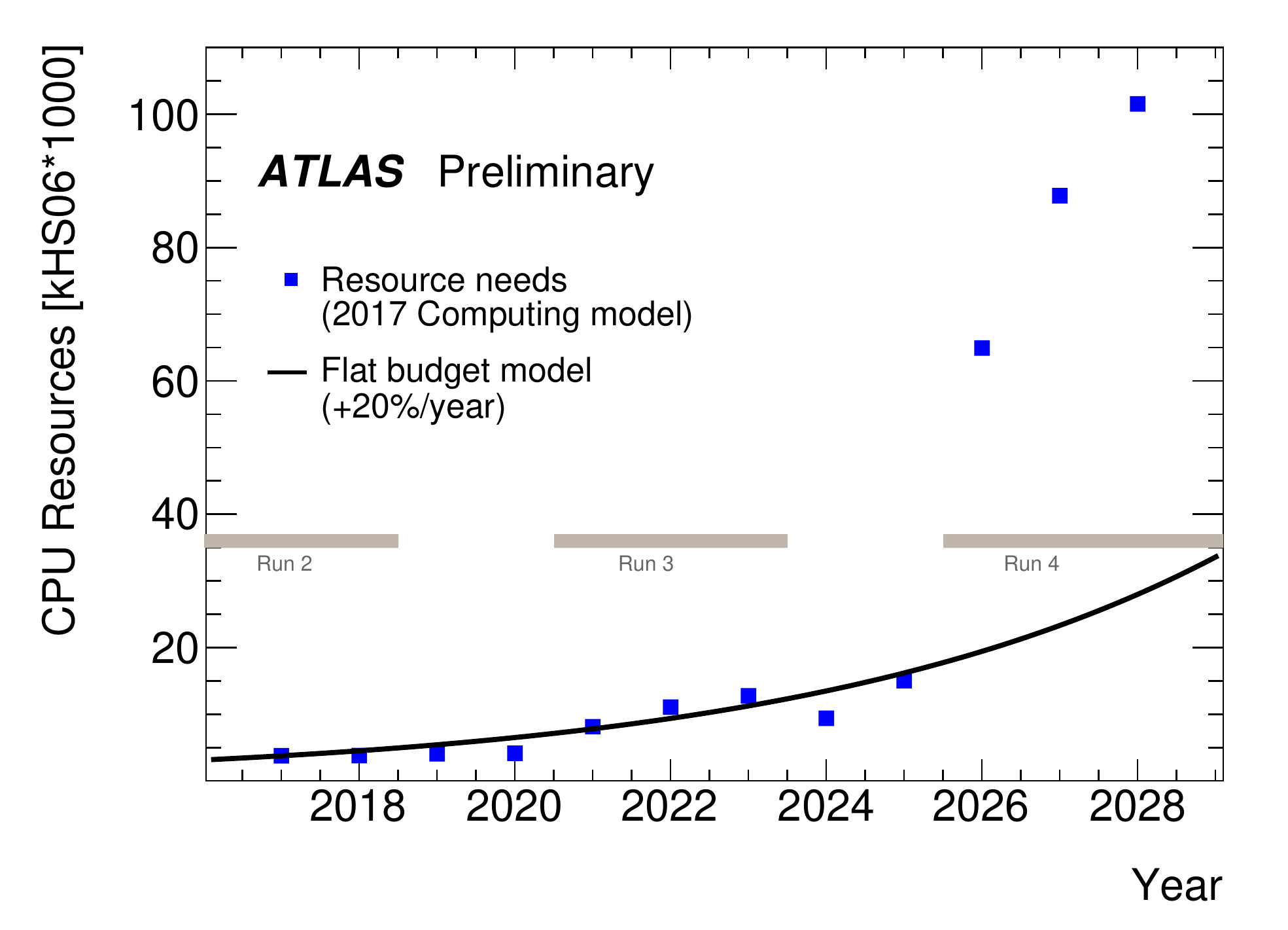}
        \caption{Estimated CPU resources (in kHS06) needed for the years 2018 to 2028 for both data and simulation processing. The blue points are estimates based on the current software performance estimates and using the ATLAS computing model parameters from 2017. The solid line shows the amount of resources expected to be available if a flat funding scenario is assumed, which implies an increase of 20\% per year, based on the current technology trends.}
        \label{fig:atlas-cpu}
    \end{subfigure}
    \quad
    \begin{subfigure}[b]{0.9\textwidth}
    	    \centering
        \includegraphics[width=0.6\textwidth]{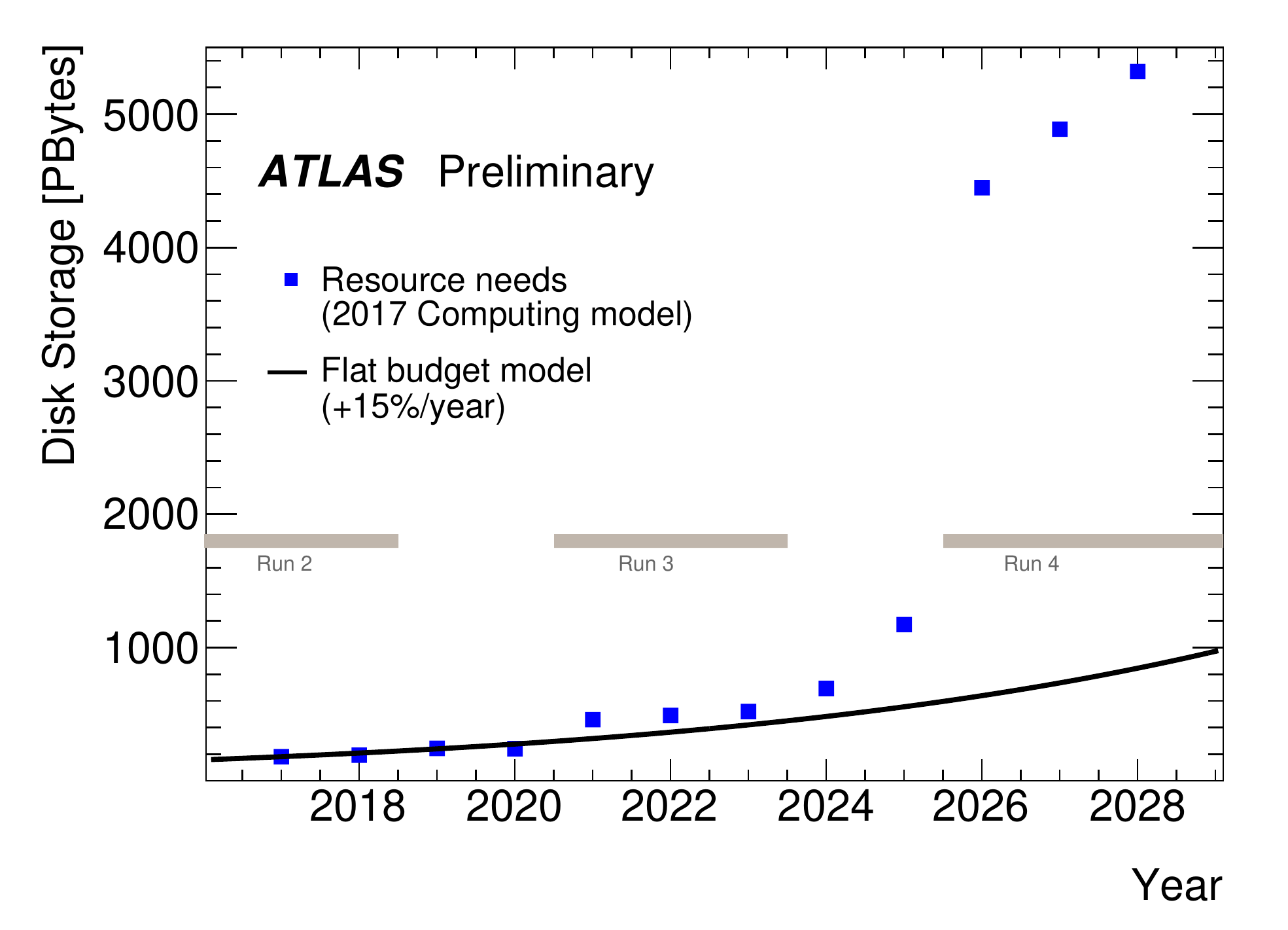}
        \caption{Estimated total disk resources (in PB) needed for the years 2018 to 2028 for both data and simulation processing. The blue points are estimates based on the current event sizes estimates and using the ATLAS computing model parameters from 2017. The solid line shows the amount of resources expected to be available if a flat funding scenario is assumed, which implies an increase of 15\% per year, based on the current technology trends.}
        \label{fig:atlas-disk}
    \end{subfigure}
    \caption{ATLAS resources required into the HL-LHC era, using the current computing model
    and software performance.\cite{ATLAS-HL-LHC-Plots}}
    \label{fig:atlas-resources}
\end{figure*}

The steep rise in resources that are then required to manage this data
can be estimated from an extrapolation of the Run 2 computing model and
is shown in Figures~\ref{fig:cms-resources} and~\ref{fig:atlas-resources}.

In general, it can be said that the amount of data that experiments can
collect and process in the future will be limited by affordable software
and computing, and therefore the physics reach during HL-LHC will be
limited by how efficiently these resources can be used.

The ATLAS numbers, in Figure~\ref{fig:atlas-resources}, are particularly interesting as they
estimate the resources that will be available to the experiment if a
flat funding profile is maintained, taking into account the expected
technology improvements given current trends~\cite{Panzer2017}. As can be
seen, the shortfall between needs and bare technology gains is
considerable: a factor 4 in CPU and a factor 7 in disk in 2027.

While the density of transistors on silicon continues to increase
following Moore's Law (albeit more slowly than in the past), power
density constraints have limited the clock speed of processors for more
than a decade~\cite{NAP12980}. This has effectively stalled any progress in the
processing capacity of a single CPU core. Instead, increases in
potential processing capacity come from increases in the core count of
CPUs and wide CPU registers. Alternative processing architectures have
become more commonplace. These range from the many-core architecture
based on standard x86\_64 cores to numerous alternatives such as GPUs.
For GPUs, the processing model is very different~\cite{Cook:2012:CPD:2430671}, allowing a much
greater fraction of the die to be dedicated to arithmetic calculations,
but at a price in programming difficulty and memory handling for the
developer that tends to be specific to each processor generation.
Further developments may even see the use of FPGAs for more
general-purpose tasks. Fully exploiting these evolutions requires a
shift in programming model to one based on \emph{concurrency}.

Even with the throttling of clock speed to limit power consumption,
power remains a major issue. Low power architectures are in huge demand.
At one level this might challenge the dominance of x86\_64 by simply
replacing it with, for example, AArch64 devices that may achieve lower power
costs for the scale of HEP computing needs than Intel has achieved with its
Xeon architecture~\cite{1742-6596-608-1-012001,1742-6596-513-5-052014,2014arXiv1404.6929A}.
More extreme is an architecture that would see
specialised processing units dedicated to particular tasks, but with
possibly large parts of the device switched off most of the time,
so-called dark silicon~\cite{martin2014,IntelCSA}.

Limitations in affordable storage also pose a major challenge, as does
the I/O rates of higher capacity hard disks. Network
bandwidth will probably continue to increase at the required level, but
the ability to use it efficiently will need a closer integration with
applications. This will require software developments to
support distributed computing (data and workload management, software
distribution and data access) and an increasing awareness of the
extremely hierarchical view of data, from long latency tape access and
medium-latency network access through to the CPU memory hierarchy.

Taking advantage of these new architectures and programming paradigms
will be critical for HEP to increase the ability of our code to deliver
physics results efficiently, and to meet the processing challenges of
the future. Some of this work will be focused on re-optimised
implementations of existing algorithms. This will be complicated by the
fact that much of our code is written for the much simpler model of
serial processing, and without the software engineering needed for
sustainability. Proper support for taking advantage of concurrent
programming techniques, such as vectorisation and thread-based
programming, through frameworks and libraries, will be essential, as the
majority of the code will still be written by physicists. Other
approaches should examine new algorithms and techniques, including
highly parallelised code that can run on GPUs or the use of machine
learning techniques to replace computationally expensive pieces of
simulation or pattern recognition. The ensemble of computing work that
is needed by the experiments must remain sufficiently flexible to take
advantage of different architectures that will provide computing to HEP
in the future. The use of high performance computing sites and
commercial cloud providers will very likely be a requirement for the
community and will bring particular constraints and demand flexibility.

These technical challenges are accompanied by significant human
challenges. Software is written by many people in the collaborations,
with varying levels of expertise, from a few experts with precious
skills to novice coders. This implies organising training in effective
coding techniques and providing excellent documentation, examples and
support. Although it is inevitable that some developments will remain
within the scope of a single experiment, tackling software problems
coherently as a community will be critical to achieving success in the
future. This will range from sharing knowledge of techniques and best
practice to establishing common libraries and projects that will provide
generic solutions to the community. Writing code that supports a wider
subset of the community than just a single experiment will almost
certainly be mandated upon HEP and presents a greater challenge, but the
potential benefits are huge. Attracting, and retaining, people with the
required skills who can provide leadership is another significant
challenge, since it impacts on the need to give adequate recognition to
physicists who specialise in software development. This is an important
issue that is treated in more detail later in the report.

Particle physics is no longer alone in facing these massive data
challenges. Experiments in other fields, from astronomy~\cite{10.1371/journal.pbio.1002195}
to genomics~\cite{ijms18020412},
will produce huge amounts of data in the future, and will need to
overcome the same challenges that we face, i.e., massive data handling and
efficient scientific programming. Establishing links with these fields
has already started. Additionally, interest from the computing science
community in solving these data challenges exists, and mutually
beneficial relationships would be possible where there are genuine
research problems that are of academic interest to that community and
provide practical solutions to ours. The efficient processing of massive
data volumes is also a challenge faced by industry, in particular the
internet economy, which developed novel and major new technologies under
the banner of \emph{Big Data} that may be applicable to our use cases
\cite{47224,Shanahan:2015:LSD:2783258.2789993,Armbrust:2015:SSR:2723372.2742797}.

Establishing a programme of investment in software for the HEP
community, with a view to ensuring effective and sustainable software
for the coming decades, will be essential to allow us to reap the
physics benefits of the multi-exabyte data to come. It was in recognition of
this fact that the HSF itself was set up and already works to promote
these common projects and community developments~\cite{HSF2015}.

\hypertarget{programme-of-work}{%
\section{Programme of Work}\label{programme-of-work}}

In the following we describe the programme of work being proposed for
the range of topics covered by the CWP working groups. We summarise the
main specific challenges each topic will face, describe current
practices, and propose a number of R\&D tasks that should be undertaken
in order to meet the challenges. R\&D tasks are grouped in two different
timescales: short term (by 2020, in time for the HL-LHC Computing Technical Design Reports of
ATLAS and CMS) and longer-term actions (by 2022, to be ready for testing
or deployment during LHC Run 3).

\hypertarget{physics-generators}{%
\subsection{Physics Generators}\label{physics-generators}}

\subsubsection*{Scope and Challenges}

Monte-Carlo event generators are a vital part of modern particle
physics, providing a key component of the understanding and
interpretation of experiment data. Collider experiments have a need for
theoretical QCD predictions at very high precision. Already in LHC Run
2, experimental uncertainties for many analyses are at the same level as,
or lower than, those from theory. Many analyses have irreducible
QCD-induced backgrounds, where statistical extrapolation into the signal
region can only come from theory calculations. With future experiment
and machine upgrades, as well as reanalysis of current data, measured
uncertainties will shrink even further, and this will increase the need
to reduce the corresponding errors from theory.

Increasing accuracy will compel the use of higher-order perturbation
theory generators with challenging computational demands. Generating
Monte Carlo events using leading order (LO) generators is only a small
part of the overall computing requirements for HEP experiments.
Next-to-leading order (NLO) event generation, used more during LHC Run
2, is already using significant resources. Higher accuracy
theoretical cross sections calculated at next-to-next-to-leading (NNLO),
already important in some Run 2 analyses, are not widely used
because of computational cost. By HL-LHC the use of NNLO event
generation will be more widely required, so these obstacles to their
adoption must be overcome.
Increasing the order of the generators increases greatly the complexity
of the phase space integration required to calculate the appropriate QCD
matrix elements. The difficulty of this integration arises from the need
to have sufficient coverage in a high-dimensional space (10-15
dimensions, with numerous local maxima); the appearance of negative event
weights; and the fact that many terms in the integration cancel, so that
a very high degree of accuracy of each term is required. Memory demands
for generators have generally been low and initialisation times have
been fast, but an increase in order means that memory consumption
becomes important and initialisation times can become a significant
fraction of the job's run time.

For HEP experiments, in many cases, meaningful predictions can only be
obtained by combining higher-order perturbative calculations with parton
showers. This procedure is also needed as high-multiplicity final states
become more interesting at higher luminosities and event rates. Matching
(N)NLO fixed-order calculations to parton shower algorithms can have a
very low efficiency, and increases further the computational load needed
to generate the necessary number of particle-level events. In addition,
many of the current models for the combination of parton-level event
generators and parton shower codes are incompatible with requirements
for concurrency on modern architectures. It is a major challenge to
ensure that this software can run efficiently on next generation
hardware and software systems.

Developments in generator software are mainly done by the HEP theory
community. Theorists typically derive career recognition and advancement from
making contributions to theory itself, rather than by making improvements
to the computational efficiency of generators \emph{per se}. So,
improving the computational efficiency of event generators, and allowing
them to run effectively on resources such as high performance computing
facilities (HPCs), will mean engaging with experts in computational
optimisation who can work with the theorists who develop generators.

The challenge in the next decade is to advance the theory and practical
implementation of event generators to support the needs of future
experiments, reaching a new level of theory precision and recognising
the demands for computation and computational efficiency that this will
bring.

\subsubsection*{Current Practice}

Extensive use of LO generators and parton shower algorithms are still
made by most HEP experiments. Each experiment has its own simulation
needs, but for the LHC experiments tens of billions of generated events
are now used each year for Monte Carlo simulations. During LHC Run 2 more and
more NLO generators were used, because of their increased theoretical
precision and stability. The raw computational complexity of NLO
amplitudes, combined with many-body phase-space evaluations and the
inefficiencies of the matching process, leads to a potentially much-increased CPU
budget for physics event simulation for ATLAS and CMS.

The use of NLO generators by the experiments today is also limited because of
the way the generators are implemented, producing significant numbers of
negative event weights. This means that the total number of events the
experiments need to generate, simulate, and reconstruct can be many
times larger for NLO than for LO samples. At the same time, the
experiments budget only a similar number of Monte Carlo simulation events as from the real data. Having large NLO samples is thus not consistent with
existing computing budgets until a different scheme is developed that
does not depend on negative event weights or produces them only at a
significantly reduced rate.

While most event generation is run on ``standard'' grid resources,
effort is ongoing to run more demanding tasks on HPC resources, e.g.,
W-boson + 5-jet events at the Argonne Mira HPC). However, scaling for
efficient running on some of the existing HPC resources is not trivial
and requires effort.

Standard HEP libraries such as LHAPDF~\cite{LHAPDF}, HepMC\cite{HepMC},
and Rivet~\cite{Rivet} are used by the generators for integration into
the experiments' event generation workflows. These require extensions
and sustained maintenance that should be considered a shared
responsibility of the theoretical and experimental communities in the
context of large-scale experiments. In practice, however, it has been
difficult to achieve the level of support that is really needed as there
has been a lack of recognition for this work. To help improve the
capabilities and performance of generators as used by the experimental
HEP programme, and to foster interaction between the communities, the
MCnet~\cite{MCnet} short-term studentship programme has been very useful.
Interested experimental PhD students can join a generator group for
several months to work on improving a physics aspect of the simulation
that is relevant to their work, or to improve the integration of the
generator into an experimental framework.

\subsubsection*{Research and Development Programme}

As the Monte Carlo projects are funded mainly to develop theoretical
improvements, and not mainly as ``suppliers'' to the experimental HEP
programme, any strong requests towards efficiency improvements from the
experimental community would need to be backed up by plausible avenues
of support that can fund contributions from software engineers with the
correct technical skills in software optimisation to work within the
generator author teams.

In a similar way to the MCnet studentships, a matchmaking scheme could
focus on the software engineering side, and transfer some of the
expertise available in the experiments and facilities teams to the
generator projects. Sustainable improvements are unlikely to be
delivered by graduate students ``learning on the job'' and then leaving
after a few months, so meeting the requirement of transferring technical
expertise and effort will likely require placements for experienced
optimisation specialists and a medium- to long-term connection to the
generator project.

HEP experiments, which are now managed by very large collaborations including many
technical experts, can also play a key role in sustaining a healthy
relationship between theory and experiment software. Effort to work on
common tools that benefit both the experiment itself and the wider
community would provide shared value that justifies direct investment
from the stakeholders. This model would also be
beneficial for core HEP tools like LHAPDF, HepMC and Rivet, where future
improvements have no theoretical physics interest anymore, putting them
in a similar situation to generator performance improvements. One
structural issue blocking such a mode of operation is that some
experiments do not currently recognise contributions to external
projects as experiment service work --- a situation deserving of review
in areas where external software tools are critical to experiment
success.

In the following we describe specific areas of R\&D for event generation
up to 2022 and beyond.

\begin{itemize}
\item
  The development of new and improved theoretical algorithms provides
  the largest potential for improving event generators. While it is not
  guaranteed that simply increasing the effort dedicated to this task
  will bring about the desired result, the long-term support of event
  generator development, and the creation of career opportunities in
  this research area, are critical given the commitment to experiments
  on multi-decade scales.
\item
  Expand development in reweighting event samples, where new physics
  signatures can be explored by updating the partonic weights according
  to new matrix elements. It is necessary that the phase space for the
  updated model be a subset of the original one, which is an important
  limitation. The procedure is more complex at NLO and can require
  additional information to be stored in the event files to properly
  reweight in different cases. Overcoming the technical issues from
  utilising negative event weights is crucial. Nevertheless, the method
  can be powerful in many cases, and would hugely reduce the time needed
  for the generation of BSM samples.
\item
  At a more technical level, concurrency is an avenue that has yet to be
  explored in depth for event generation. As the calculation of matrix
  elements requires VEGAS-style integration, this work would be helped
  by the development of a new Monte-Carlo integrator. For multi-particle
  interactions, factorising the full phase space integration into lower
  dimensional integrals would be a powerful method of parallelising,
  while the interference between different Feynman graphs can be handled
  with known techniques.
\item
  For many widely used generators, basic problems of concurrency and
  thread hostility need to be tackled, to make these packages suitable
  for efficient large scale use on modern processors and within modern
  HEP software frameworks. Providing appropriate common tools for
  interfacing, benchmarking and optimising multithreaded code would
  allow expertise to be shared effectively~\cite{HSF-CWP-2017-13}.
\item
  In most generators, parallelism was added post-facto, which leads to
  scaling problems when the level of parallelism becomes very large,
  e.g., on HPC machines. These HPC machines will be part of the
  computing resource pool used by HEP, so solving scaling issues on
  these resources for event generation is important, particularly as the
  smaller generator code bases can make porting to non-x86\_64
  architectures more tractable. The problem of long and inefficient
  initialisation when a job utilises hundreds or thousands of cores on
  an HPC needs to be tackled. While the memory consumption of event
  generators is generally modest, the generation of tree-level
  contributions to high multiplicity final states can use significant
  memory, and gains would be expected from optimising here.
\item
  An underexplored avenue is the efficiency of event generation as
  used by the experiments. An increasingly common usage is to generate
  very large inclusive event samples, which are filtered on event
  final-state criteria to decide which events are to be retained and
  passed onto detector simulation and reconstruction. This naturally
  introduces a large waste of very CPU-expensive event generation, which
  could be reduced by developing filtering tools within the generators
  themselves, designed for compatibility with the experiments'
  requirements. A particularly wasteful example is where events are
  separated into orthogonal subsamples by filtering, in which case the
  same large inclusive sample is generated many times, with each stream
  filtering the events into a different group: allowing a single
  inclusive event generation to be filtered into several orthogonal
  output streams would improve efficiency.
\end{itemize}

\hypertarget{detector-simulation}{%
\subsection{Detector Simulation}\label{detector-simulation}}

\subsubsection*{Scope and Challenges}

For all its success so far, the challenges faced by the HEP field in the
simulation domain are daunting. During the first two runs, the LHC
experiments produced, reconstructed, stored, transferred, and analysed
tens of billions of simulated events. This effort required more than
half of the total computing resources allocated to the experiments. As
part of the HL-LHC physics programme, the upgraded
experiments expect to collect 150 times more data than in Run 1; demand
for larger simulation samples to satisfy analysis needs will grow
accordingly. In addition, simulation tools have to serve diverse
communities, including accelerator-based particle physics research
utilising proton-proton colliders, neutrino, dark matter, and muon
experiments, as well as the cosmic frontier. The complex detectors of
the future, with different module- or cell-level shapes, finer
segmentation, and novel materials and detection techniques, require
additional features in geometry tools and bring new demands on physics
coverage and accuracy within the constraints of the available computing
budget. The diversification of the physics programmes also requires new
and improved physics models. More extensive use of Fast Simulation is a
potential solution, under the assumption that it is possible to improve
time performance without an unacceptable loss of physics accuracy.

The gains that can be made by speeding up critical elements of the
Geant4 simulation toolkit can be leveraged for all applications that use
it, and it is therefore well worth the investment in effort needed to
achieve it. The main challenges to be addressed if the required physics
and software performance goals are to be achieved are:

\begin{itemize}
\item
  Review the implementations of physics models, including the
  assumptions, approximations, and
  limitations. In the best cases this can achieve higher precision and
  improve runtime performance through code modernisation~\cite{Novak2018}.
  The extension of the validity of models up to energies of 100 TeV is foreseen
  for the Future Circular Collider (FCC) project~\cite{FCC} and provides
  a good opportunity for this modernisation.
\item
  Redesigning, developing, and commissioning detector simulation
  toolkits to be more efficient when executed on current vector CPUs and
  emerging new architectures, including GPUs, where use of SIMD
  vectorisation is vital; this includes porting and optimising the
  experiments' simulation applications to allow exploitation of large
  HPC facilities.
\item
  Exploring different Fast Simulation options, where the full detector
  simulation is replaced, in whole or in part, by computationally
  efficient techniques. An area of investigation is common frameworks
  for fast tuning and validation.
\item
  Developing, improving and optimising geometry tools that can be shared
  am\-ong experiments to make the modeling of complex detectors
  computationally more efficient, modular, and transparent.
\item
  Developing techniques for background modeling, including contributions
  of multiple hard interactions overlapping the event of interest in
  collider experiments (pileup).
\item
  Revisiting digitisation algorithms to improve performance and
  exploring opportunities for code sharing among experiments.
\end{itemize}

It is obviously of critical importance that the whole community of
scientists working in the simulation domain continue to work together in
as efficient a way as possible in order to deliver the required
improvements. Very specific expertise is required across all simulation
domains, such as physics modeling, tracking through complex geometries
and magnetic fields, and building realistic applications that accurately
simulate highly complex detectors. Continuous support is needed to
recruit, train, and retain people with a unique set of skills needed to
guarantee the development, maintenance, and support of simulation codes
over the long timeframes foreseen in the HEP experimental programme.

\subsubsection*{Current Practices}

The Geant4 detector simulation toolkit is at the core of simulation in
almost every HEP experiment. Its continuous development, maintenance,
and support for the experiments is of vital importance. New or refined
functionality in physics coverage and accuracy continues to be delivered
in the ongoing development programme and software
performance improvements are introduced whenever possible.

Physics models are a critical part of the detector simulation, and are
continuously being reviewed, and in some cases reimplemented, in order
to improve accuracy and software performance. Electromagnetic (EM)
transport simulation is challenging as it occupies a large part of the
computing resources used in full detector simulation. Significant
efforts have been made in the recent past to better describe the
simulation of electromagnetic shower shapes, in particular to model the
$H \to \gamma \gamma$ signal and background accurately at the LHC. This
effort is being continued with an emphasis on reviewing the models'
assumptions, approximations, and limitations, especially at very high
energy, with a view to improving their respective software
implementations. In addition, a new ``theory-based'' model
(Goudsmit-Saunderson), for describing the \emph{multiple scattering} of
electrons and positrons, has been developed that has been demonstrated
to outperform, in terms of physics accuracy and speed, the current
models in Geant4. The models used to describe the \emph{bremsstrahlung}
process have also been reviewed, and recently an improved theoretical
description of the Landau-Pomeranchuk-Migdal effect was introduced
that plays a significant role at high energies. Theoretical review of
all electromagnetic models, including those of hadrons and ions, is
therefore of high priority both for HL-LHC and for FCC studies.

Hadronic physics simulation covers purely hadronic interactions. It is
not possible for a single model to describe all the physics encountered
in a simulation due to the large energy range that needs to be covered
and the simplified approximations that are used to overcome the
difficulty of solving the full theory (QCD). Currently the most-used
reference physics list for high energy and space applications is
FTFP\_BERT. It uses the Geant4 Bertini cascade for hadron--nucleus
interactions from 0 to 12 GeV incident hadron energy and the FTF parton
string model for hadron--nucleus interactions from 3 GeV upwards.
QGSP\_BERT is a popular alternative which replaces the FTF model with
the QGS model over the high energy range. The existence of more than one
model (for each energy range) is very valuable in order to be able to
determine the systematics effects related to the approximations used.
The use of highly granular calorimeters, such as the ones being designed
by the CALICE collaboration for future linear colliders, allows a
detailed validation of the development of hadronic showers with
test-beam data. Preliminary results suggest that the lateral profiles of
Geant4 hadronic showers are too narrow. Comparisons with LHC test-beam
data have shown that a fundamental ingredient for improving the
description of the lateral development of showers is the use of
intermediate and low energy models that can describe the cascading of
hadrons in nuclear matter. Additional work is currently being invested
in the further improvement of the QGS model, which is a more
theory-based approach than the phenomenological FTF model, and therefore
offers better confidence at high energies, up to a few TeV. This again
is a large endeavour and requires continuous effort over a long time.

The Geant4 collaboration is working closely with user communities to
enrich the physics models' validation system with data acquired during
physics runs and test beam campaigns. In producing new models of physics
interactions and improving the fidelity of the models that exist, it is
absolutely imperative that high-quality data are available. Simulation
model tuning often relies on test beam data, and a program to improve
the library of available data could be invaluable to the community. Such
data would ideally include both thin-target test beams for improving
interaction models and calorimeter targets for improving shower models.
This data could potentially be used for directly tuning Fast Simulation
models as well.

There are specific challenges associated with the Intensity Frontier
experimental programme, in particular simulation of the beamline and the
neutrino flux. Neutrino experiments rely heavily on detector simulations
to reconstruct neutrino energy, which requires accurate modelling of
energy deposition by a variety of particles across a range of energies.
Muon experiments such as Muon g-2 and Mu2e also face large simulation
challenges; since they are searching for extremely rare effects, they
must grapple with very low signal to background ratios and the modeling
of low cross-section background processes. Additionally, the size of the
computational problem is a serious challenge, as large simulation runs
are required to adequately sample all relevant areas of experimental
phase space, even when techniques to minimise the required computations
are used. There is also a need to simulate the effects of low energy
neutrons, which requires large computational resources. Geant4 is the
primary simulation toolkit for all of these experiments.

Simulation toolkits do not include effects like charge drift in an
electric field or models of the readout electronics of the experiments.
Instead, these effects are normally taken into account in a separate
step called digitisation. Digitisation is inherently local to a given
sub-detector and often even to a given readout element, so that there
are many opportunities for parallelism in terms of vectorisation and
multiprocessing or multithreading, if the code and the data objects are
designed optimally. Recently, both hardware and software projects have
benefitted from an increased level of sharing among experiments. The
LArSoft Collaboration develops and supports a shared base of physics
software across Liquid Argon (LAr) Time Projection Chamber (TPC)
experiments, which includes providing common digitisation code.
Similarly, an effort exists among the LHC experiments to share code for
modeling radiation damage effects in silicon. As ATLAS and CMS expect
to use similar readout chips in their future trackers, further code
sharing might be possible.

The Geant4 simulation toolkit will also evolve over the next decade to
include contributions from various R\&D projects, as described in the
following section. This is required to ensure the support of experiments
through continuous maintenance and improvement of the Geant4 simulation
toolkit. This is necessary until production versions of potentially
alternative engines, such as those resulting from ongoing R\&D work,
become available, integrated, and validated by experiments. The agreed
ongoing strategy to make this adoption possible is to ensure that new
developments resulting from the R\&D programme can be tested with
realistic prototypes and then be integrated, validated, and deployed in
a timely fashion in Geant4.

\subsubsection*{Research and Development Programme}

To meet the challenge of improving the performance by a large factor, an
ambitious R\&D programme is underway to investigate each component of
the simulation software for the long term. In the following we describe
in detail some of the studies to be performed in the next 3-5 years.

\begin{itemize}
\item
  Particle Transport and Vectorisation: the study of an efficient
  transport of particles (tracks) in groups so as to maximise the
  benefit of using SIMD operations.
\item
  Modularisation: improvement of Geant4 design to allow for a tighter
  and easier integration of single sub-packages of the code into
  experimental frameworks.
\item
  Physics Models: extensions and refinements of the physics algorithms
  to provide new and more performant physics capabilities.
\item
  Other activities: integration of multi-threading capabilities in
  experiment applications; experiment-agnostic software products to cope
  with increased pile\-up, fast simulation, digitisation, and efficient
  production of high-quality ran\-dom numbers.
\end{itemize}

\paragraph{Particle Transport and Vectorisation} One of the most
ambitious elements of the simulation R\&D programme is a new approach to
managing particle transport, which has been introduced by the GeantV
project. The aim is to deliver a multithreaded vectorised transport
engine that has the potential to deliver large performance benefits. Its
main feature is track-level parallelisation, bundling particles with
similar properties from different events to process them in a single
thread. This approach, combined with SIMD vectorisation coding
techniques and improved data locality, is expected to yield significant
speed-ups, which are to be measured in a realistic prototype currently
under development. For the GeantV transport engine to display its best
computing performance, it is necessary to vectorise and optimise the
accompanying modules, including geometry, navigation, and the physics
models. These are developed as independent libraries so that they can
also be used together with the current Geant4 transport engine. Of
course, when used with the current Geant4 they will not expose their
full performance potential, since transport in Geant4 is currently
sequential, but this allows for a preliminary validation and comparison
with the existing implementations. The benefit of this approach is that
new developments can be delivered as soon as they are available. The new
vectorised geometry package (VecGeom), developed as part of GeantV R\&D
and successfully integrated into Geant4, is an example that demonstrated
the benefit of this approach. By the end of 2018 it is intended to have
a proof-of-concept for the new particle transport engine that includes
vectorised EM physics, vectorised magnetic field propagation and that
uses the new vectorised geometry package. This will form a sound basis
for making performance comparisons for simulating EM showers in a
realistic detector.

\begin{itemize}
\item
  2019: the \emph{beta} release of the GeantV transport engine will
  contain enough functionality to build the first real applications.
  This will allow performance to be measured and give sufficient time to
  prepare for HL-LHC running. It should include the use of vectorisation
  in most of the components, including physics modelling for electrons,
  gammas and positrons, whilst still maintaining simulation
  reproducibility, and I/O in a concurrent environment and multi-event
  user data management.
\end{itemize}

\paragraph{Modularisation} Starting from the next release, a
modularisation of Geant4 is being pursued that will allow an easier
integration in experimental frameworks, with the possibility to include
only the Geant4 modules that are actually used. A further use case is
the possibility to use one of the Geant4 components in isolation, e.g.,
to use hadronic interaction modeling without kernel components from a
fast simulation framework. As a first step a preliminary review of
libraries' granularity is being pursued, which will be followed by a
review of intra-library dependencies with the final goal of reducing
their dependencies.

\begin{itemize}
\item
  2019: Redesign of some Geant4 kernel components to improve the
  efficiency of the simulation on HPC systems, starting from improved
  handling of Geant4 \emph{database}s on large core-count systems. A
  review will be made of the multithreading design to be closer to
  task-based frameworks, such as Intel's Thread\-ed Building Blocks
  (TBB)~\cite{TBB}.
\end{itemize}

\paragraph{Physics Models} It is intended to develop
new and extended physics models to
cover extended energy and physics processing of present and future
colliders, Intensity Frontier experiments, and direct dark matter search
experiments. The goal is to extend the missing models (e.g., neutrino
interactions), improve models' physics accuracy and, at the same time,
improve CPU and memory efficiency. The deliverables of these R\&D
efforts include physics modules that produce equivalent quality physics,
and will therefore require extensive validation in realistic
applications.

\begin{itemize}
\item
  2020: Improved implementation of hadronic cascade models for LHC and,
  in particular, Liquid Argon detectors. Improved accuracy models of
  EM interactions of photons and electrons. To address the needs of
  cosmic frontier experiments, optical photon transport must be improved
  and made faster.
\item
  2022: Implementation of EPOS string model for multi-GeV to multi-TeV
  interactions, for FCC detector simulation and systematic studies of
  HL-LHC detectors.
\end{itemize}

\paragraph{Experiment Applications} The experiment applications are
essential for validating the software and physics performance of new
versions of the simulation toolkit. ATLAS and CMS have already started
to integrate Geant4 multithreading capability in their simulation
applications; in the case of CMS the first Full Simulation production in
multithreaded mode was delivered in the autumn of 2017. Specific
milestones are as follows:

\begin{itemize}
\item
  2020: LHC, Neutrino, Dark Matter, and Muon experiments to demonstrate
  the ability to run their detector simulation in multithreaded mode,
  using the improved navigation and electromagnetic physics packages.
  This should bring experiments more accurate physics and improved
  performance.
\item
  2020: Early integration of the beta release of the GeantV transport
  engine in the experiments' simulation, including the implementation of
  the new user interfaces, which will allow the first performance
  measurements and physics validation to be made.
\item
  2022: The availability of a production version of the new track-level
  parallelisation and fully vectorised geometry, navigation, and physics
  libraries will offer the experiments the option to finalise
  integration into their frameworks; intensive work will be needed in
  physics validation and computing performance tests. If successful, the
  new engine could be in production on the timescale of the start of the
  HL-LHC run in 2026.
\end{itemize}

\paragraph{Pileup} Backgrounds to hard-scatter events have many
components including in-time pileup, out-of-time pileup, cavern
background and beam-gas collisions. All of these components can be
simulated, but they present storage and I/O challenges related to the
handling of the large simulated minimum bias samples used to model the
extra interactions. An R\&D programme is needed to study different
approaches to managing these backgrounds within the next 3 years:

\begin{itemize}
\item
  Real zero-bias events can be collected, bypassing any zero
  suppression, and overlaid on the fully simulated hard scatters. This
  approach faces challenges related to the collection of
  non-zero-suppressed samples or the use of suppressed events,
  non-linear effects when adding electronic signals from different
  samples, and sub-detector misalignment consistency between the
  simulation and the real experiment. Collecting calibration and
  alignment data at the start of a new Run would necessarily incur
  delays such that this approach is mainly of use in the final analyses.
  The experiments are expected to invest in the development of the
  zero-bias overlay approach by 2020.
\item
  The baseline option is to ``pre-mix'' together the minimum bias
  collisions into individual events that have the full background
  expected for a single collision of interest. Experiments will invest
  effort on improving their pre-mixing techniques, which allow the
  mixing to be performed at the digitisation level, reducing the disk and
  network usage for a single event.
\end{itemize}

\paragraph{Fast Simulation} The work on Fast Simulation is also
accelerating with the objective of producing a flexible framework that
permits Full and Fast simulation to be combined for different particles
in the same event. Various approaches to Fast Simulation are being tried
all with the same goal of saving computing time, under the assumption
that it is possible to improve time performance without an unacceptable
loss of physics accuracy. There has recently been a great deal of
interest in the use of Machine Learning in Fast Simulation, most of
which has focused on the use of multi-objective regression and
generative adversarial networks (GANs)~\cite{NIPS2014_5423}. Since use of GANs allows for
non-parametric learning in cases such as calorimetric shower
fluctuations, it is a promising avenue for generating non-Gaussian and
highly correlated physical effects. This is an obvious area for future
expansion and development, as it is currently in its infancy.

\begin{itemize}
\item
  2018: Assessment of the benefit of machine learning approach for Fast
  Simulation.
\item
  2019: ML-based Fast Simulation for some physics observables.
\item
  2022: Demonstrate the potential of a common Fast Simulation
  infrastructure applicable to the variety of detector configurations.
\end{itemize}

\paragraph{Digitisation} It is expected that, within the next 3
years, common digitisation efforts are well-established among
experiments, and advanced high-performance gener\-ic digitisation
examples, which experiments could use as a basis to develop their own
code, become available. For example, the development of next generation
silicon detectors requires realistic simulation of the charge collection
and digitisation processes. Owing to the large variety of technologies,
common software frameworks need to be flexible and modular to cater for
the different needs.

\begin{itemize}
\item
  2020: Deliver advanced high-performance, SIMD-friendly generic
  digitisation examples that experiments can use as a basis to develop
  their own code.
\item
  2022: Fully tested and validated optimised digitisation code that can
  be used by the HL-LHC and DUNE experiments.
\end{itemize}

\paragraph{Pseudorandom Number Generation} The selection of
pseudorandom number generators (PRNGs) presents challenges when running
on infrastructures with a large degree of parallelism, as
reproducibility is a key requirement. HEP will collaborate with
researchers in the development of PRNGs, seeking to obtain generators
that address better our challenging requirements. Specific milestones
are:

\begin{itemize}
\item
  2020: Develop a single library containing sequential and vectorised
  implementations of the set of state-of-the-art PRNGs, to replace the
  existing ROOT and CLHEP implementations. Potential use of C++11 PRNG
  interfaces and implementations, and their extension for our further
  requirements (output of multiple values, vectorisation) will be
  investigated.
\item
  2022: Promote a transition to the use of this library to replace
  existing implementations in ROOT and Geant4.
\end{itemize}

\hypertarget{software-trigger-and-event-reconstruction}{%
\subsection{Software Trigger and Event
Reconstruction}\label{software-trigger-and-event-reconstruction}}

\subsubsection*{Scope and Challenges}

The reconstruction of raw detector data and simulated data, and its
processing in real time, represent a major component of today's
computing requirements in HEP. Advances in the capabilities of
facilities and future
experiments bring the potential for a dramatic increase in physics
reach, at the price of increased event complexities and rates. It is
therefore essential that event reconstruction algorithms and software
triggers continue to evolve so that they are able to efficiently exploit
future computing architectures, and deal with the increase in data rates
without loss of physics. Projections into future, e.g., at HL-LHC conditions,
show that without significant changes in approach or algorithms
the increase in resources needed would be incompatible with
the expected budget.

At the HL-LHC, the central challenge for object reconstruction is to
maintain excellent efficiency and resolution in the face of high pileup
values, especially at low transverse momentum ($p_T$).
Detector upgrades, such as increases in channel density, high-precision
timing, and improved detector geometric layouts, are essential to
overcome these problems. In many cases these new technologies bring
novel requirements to software trigger and/or event reconstruction
algorithms, or require new algorithms to be developed. Ones of
particular importance at the HL-LHC include high-granularity
calorimetry, precision timing detectors, and hardware triggers based on
tracking information, which may seed later software trigger and
reconstruction algorithms.

At the same time, trigger systems for next-generation experiments are
evolving to be more capable, both in their ability to select a wider
range of events of interest for the physics programme, and their ability
to stream a larger rate of events for further processing. ATLAS and CMS
both target systems where the output of the hardware trigger system is
increased by an order of magnitude over the current capability, up to 1
MHz~\cite{CERN-LHCC-2015-020,CMSCollaboration:2015zni}. In LHCb~\cite{CERN-LHCC-2014-016}
and ALICE~\cite{Buncic:2011297}, the full collision
rate (between 30 to 40 MHz for typical LHC proton-proton operations)
will be streamed to real-time or near-real-time software trigger
systems. The increase in event complexity also brings a ``problem'' of
an overabundance of signals to the experiments, and specifically to the
software trigger algorithms. The evolution towards a genuine real-time
analysis of data has been driven by the need to analyse more signal than
can be written out for traditional processing, and technological
developments that enable this without reducing the analysis sensitivity
or introducing biases.

Evolutions in computing technologies are an opportunity to move
beyond commodity x86\_64 technologies, which HEP has used very
effectively over the past 20 years, but also represent a significant challenge
if we are to
derive sufficient event processing throughput per cost to reasonably
enable our physics programmes~\cite{Bird:1695401}.
Among these challenges, important items identified include the increase
of SIMD capabilities, the evolution towards multi- or many-core
architectures, the slow increase in memory bandwidth relative to CPU
capabilities, the rise of heterogeneous hardware, and the possible
evolution in facilities available to HEP production systems.

The move towards open source software development and continuous
integration systems brings opportunities to assist developers of
software trigger and event reconstruction algorithms. Continuous
integration systems based on standard open-source tools have already
allowed automated code quality and performance checks, both for
algorithm developers and code integration teams. Scaling these up to
allow for sufficiently high-statistics checks is still an
outstanding challenge. Also, code quality demands increase as
traditional offline analysis components migrate into trigger systems,
where algorithms can only be run once, and any problem means losing data
permanently.

\subsubsection*{Current Practices}

Substantial computing facilities are in use for both online and offline
event processing across all experiments surveyed. In most experiments,
online facilities are dedicated to the operation of the software
trigger, but a recent trend has been to use them opportunistically
for offline processing too, when the software trigger does not make them
100\% busy. On the other hand, offline facilities are shared with
event reconstruction, simulation, and analysis. CPU in use by
experiments is typically measured at the scale of tens or hundreds of
thousands of x86\_64 processing cores.

The CPU needed for event reconstruction tends to be dominated by charged
particle reconstruction (tracking), especially when the number of
collisions per bunch crossing is high and an efficient
reconstruction low p\textsubscript{T} particles is required.
Calorimetric reconstruction, particle flow reconstruction, and particle
identification algorithms also make up significant parts of the CPU
budget in some experiments. Disk storage is typically 10s to 100s of PBs
per experiment. It is dominantly used to make the output of the event
reconstruction, both for real data and simulation, available for
analysis.

Current experiments have moved towards smaller, but still
flexible, tiered data formats. These tiers are typically based on the
ROOT file format and constructed to
facilitate both skimming of interesting events and the selection of
interesting pieces of events by individual analysis groups or through
centralised analysis processing systems. Initial implementations of
real-time analysis systems are in use within several experiments. These
approaches remove the detector data that typically makes up the raw data
tier kept for offline reconstruction, and keep only final analysis
objects~\cite{Aaij2016,ATL-DAQ-PUB-2017-003,Khachatryan:2149625}.

Systems critical for reconstruction, calibration, and alignment
generally implement a high level of automation in all experiments.
They are an integral part of the data taking and data reconstruction processing chain, both in the online systems as well as the offline processing setup.

\subsubsection*{Research and Development Programme}

Seven key areas, itemised below, have been identified where research and
development is necessary to enable the community to exploit the full
power of the enormous datasets that we will be collecting. Three of
these areas concern the increasingly parallel and heterogeneous
computing architectures that we will have to write our code for. In
addition to a general effort to vectorise our codebases, we must
understand what kinds of algorithms are best suited to what kinds of
hardware architectures. It is an area where collaboration with the
computer science community is required. We also need to develop
benchmarks that allow us to compare the
physics-per-dollar-per-watt performance of different algorithms across a
range of potential architectures, and find ways to optimally utilise
heterogeneous processing centres. The consequent increase in the
complexity and diversity of our codebase will necessitate both a
determined push to educate physicists in modern algorithmic approaches and coding practices,
and a development of more sophisticated and automated quality
assurance and control. The increasing granularity of
our detectors, and the addition of timing information, which seems
mandatory to cope with the extreme pileup conditions at the HL-LHC, will
require new kinds of reconstruction algorithms that are sufficiently
fast for use in real-time. Finally, the increased
signal rates will mandate a push towards real-time analysis in many
areas of HEP, in particular those with low-p\textbf{\textsubscript{T}}
signatures.

\begin{itemize}
\item
  HEP developed toolkits and algorithms typically make poor use of
  vector units on commodity computing systems. Improving this will bring
  speedups to applications running on both current computing systems and
  most future architectures. The goal for work in this area is to evolve
  current toolkit and algorithm implementations, and best programming
  techniques, to better use SIMD capabilities of current and future
  CPU architectures.
\item
  Computing platforms are generally evolving towards having more cores
  in order to increase processing capability. This evolution has
  resulted in multithreaded frameworks in use, or in development, across
  HEP. Algorithm developers can improve throughput by being thread-safe
  and enabling the use of fine-grained parallelism. The goal is to
  evolve current event models, toolkits and algorithm implementations,
  and best programming techniques, to improve the throughput of
  multithreaded software trigger and event reconstruction applications.
\item
  Computing architectures using technologies beyond CPUs offer an
  interesting alternative for increasing throughput of the most
  time-consuming trigger or reconstruction algorithms. Examples
  such as GPUs and FPGAs could be integrated into dedicated
  trigger or specialised reconstruction processing facilities,
  in particular
  online computing farms. The goal is to demonstrate how the throughput
  of toolkits or algorithms can be improved
  in a production environment and to understand
  how much these new architectures require rethinking the algorithms
  used today. In addition, it is necessary to assess and minimise
  possible additional costs coming from the maintenance of multiple
  implementations of the same algorithm on different architectures.
\item
  HEP experiments have extensive continuous integration systems,
  including varying code regression checks that have enhanced the
  quality assurance (QA) and quality control (QC) procedures for
  software development in recent years. These are typically maintained
  by individual experiments and have not yet reached the point where
  statistical regression, technical, and physics performance checks can
  be performed for each proposed software change. The goal is to enable
  the development, automation, and deployment of extended QA and QC
  tools and facilities for software trigger and event reconstruction
  algorithms.
\item
  Real-time analysis techniques are being adopted to enable a wider
  range of physics signals to be saved by the trigger for final
  analysis. As rates increase, these techniques can become more
  important and widespread by enabling only the parts of an event
  associated with the signal candidates to be saved, reducing the
  disk space requirement. The goal is to evaluate and demonstrate the tools
  needed to facilitate real-time analysis techniques. Research topics
  include the study of compression and custom data formats, toolkits for real-time
  detector calibration and validation that enable full offline analysis
  chains to be ported into real-time, and frameworks that allow
  non-expert offline analysts to design and deploy real-time analyses
  without compromising data taking quality.
\item
  The central challenge for object reconstruction at the HL-LHC is to
  maintain excellent efficiency and resolution in the face of high
  pileup, especially at low object p\textsubscript{T}.
  Trigger systems and reconstruction software need to exploit new techniques
  and higher granularity detectors to maintain or even improve physics
  measurements in the future. It is also becoming increasingly clear
  that reconstruction in very high pileup environments, such as the
  HL-LHC or FCC-hh, will not be possible without adding some timing
  information to our detectors, in order to exploit the finite time
  during which the beams cross and the interactions are produced. The
  goal is to develop and demonstrate efficient techniques for physics
  object reconstruction and identification in complex environments.
\item
  Future experimental facilities will bring a large increase in event
  complexity. The performance scaling of current-generation algorithms with this
  complexity must be improved to avoid a large increase in resource
  needs. In addition, it may become necessary to deploy
  new algorithms in order to solve these problems, including advanced
  machine learning techniques. The goal is to evolve or rewrite existing
  toolkits and algorithms focused on their physics and technical
  performance at high event complexity, e.g., high pileup at HL-LHC.
  Most important targets are those which limit expected throughput
  performance at future facilities, most significantly
  charged-particle tracking. A
  number of efforts in this area are already in progress~\cite{recocwp}.
\end{itemize}

\hypertarget{data-analysis-and-interpretation}{%
\subsection{Data Analysis and
Interpretation}\label{data-analysis-and-interpretation}}

\subsubsection*{Scope and Challenges}

Scientific questions are answered by analysing the data obtained from
suitably designed experiments and comparing measurements with
predictions from models and theories. Such comparisons are typically
performed long after data taking, but can sometimes also be executed in
near-real time on selected samples of reduced size.

The final stages of analysis are undertaken by small groups or even
individual researchers. The baseline analysis model utilises successive
stages of data reduction, finally reaching a compact dataset for quick
real-time iterations. This approach aims at exploiting the maximum
possible scientific potential of the data, whilst minimising the ``time
to insight'' for a large number of different analyses performed in
parallel. It is a complicated combination of diverse criteria, ranging
from the need to make efficient use of computing resources to the
management styles of the experiment collaborations. Any analysis system
has to be flexible enough to cope with deadlines imposed by conference
schedules. Future analysis models must adapt to the massive increases in
data taken by the experiments, while retaining this essential ``time to
insight'' optimisation.

Over the past 20 years the HEP community has developed and gravitated
around a single analysis ecosystem based on ROOT.

ROOT is a general-purpose object
oriented framework that addresses the selection, integration,
development, and support of a number of foundation and utility class
libraries that can be used as a basis for developing HEP application
codes. The added value to the HEP community is that it provides an
integrated and validated toolkit, and its use encompasses the full event
processing chain; it has a major impact on the way HEP analysis is
performed. This lowers the hurdle to start an analysis, enabling the
community to communicate using a common analysis language, as well as
making common improvements as additions to the toolkit quickly become
available. The ongoing ROOT programme of work addresses important new
requirements, in both functionality and performance, and this is given
a high priority by the HEP community.

An important new development in the analysis domain has been the
emergence of new analysis tools coming from industry and open-source
projects (e.g. Jupyter notebooks~\cite{Jupyter}, the scikit-learn package~\cite{Pedregosa:2011:SML:1953048.2078195}), and this presents new opportunities for improving the HEP
analysis software ecosystem. The HEP community is very interested in
using these software tools, together with established components, in an
interchangeable way. The main challenge will be to enable new
open-source tools to be plugged in dynamically to the existing ecosystem
and to provide mechanisms that allow the existing and new components to
interact and exchange data efficiently. To improve our ability to
analyse much larger datasets, R\&D will be needed to investigate file
formats, compression algorithms, and new ways of storing and accessing
data for analysis and to adapt workflows to run on future computing
infrastructures.

Reproducibility is the cornerstone of scientific results. It is
currently difficult to repeat most HEP analyses in exactly the manner
they were originally performed. This difficulty mainly arises due to the
number of scientists involved, the large number of steps in a typical
HEP analysis workflow, and the complexity of the analyses themselves. A
challenge specific to data analysis and interpretation is tracking the
evolution of relationships between all the different components of an
analysis, i.e. the provenance of each step.

Reproducibility of scientific results goes in hand with the need to preserve both the data and the software.
Section~\ref{data-and-software-preservation} develops this latter topic where the FAIR principles of data management are embraced.

Robust methods for data reinterpretation are also critical.
Collaborations typically interpret results in the context of specific
models for new physics searches and sometimes reinterpret those same
searches in the context of alternative theories. However, understanding
the full implications of these searches requires the interpretation of
the experimental results in the context of many more theoretical models
than are currently explored at the time of publication. Analysis
reproducibility and reinterpretation strategies need to be considered in
all new approaches under investigation, so that they become a
fundamental component of the system as a whole.

Adapting to the rapidly evolving landscape of software tools, as well as
to methodological approaches to data analysis, requires effort in
continuous training, both for novices as well as for experienced
researchers, as detailed in the Section
\ref{training-and-careers}. The
maintenance and sustainability of the current analysis ecosystem also
present a major challenge, as currently this effort is provided by just
a few institutions. Legacy and less-used parts of the ecosystem need to
be managed appropriately. New policies are needed to retire little used
or obsolete components and free up effort for the development of new
components. These new tools should be made attractive and useful to a
significant part of the community to attract new contributors.

\subsubsection*{Current Practices}

Methods for analysing HEP data have been developed over many years and
successfully applied to produce physics results, including more than
2000 publications, during LHC Runs 1 and 2. Analysis at the LHC
experiments typically starts with users running code over centrally
managed data that is of O(100kB/event) and contains all of the
information required to perform a typical analysis leading to
publication. The most common approach is through a campaign of
\emph{data reduction} and \emph{refinement}, ultimately producing
simplified data structures of arrays of simple data types (``flat
ntuples'') and histograms used to make plots and tables, from which
physics results can be derived.

The current centrally-managed data typically used by a Run 2 data
analysis at the LHC (hundreds of TB) is far too large to be delivered
locally to the user. An often-stated requirement of the data reduction
steps is to arrive at a dataset that ``can fit on a laptop'', in order
to facilitate low-latency, high-rate access to a manageable amount of
data during the final stages of an analysis. Creating and retaining
intermediate datasets produced by data reduction campaigns, bringing and
keeping them ``close'' to the analysers, is designed to minimise latency
and the risks related to resource contention. At the same time, disk
space requirements are usually a key constraint of the experiment
computing models. The
LHC experiments have made a continuous effort to produce optimised
analysis-oriented data formats with enough information to avoid the need
to use intermediate formats. Another effective strategy has been to
combine analyses from different users and execute them within the same
batch jobs (so-called ``analysis trains''), thereby reducing the number
of times data must be read from the storage systems. This has improved
performance and usability, and simplified the task of the bookkeeping.

There has been a huge investment in using C++ for performance-critical
code, in particular in event reconstruction and simulation, and this
will continue in the future. However, for analysis applications, Python
has emerged as the language of choice in the data science community, and
its use continues to grow within HEP. Python is highly
appreciated for its ability to support fast development cycles, for its
ease-of-use, and it offers an abundance of well-maintained and advanced
open source software packages. Experience shows that the simpler
interfaces and code constructs of Python could reduce the complexity of
analysis code, and therefore contribute to decreasing the ``time to
insight'' for HEP analyses, as well as increasing their sustainability.
Increased HEP investment is needed to allow Python to become a first
class supported language.

One new model of data analysis, developed outside of HEP, maintains the
concept of sequential reduction, but mixes interactivity with
batch processing. These exploit new cluster management systems, most
notably Apache Spark~\cite{Spark,Armbrust:2015:SSR:2723372.2742797}, which uses open-source tools contributed both by
industry and the data-science community. Other products implementing the
same analysis concepts and workflows are emerging, such as TensorFlow~\cite{tensorflow2015-whitepaper},
Dask~\cite{rocklin2015dask,dask_ref}, Pachyderm~\cite{thepachydermteam},
Blaze~\cite{wiebe2014blaze}, Parsl~\cite{babuji_yadu_2017_853492},
and Thrill~\cite{bingmann2016thrill}. This approach can complement
the present and widely adopted Grid processing of datasets. It may

potentially simplify the access to data and the expression of
parallelism, thereby improving the exploitation of cluster resources.

An alternative approach, which was pioneered in astronomy but has become
more widespread throughout the Big Data world, is to perform \emph{fast
querying} of centrally managed data and compute remotely on the queried
data to produce the analysis products of interest. The analysis workflow
is accomplished without focus on persistence of data traditionally
associated with data reduction, although transient data may be generated
in order to efficiently accomplish this workflow and optionally can be
retained to facilitate an analysis ``checkpoint'' for subsequent
execution. In this approach, the focus is on obtaining the analysis
end-products in a way that does not necessitate a data reduction
campaign. It is of interest to understand the role
that such an approach could have in the global analysis infrastructure,
and if it can bring an optimisation of the global storage and computing
resources required for the processing of raw data to analysis.

Another active area regarding analysis in the world outside HEP is the
switch to a functional or declarative programming model, as for example
provided by Scala~\cite{Scala} in the Spark environment. This allows scientists to
express the intended data transformation as a query on data. Instead of
having to define and control the ``how'', the analyst declares the
``what'' of their analysis, essentially removing the need to define the
event loop in an analysis, and leave it to underlying services and
systems to optimally iterate over events. It appears that these
high-level approaches will allow abstraction from the underlying
implementations, allowing the computing systems more freedom in
optimising the utilisation of diverse forms of computing resources. R\&D
is already under way, e.g., TDataFrame~\cite{TDataFrame} in ROOT, and
this needs to be continued with the ultimate goal of establishing a
prototype functional or declarative programming paradigm.

\subsubsection*{Research and Development Programme}

Towards HL-LHC, we envisage dedicated data analysis facilities for
experimenters, offering an extendable environment that can provide fully
functional analysis capabilities, integrating all these technologies
relevant for HEP. Initial prototypes of such analysis facilities are
currently under development. On the time scale of HL-LHC, such dedicated
analysis facilities would provide a complete system engineered for
latency optimisation and stability.

The following R\&D programme lists the tasks that need to be
accomplished. By 2020:

\begin{itemize}
\item
  Enable new open-source software tools to be plugged in dynamically to
  the existing ecosystem, and provide mechanisms to dynamically exchange
  parts of the ecosystem with new components.
  \item
    Prototype a comprehensive set of mechanisms for interacting and
    exchanging data between new open-source tools and the existing
    analysis ecosystem.
\item
  Complete an advanced prototype of a low-latency response,
  high-capacity analysis facility, incorporating fast caching
  technologies to explore a query-based analysis approach and
  open-source cluster-management tools. It should in particular include
  an evaluation of additional storage layers, such as SSD storage and
  NVRAM-like storage, and cloud and Big Data orchestration systems.
\item
  Expand support of Python in our ecosystem with a strategy for ensuring
  long-term maintenance and sustainability. In particular in ROOT, the
  current Python bindings should evolve to reach the ease of use of native Python modules.
\item
  Develop a prototype based on a functional or declarative programming
  model for data analysis.
\item
  Conceptualise and prototype an analysis ``Interpretation Gateway'',
  including data repositories, e.g., HEPData~\cite{HEPData,HEPDataRepo}, and
  analysis preservation and reinterpretation tools.
\end{itemize}

By 2022:

\begin{itemize}
\item
  Evaluate chosen architectures for analysis facilities, verify their
  design and provide input for corrective actions to test them on a
  larger scale during Run 3.
\item
  Develop a blueprint for remaining analysis facility developments,
  system design and support model.
\end{itemize}

\hypertarget{machine-learning}{%
\subsection{Machine Learning}\label{machine-learning}}

Machine Learning (ML) is a rapidly evolving approach to characterising
and describing data with the potential to radically change how data is
reduced and analysed. Some applications will qualitatively improve the
physics reach of datasets. Others will allow much more efficient use of
processing and storage resources, effectively extending the physics
reach of experiments. Many of the activities in this area will
explicitly overlap with those in the other focus areas, whereas others
will be more generic. As a first approximation, the HEP community will
build domain-specific applications on top of existing toolkits and ML
algorithms developed by computer scientists, data scientists, and
scientific software developers from outside the HEP world. Work will
also be done to understand where problems do not map well onto existing
paradigms and how these problems can be recast into abstract
formulations of more general interest.

\subsubsection*{Scope and Challenges}

The Machine Learning, Statistics, and Data Science communities have
developed a variety of powerful ML approaches for classification (using
pre-defined categories), clustering (where categories are discovered),
regression (to produce continuous outputs), density estimation,
dimensionality reduction, etc. Some of these have been used productively
in HEP for more than 20 years, others have been introduced relatively
recently. The portfolio of ML techniques and tools is in constant
evolution, and a benefit is that many have well-documented open source
software implementations. ML has already become ubiquitous in some
HEP applications, most notably in classifiers used to discriminate
between signals and backgrounds in final offline analyses. It is
also increasingly used in both online and offline reconstruction and
particle identification algorithms, as well as the classification of
reconstruction-level objects, such as jets.

The abundance of, and advancements in, ML algorithms and implementations
present both opportunities and challenges for HEP.
The community
needs to understand which are most appropriate for our use,
tradeoffs for using one tool compared to another, and
the tradeoffs of using ML algorithms compared to using more traditional software.
These issues are not necessarily ``factorisable'', and a key goal will
be to ensure that, as HEP research teams investigate the numerous
approaches at hand, the expertise acquired and lessons learned, get
adequately disseminated to the wider community. In general, each
\emph{team}, typically a small group of scientists from a collaboration,
will serve as a source of expertise, helping others develop and
deploy experiment-specific ML-based algorithms in their software stacks.
It should provide training to those developing new ML-based algorithms,
as well as those planning to use established ML tools.

With the advent of more powerful hardware, particularly GPUs and ML dedicated
processors, as well as more performant ML
algorithms, the ML toolset will be used to develop application software
that could potentially, amongst other things:

\begin{itemize}
\item
  Replace the most computationally expensive parts of pattern
  recognition algorithms and parameter extraction algorithms for
  characterising reconstructed objects. For example, investigating how
  ML algorithms could improve the physics performance or execution speed
  of charged track and vertex reconstruction, one of the most CPU
  intensive elements of our current software.
\item
  Extend the use of ML algorithms for real-time event classification and
  analysis, as discussed in more detail in Section \ref{software-trigger-and-event-reconstruction}.
\item
  Extend the physics reach of experiments by extending the role of ML at
  the analysis stage: handling data/MC or control/signal region
  differences, interpolating between mass points, training in a
  systematics-aware way, etc.
\item
  Compress data significantly with negligible loss of fidelity in terms
  of physics utility.
\end{itemize}

As already discussed, many particle physics detectors produce much more
data than can be moved to permanent storage. The process of reducing the
size of the datasets is managed by the trigger system. ML algorithms
have already been used very successfully for triggering, to rapidly
characterise which events should be selected for additional
consideration and eventually saved to long-term storage. In the era
of the HL-LHC, the challenges will increase both quantitatively and
qualitatively as the number of proton-proton collisions per bunch
crossing increases. The scope of ML applications in the trigger will
need to expand in order to tackle the challenges to come.

\subsubsection*{Current Practices}

The use of ML in HEP analyses has become commonplace over the past two
decades, and the most common use case has been in signal/background
classification. The vast majority of HEP analyses published in recent
years have used the HEP-specific software package TMVA~\cite{TMVA} included in ROOT. Recently, however, many HEP analysts have begun
migrating to non-HEP ML packages such as scikit-learn~\cite{Pedregosa:2011:SML:1953048.2078195}
and Keras~\cite{Keras}, although these efforts have yet to result
in physics publications from major collaborations. Data scientists at
Yandex created a Python package that provides a consistent API to most
ML packages used in HEP~\cite{REP}. Packages like Spearmint~\cite{Spearmint} and scikit-optimize~\cite{scikit-optimize} perform Bayesian optimisation and can improve HEP Monte
Carlo work.

This shift in the set of ML techniques and packages utilised is
especially strong in the neutrino physics community, where new
experiments such as DUNE place ML at the very heart of their
reconstruction algorithms and event selection. The shift is also
occurring among LHC collaborations, where ML is becoming more and more
commonplace in reconstruction and real-time applications. Examples where
ML has already been deployed in a limited way include charged and
neutral particle reconstruction and identification, jet reconstruction
and identification, and determining a particle's production properties
(flavour tagging), based on information from the rest of the event. In
addition, ML algorithms have been developed that are
insensitive to changing detector performance, for use in real-time
applications, and algorithms that are minimally biased with respect to
the physical observables of interest.

At present, much of this development has happened in specific
collaborations. While each experiment has, or is likely to have,
different specific use cases, we expect that many of these will be
sufficiently similar to each other that R\&D can be done in common. Even
when this is not possible, experience with one type of problem will
provide insights into how to approach other types of problem. This is
why the Inter-experiment Machine Learning forum (IML~\cite{IML}) was
created at CERN in 2016, as a compliment to experiment specific ML R\&D groups.
It has already fostered closer collaboration between LHC and non-LHC
collaborations in the ML field.

\subsubsection*{Research and Development Roadmap and Goals}

The R\&D roadmap presented here is based on the preliminary work done in
recent years, coordinated by the IML, which will remain the main
forum to coordinate work in ML in HEP and ensure the proper links
with the data science communities. The following programme of work is
foreseen.

By 2020:
\begin{itemize}
\item
  \emph{Particle identification and particle properties:} in calorimeters or
  time projection chambers (TPCs), where the data can be represented as
  a 2D or 3D image (or even in 4D, including timing information), the
  problems can be cast as a computer vision task. Deep Learning (DL),
  one class of ML algorithm, in which neural networks are used to
  reconstruct images from pixel intensities, is a good candidate to
  identify particles and extract many parameters. Promising DL
  architectures for these tasks include convolutional, recurrent, and
  adversarial neural networks. A particularly important application is
  to Liquid Argon TPCs (LArTPCs), which is the chosen detection
  technology for DUNE, the new flagship experiment in the neutrino programme.
  A proof of
  concept and comparison of DL architectures should be finalised by
  2020. Particle identification can also be explored to tag the flavour
  of jets in collider experiments (e.g., so-called b-tagging). The
  investigation of these concepts, which connect to Natural Language
  Processing~\cite{Collobert:2011:NLP:1953048.2078186}, has started at the LHC and is to be pursued on the same
  timescale.
\item
  \emph{ML middleware and data formats for offline usage:} HEP relies
  on the ROOT format for its data, wheras the ML community
  has developed several other formats, often associated with specific ML
  tools. A desirable data format for ML applications should have the
  following attributes: high read-write speed for efficient training,
  sparse readability without loading the entire dataset into RAM,
  compressibility, and widespread adoption by the ML community. The thorough
  evaluation of the different data formats and their impact on ML
  performance in the HEP context must be continued, and it is necessary
  to define a strategy for bridging or migrating HEP formats to the
  chosen ML format(s), or vice-versa.
\item
  \emph{Computing resource optimisations:} managing large
  volume data transfers is one
  of the challenges facing current computing facilities. Networks
  play a crucial role in data exchange and so a
  network-aware application layer may significantly improve experiment
  operations. ML is a promising technology to identify anomalies in
  network traffic, to predict and prevent network congestion, to detect
  bugs via analysis of self-learning networks, and for WAN path
  optimisation based on user access patterns.
\item
  \emph{ML as a Service (MLaaS):} current cloud providers rely on a MLaaS model
  exploiting interactive
  machine learning tools in order to make efficient use of resources, however,
  this is not yet widely used in HEP. HEP
  services for interactive analysis, such as CERN's Service for
  Web-based Analysis, SWAN~\cite{SWAN}, may play an important role in adoption
  of machine learning tools in HEP workflows. In order to use these tools
  more efficiently, sufficient and appropriately tailored hardware and
  instances other than SWAN will be identified.
\end{itemize}

By 2022:
\begin{itemize}
\item
  \emph{Detector anomaly detection:} data taking is
  continuously monitored by physicists taking shifts to monitor and
  assess the quality of the incoming data, largely using reference
  histograms produced by experts. A whole class of ML algorithms called anomaly detection
  can be useful for automating this important task. Such unsupervised algorithms are able
  to learn from data and produce an alert when deviations are observed.
  By monitoring many variables at the same time, such algorithms are
  sensitive to subtle signs forewarning of imminent failure, so that
  pre-emptive maintenance can be scheduled. These techniques are already
  used in industry.
\item
  \emph{Simulation:} recent progress in high fidelity fast generative models,
  such as Generative Adversarial Networks (GANs)~\cite{NIPS2014_5423} and Variational
  Autoencoders (VAEs)~\cite{2013arXiv1312.6114K},
  which are able to sample high dimensional feature
  distributions by learning from existing data samples, offer a
  promising alternative for Fast Simulation. A simplified first attempt at
  using such techniques in simulation saw orders of magnitude increase
  in speed over existing Fast Simulation techniques, but has not yet
  reached the required accuracy~\cite{CaloGAN}.
\item
  \emph{Triggering and real-time analysis:} one of the challenges is the
  trade-off in algorithm complexity and performance under strict
  inference time constraints. To deal with the increasing event
  complexity at HL-LHC, the use of sophisticated ML
  algorithms will be explored at all trigger levels, building on the pioneering work of
  the LHC collaborations. A critical part of this work will be to
  understand which ML techniques allow us to maximally exploit
  future computing architectures.
\item
  \emph{Sustainable Matrix Element Method (MEM):} MEM is a powerful
  technique that can be utilised for making measurements of physical model
  parameters and direct searches for new phenomena. As it is
  very computationally intensive its use in HEP is limited.
  Although the use of neural networks for numerical integration is not new,
  it is a
  technical challenge to design a network sufficiently rich
  to encode the complexity of the ME calculation for a given process
  over the phase space relevant to the signal process. Deep Neural
  Networks (DNNs) are good candidates~\cite{Bendavid2017,BendavidIML2017}.
\item
  \emph{Tracking:} pattern recognition is always a computationally challenging
  step. It becomes a huge challenge in the HL-LHC environment.
  Adequate ML techniques may provide a solution that scales linearly
  with LHC intensity. Several efforts in the HEP community have started
  to investigate ML algorithms for track pattern recognition on
  many-core processors.
\end{itemize}

\hypertarget{data-organisation-management-and-access}{%
\subsection{Data Organisation, Management and
Access}\label{data-organisation-management-and-access}}

The scientific reach of data-intensive experiments is limited by how
fast data can be accessed and digested by computational resources. Changes in
computing technology and large increases in data volume require new
computational models~\cite{Butler2013}, compatible with
budget constraints. The integration of newly emerging data
analysis paradigms into our computational model has the potential to
enable new analysis methods and increase scientific output. The field,
as a whole, has a window in which to adapt our data access and data
management schemes to ones that are more suited and optimally matched to
advanced computing models and a wide range of analysis applications.

\subsubsection*{Scope and Challenges}

The LHC experiments currently provision and manage about an exabyte of
storage, approximately half of which is archival, and half is
traditional disk storage. Other experiments that will soon start data taking have
similar needs, e.g., Belle II has the same data volumes as ATLAS. The
HL-LHC storage requirements per year are expected to jump by a factor
close to 10, which is a growth rate faster than can be
accommodated by projected technology gains. Storage will remain one of the major
cost drivers for HEP computing, at a level roughly equal to the cost
of the computational resources. The combination of storage and analysis
computing costs may restrict scientific output and the potential physics
reach of the experiments, so new techniques and algorithms are likely to
be required.

In devising experiment computing models for this era many factors have
to be taken into account. In particular, the increasing availability of
very high-speed networks may reduce the need for CPU and data
co-location. Such networks may allow for more
extensive use of data access over the wide-area network (WAN), which may
provide failover capabilities, global and federated data namespaces, and
will have an impact on data caching. Shifts in data presentation and
analysis models, such as the use of event-based data streaming along
with more traditional dataset-based or file-based data access, will be
particularly important for optimising the utilisation of opportunistic
computing cycles on HPC facilities, commercial cloud resources, and
campus clusters. This can potentially resolve currently limiting factors
such as job eviction.

The three main challenges for data management in the HL-LHC follow:

\begin{itemize}
\item
  The experiments will significantly increase both
  the data rate and the data volume. The computing systems will need to
  handle this with as small a cost increase as possible and within
  evolving storage technology limitations.
\item
  The significantly increased computational requirements for the HL-LHC
  era will also place new requirements on data access. Specifically, the use of
  new types of computing resources (cloud, HPC) that have different dynamic
  availability and characteristics will require more dynamic data
  management and access systems.
\item
  Applications employing new techniques, such as
  training for machine learning or high rate data query systems, will likely be employed to
  meet the computational constraints and to extend physics reach.
  These new applications will place new requirements
  on how and where data is accessed and produced. Specific applications,
  such as training for machine learning, may require use of specialised
  processor resources, such as GPUs, placing further requirements on
  data.
\end{itemize}

The projected event complexity of data from future HL-LHC
runs with high pileup and from high resolution Liquid Argon detectors at
DUNE will require advanced reconstruction algorithms and analysis tools
to interpret the data. The precursors of these tools, in the form of
new pattern recognition and tracking algorithms,
are already proving to be drivers for the
compute needs of the HEP community. The storage systems that are
developed, and the data management techniques that are employed,
will need to be matched to these changes in computational work, so
as not to hamper potential improvements.

As with computing resources, the landscape of storage solutions
is trending towards heterogeneity. The ability to
leverage new storage technologies as they become available into existing
data delivery models is a challenge that we must be prepared for. This
also implies the need to leverage
``tactical storage'', i.e., storage that becomes more cost-effective as
it becomes available (e.g., from a cloud provider), and have a data
management and provisioning system that can exploit such resources at
short notice. Volatile data sources would impact many aspects of the
system: catalogues, job brokering, monitoring and alerting, accounting,
the applications themselves.

On the hardware side, R\&D is needed in alternative approaches to data
archiving to determine the possible cost/performance tradeoffs.
Currently, tape is extensively used to hold data that cannot be
economically made available online. While the data is still accessible,
it comes with a high latency penalty, limiting effective data access. We
suggest investigating either separate direct access-based archives
(e.g., disk or optical) or new models that hierarchically overlay online
direct access volumes with archive space. This is especially relevant
when access latency is proportional to storage density. Either approach
would need to also evaluate reliability risks and the effort needed to
provide data stability. For this work, we should exchange experiences
with communities that rely on large tape archives for their primary
storage.

Cost reductions in the maintenance and operation of storage
infrastructure can be realised through convergence of the major
experiments and resource providers on shared solutions. This does not
necessarily mean promoting a monoculture, as different solutions will be
adapted to certain major classes of use cases, type of site, or funding
environment. There will always be a judgement to make on the
desirability of using a variety of specialised systems, or of
abstracting the commonalities through a more limited, but common,
interface. Reduced costs and improved sustainability will be further
promoted by extending these concepts of convergence beyond HEP and into
the other large-scale scientific endeavours that will share the
infrastructure in the coming decade (e.g., the SKA and CTA experiments).
Efforts must be made as early as possible, during the formative design
phases of such projects, to create the necessary links.

Finally, all changes undertaken must not make the ease of access to data
any worse than it is under current computing models. We must also be
prepared to accept the fact that the best possible solution may require
significant changes in the way data is handled and analysed. What is
clear is that current practices will not scale to the needs of HL-LHC
and other major HEP experiments of the coming era.

\subsubsection*{Current Practices}

The original LHC computing models were based on simpler models used
before distributed computing was a central part of HEP computing. This
allowed for a reasonably clean separation between four different aspects
of interacting with data, namely data organisation, data management,
data access, and data granularity. The meaning of these terms may be
summarised in what follows.

\begin{itemize}
\item
  \emph{Data organisation} is essentially how data is structured as it
  is written. Most data is written in files, in ROOT format, typically
  with a column-wise organisation of the data. The records corresponding
  to these columns are compressed. The internal details of this
  organisation are visible only to individual software applications.
\item
  In the past, the key challenge for \emph{data management} was the transition to
  use distributed computing in the form of the grid. The experiments
  developed dedicated data transfer and placement systems, along with
  catalogues, to move data between computing centres. Originally,
  computing models were rather static: data was placed at sites, and the
  relevant compute jobs were sent to the right locations. Since LHC
  startup, this model has been made more flexible to limit non-optimal
  pre-placement and to take into account data popularity. In addition,
  applications might interact with catalogues or, at times, the workflow
  management system does this on behalf of the applications.
\item
  \emph{Data access}: historically, various protocols have been used for
  direct reads (rfio, dcap, xrootd, etc.) where jobs are reading data
  explicitly staged-in or cached by the compute resource used or the
  site it belongs to. A recent move has been the convergence towards
  xrootd as the main protocol for direct access. With direct access,
  applications may use alternative protocols to those used by data
  transfers between sites. In addition, LHC experiments have been
  increasingly using remote access to the data, without any stage-in
  operations, using the possibilities offered by protocols such as xrootd
  or http.
\item
  \emph{Data granularity}: the data is split into datasets, as defined
  by physics selections and use cases, consisting of a set of individual
  files. While individual files in datasets can be processed in
  parallel, the files themselves are usually processed as a whole.
\end{itemize}

Before LHC turn-on, and in the first years of the LHC, these four
areas were to first order optimised independently. As LHC computing
matured, interest has turned to optimisations spanning multiple areas.
For example, the recent use of ``Data Federations'' mixes up Data
Management and Access. As we will see below, some of the foreseen
opportunities towards HL-LHC may require global optimisations.

Thus, in this section we take a broader view than traditional data
management and consider the combination of ``Data Organisation,
Management and Access'' (DOMA) together. We believe that this fuller
picture will provide important opportunities for improving
efficiency and scaleability, as we enter the many-exabyte era.

\subsubsection*{Research and Development Programme}

In the following, we describe
tasks that will need to be carried out in order to demonstrate that the
increased volume and complexity of data expected over the coming decade
can be stored, accessed, and analysed at an affordable cost.

\begin{itemize}
\item
  Sub-file granularity, e.g.,
  event-based, will be studied to see whether it can be implemented
  efficiently, and in a scalable, cost-effective manner, for all
  applications making use of event selection, to see whether it offers
  an advantage over current file-based granularity. The following tasks
  should be completed by 2020:
  \begin{itemize}
  \item
    Quantify the impact on
    performance and resource utilisation  of the storage and network
    for the main
    access patterns, i.e., simulation, reconstruction, analysis.
  \item
    Assess the impact on
    catalogues and data distribution.
  \item
    Assess whether
    event-granularity makes sense in object stores that tend to require
    large chunks of data for efficiency.
  \item
    Test for improvement in
    recoverability from preemption, in particular when using cloud spot
    resources and/or dynamic HPC resources.
  \end{itemize}
\item
  We will seek to derive benefits
  from data organisation and analysis technologies adopted by other big
  data users. A proof-of-concept that involves the following tasks needs
  to be established by 2020 to allow full implementations to be made in
  the years that follow.
  \begin{itemize}
  \item
    Study the impact of
    column-wise, versus row-wise, organisation of data on the
    performance of each kind of access.
  \item
    Investigate efficient data
    storage and access solutions that support the use of map-reduce or
    Spark-like analysis services.
  \item
    Evaluate just-in-time
    decompression schemes and mappings onto hardware architectures
    considering the flow of data, from spinning disk to memory and
    application.
  \end{itemize}
\item
  Investigate the role data
  placement optimisations can play, such as caching, in order to use
  computing resources effectively, and the technologies that can be used
  for this. The following tasks should be completed by 2020:
  \begin{itemize}
  \item
   Quantify the benefit of
    placement optimisation for reconstruction,
    analysis, and simulation.
  \item
    Assess the benefit of caching
    for Machine Learning-based applications, in particular for the
    learning phase, and follow-up the evolution of technology outside
    HEP.
  \end{itemize}
  In the longer term the benefits
  that can be derived from using different approaches to the way HEP is
  currently managing its data delivery systems should be studied. Two
  different content delivery methods will be looked at, namely Content
  Delivery Networks (CDN) and Named Data Networking (NDN).
\item
  Study how to minimise HEP
  infrastructure costs by exploiting varied quality of service from
  different storage technologies. In particular, study the role that
  opportunistic/tactical storage can play, as well as different archival
  storage solutions. A proof-of-concept should be made by 2020, with a
  full implementation to follow in the following years.
\end{itemize}

\begin{itemize}
\item
  Establish how to globally optimise data access latency, with respect
  to the efficiency of using CPU, at a sustainable cost. This involves
  studying the impact of concentrating data in fewer, larger locations
  (the ``data-lake'' approach), and making increased use of opportunistic
  compute resources located further from the data. Again, a
  proof-of-concept should be made by 2020, with a full implementation in
  the following years, if successful. This R\&D will be done in common
  with the related actions planned as part of Facilities and Distributed
  Computing.
\end{itemize}

\hypertarget{facilities-and-distributed-computing}{%
\subsection{Facilities and Distributed
Computing}\label{facilities-and-distributed-computing}}

\subsubsection*{Scope and Challenges}

As outlined in Section \ref{software-and-computing-challenges},
huge resource requirements are anticipated for
HL-LHC running. These need to be deployed and managed across the WLCG
infrastructure, which has evolved from the original ideas on deployment
before LHC data-taking started~\cite{Aderholz:510694}, to be a mature and
effective infrastructure that is now exploited by LHC experiments.
Currently, hardware costs are dominated by disk storage, closely
followed by CPU, followed by tape and networking. Naive estimates of
scaling to meet HL-LHC needs indicate that the current system would need
almost an order of magnitude more resources than will be available from
technology evolution alone. In addition, other initiatives such as Belle
II and DUNE in particle physics, but also other science projects such as
SKA, will require a comparable amount of resources on the same
infrastructure. Even anticipating substantial software improvements, the
major challenge in this area is to find the best configuration for
facilities and computing sites that make HL-LHC computing feasible. This
challenge is further complicated by substantial regional differences in
funding models, meaning that any solution must be sensitive to these
local considerations to be effective.

There are a number of changes that can be anticipated on the timescale
of the next decade that must be taken into account. There is an
increasing need to use highly heterogeneous resources, including the
use of HPC infrastructures (which can often have very particular setups
and policies that make their exploitation challenging); volunteer
computing (which is restricted in scope and unreliable, but can be a
significant resource); and cloud computing,
both commercial and research. All of these offer different resource
provisioning interfaces and can be significantly more dynamic than
directly funded HEP computing sites. In addition, diversity of computing
architectures is expected to become the norm, with different CPU
architectures, as well as more specialised GPUs and FPGAs.

This increasingly dynamic environment for resources, particularly CPU,
must be coupled with a highly reliable system for data storage and a
suitable network infrastructure for delivering this data to where it
will be processed. While CPU and disk capacity is expected to increase
by respectively 15\% and 25\% per year for the same cost~\cite{CERN-COST-EST}, the trends of
research network capacity increases show a much steeper growth, such as
two orders of magnitude from now to HL-LHC times~\cite{Roberts:17}. Therefore, the
evolution of the computing models would need to be more network centric.

In the network domain, there are new technology developments, such as
Software Defined Networks (SDNs), which enable user-defined high
capacity network paths to be controlled via experiment software, and
which could help manage these data flows~\cite{Blikra:2221659}.
Some projects already started
to explore the potential of these technologies~\cite{OSiRIS} but a
considerable R\&D is required to prove their utility and practicality.
In addition,
the networks used by HEP are likely to see large increases in traffic
from other science domains.

Underlying storage system technology will continue to evolve, for
example towards object stores, and, as proposed in
Data Organisation, Management and Access
(Section \ref{data-organisation-management-and-access}),
R\&D is also necessary to understand their usability and their role in
the HEP infrastructures. There is also the continual challenge of
assembling inhomogeneous systems and sites into an effective widely
distributed worldwide data management infrastructure that is usable by
experiments. This is particularly compounded by the scale increases for
HL-LHC where multiple replicas of data (for redundancy and availability)
will become extremely expensive.

Evolutionary change towards HL-LHC is required, as the experiments will
continue to use the current system. Mapping out a path for migration
then requires a fuller understanding of the costs and benefits of the
proposed changes. A model is needed in which the benefits of such
changes can be evaluated, taking into account hardware and human costs,
as well as the impact on software and workload performance that in turn
leads to physics impact. Even if HL-LHC is the use case used to build
this cost and performance model, because the ten years of experience
running large-scale experiments helped to define the needs, it is
believed that this work, and the resulting model, will be valuable for
other upcoming data intensive scientific initiatives. This includes
future HEP projects, such as Belle II, DUNE and possibly ILC
experiments, but also non-HEP projects, such as SKA.

\subsubsection*{Current Practices}

While there are many particular exceptions, most resources incorporated
into the current WLCG are done so in independently managed sites,
usually with some regional organisation structure, and mostly offering
both CPU and storage. The sites are usually funded directly to provide
computing to WLCG, and are in some sense then ``owned'' by HEP, albeit
often shared with others. Frequently substantial cost contributions are
made indirectly, for example through funding of energy costs or
additional staff effort, particularly at smaller centres. Tape is found
only at CERN and at large national facilities, such as the WLCG Tier-1s~\cite{Bird:1695401}.

Interfaces to these computing resources are defined by technical
operations in WLCG. Frequently there are choices that sites can make
among some limited set of approved options for interfaces. These can
overlap in functionality. Some are very HEP specific and recognised as
over-complex: work is in progress to get rid of them. The acceptable
architectures and operating systems are also defined at the WLCG level
(currently x86\_64, running Scientific Linux 6 and compatible), and
sites can deploy these either directly onto ``bare metal'' or can use an
abstraction layer, such as virtual machines or containers.

There are different logical networks being used to connect sites: LHCOPN
connects CERN with the Tier-1 centres and a mixture of LHCONE and
generic academic networks connect other sites.

Almost every experiment layers its own customised workload and data
management system on top of the base WLCG provision, with several
concepts, and a few lower-level components, in common. The pilot job
model for workloads is ubiquitous, where a real workload is dispatched
only once a job slot is secured. Data management layers aggregate files
in the storage systems into datasets and manage experiment-specific
metadata. In contrast to the MONARC model, sites are generally used more
flexibly and homogeneously by experiments, both in workloads and in data
stored.

In total, WLCG currently provides experiments with resources distributed
at about 170 sites, in 42 countries, which pledge every year the amount
of CPU and disk resources they are committed to delivering. The pledge
process is overseen by the Computing Resource Scrutiny Group (CRSG), mandated by
the funding agencies to validate the experiment requests, and to
identify mismatches with site pledges. These sites are connected by
10-100 Gb links, and deliver approximately 500k CPU cores and 1 EB of
storage, of which 400 PB is disk. More than 200M jobs are executed each day~\cite{Bird2017}.

\subsubsection*{Research and Development programme}

The following areas of study are ongoing, and will involve technology
evaluations, prototyping, and scale tests. Several of the items below
require some coordination with other topical areas discussed in this
document, and some work is still needed to finalise the detailed action
plan. These actions will need to be structured to meet the common
milestones of informing the HL-LHC Computing Technical Design Reports (TDRs), and deploying
advanced prototypes during LHC Run 3.

\begin{itemize}
\item
  Understand better the relationship between the performance and costs
  of the WLCG system, and how it delivers the necessary functionality to
  support LHC physics. This will be an ongoing process, started by the
  recently formed System Performance and Cost Modeling Working Group~\cite{SPCMWG},
  and aims to provide a quantitative assessment for any proposed
  changes.
\item
  Define the functionality needed to implement a federated data centre
  concept (``data lake'') that aims to reduce the operational cost of
  storage for HL-LHC, and at the same time better manage network
  capacity, whilst maintaining the overall CPU efficiency. This would
  include the necessary qualities of service, and options for regionally
  distributed implementations, including the ability to flexibly respond
  to model changes in the balance between disk and tape. This work
  should be done in conjunction with the existing Data Organisation,
  Management and Access Working Group~\cite{WLCG-DOMA} to evaluate the impact of the
  different access patterns and data organisations envisaged.
\item
  Building upon the experience of projects currently exploring SDN
  potential, define the role for this technology in managing data transfers and
  access and the integration strategy into experiment frameworks.
\item
  Establish an agreement on the common data management functionality
  that is required by experiments, targeting a consolidation and a lower
  maintenance burden. The intimate relationship between the management
  of elements in storage systems and metadata must be recognised. This
  work requires coordination with the Data Processing Frameworks
  Working Group. It
  needs to address at least the following use cases:

  \begin{itemize}
  \item
    processing sites that may have some small disk cache, but do not
    manage primary data;
  \item
    fine grained processing strategies that may enable processing of
    small chunks of data, with appropriate bookkeeping support;
  \item
    integration of heterogeneous processing resources, such as HPCs and
    clou\-ds.
  \end{itemize}
\item
  Explore scalable and uniform means of workload scheduling, which
  incorporate dynamic heterogenous resources, and the capabilities of
  finer grained processing that increases overall efficiency. The
  optimal scheduling of special workloads that require particular
  resources is clearly required.
\item
  Contribute to the prototyping and evaluation of a quasi-interactive
  analysis facility that would offer a different model for physics
  analysis, but would also need to be integrated into the data and
  workload management of the experiments. This is work to be done in
  collaboration with groups working on new data analysis models.
\end{itemize}

\hypertarget{data-flow-processing-framework}{%
\subsection{Data-Flow Processing
Framework}\label{data-flow-processing-framework}}

\subsubsection*{Scope and Challenges}

Frameworks in HEP are used for the collaboration-wide
data processing tasks of triggering, reconstruction, and simulation, as
well as other tasks that subgroups of the collaboration are responsible
for, such as detector alignment and calibration. Providing framework
services and libraries that will satisfy the computing and data needs
for future HEP experiments in the next decade, while maintaining our
efficient exploitation of increasingly heterogeneous resources, is a
huge challenge.

To fully exploit the potential of modern processors, HEP data processing
frameworks need to allow for the parallel execution of reconstruction or
simulation algorithms on multiple events simultaneously. Frameworks face
the challenge of handling the massive parallelism and heterogeneity that
will be present in future computing facilities, including multi-core and
many-core systems, GPUs, Tensor Processing Units (TPUs), and tiered
memory systems, each integrated with storage and high-speed network
interconnections. Efficient running on heterogeneous resources will
require a tighter integration with the computing models' higher-level
systems of workflow and data management. Experiment frameworks must also
successfully integrate and marshall other HEP software that may have its
own parallelisation model, such as physics generators and detector
simulation.

Common developments across experiments are desirable in this area, but
are hampered by many decades of legacy work. Evolving our frameworks
also has to be done recognising the needs of the different stakeholders
in the system. This includes physicists, who are writing processing
algorithms for triggering, reconstruction or analysis; production
managers, who need to define processing workflows over massive datasets;
and facility managers, who require their infrastructures to be used
effectively. These frameworks are also constrained by security
requirements, mandated by the groups and agencies in charge of it.

\subsubsection*{Current Practices}

Although most frameworks used in HEP share common concepts, there are,
for mainly historical reasons, a number of different implementations;
some of these are shared between experiments. The Gaudi framework~\cite{Barrand:2001ny} was
originally developed by LHCb, but is also used by ATLAS and various
non-LHC experiments. CMS uses its own CMSSW framework~\cite{Bayatian:2006nff}, which was forked
to provide the art framework for the Fermilab Intensity Frontier experiments~\cite{Green:2012gv}.
Belle II uses basf2~\cite{1742-6596-331-3-032024}. The linear collider community developed and uses
Marlin~\cite{Gaede:2006pj}. The FAIR experiments use FairROOT, closely related to ALICE's
AliROOT. The FAIR experiments and ALICE are now developing a new
framework, which is called O2~\cite{O2}. At the time of writing, most
major frameworks support basic parallelisation, both within and across
events, based on a task-based model~\cite{Jones:2015soc}\cite{Clemencic:2015paa}.

Each framework has a processing model, which provides the means to
execute and apportion work. Mechanisms for this are threads, tasks,
processes, and inter-process communication. The different strategies used
reflect different trade-offs between constraints in the programming
model, efficiency of execution, and ease of adapting to inhomogeneous
resources. These concerns also reflect two different behaviours:
firstly, maximising throughput, where it is most important to maximise
the number of events that are processed by a given resource; secondly,
minimising latency, where the primary constraint is on how long it takes
to calculate an answer for a particular datum.

Current practice for throughput maximising system architectures have
con\-strain\-ed the scope of framework designs. Framework applications have
largely been viewed by the system as a batch job with complex
configuration, consuming resources according to rules dictated by the
computing model: one process using one core on one node, operating
independently with a fixed size memory space on a fixed set of files
(streamed or read directly). Only recently has CMS broken this tradition
starting at the beginning of Run 2, by utilising all available cores in one process space using threading. ATLAS is currently
using a multi-process fork-and-copy-on-write solution to remove the
constraint of one core/process. Both experiments were driven to solve
this problem by the ever-growing need for more memory per process
brought on by the increasing complexity of LHC events. Current practice
manages systemwide (or facility-wide) scaling by dividing up datasets,
generating a framework application configuration, and scheduling jobs on
nodes/cores to consume all available resources. Given anticipated
changes in hardware (heterogeneity, connectivity, memory, storage)
available at computing facilities, the interplay between
workflow and workload management systems and framework applications need to
be carefully examined. It may be advantageous to permit framework
applications (or systems) to span multi-node resources, allowing them
to be first-class participants in the business of scaling within a
facility. In our community some aspects of this approach, which maps
features with microservices or function as a service, is being pioneered
by the O2 framework.

\subsubsection*{Research and Development programme}

By the end of 2018: review the existing technologies that are the
important building blocks for data processing frameworks and reach
agreement on the main architectural concepts for the next generation of
frameworks. Community meetings and workshops, along the lines of the
original Concurrency Forum, are envisaged in order to foster collaboration in this work~\cite{ConcurrencyForum}. This includes the following:

\begin{itemize}
\item
  Libraries used for concurrency, their likely evolution and the issues
  in integrating the models used by detector simulation and physics
  generators into the frameworks.
\item
  Functional programming, as well as domain specific languages, as a way
  to describe the physics data processing that has to be undertaken
  rather than how it has to be implemented. This approach is based on
  the same concepts as the idea for functional approaches for
  (statistical) analysis as described in Section \ref{data-analysis-and-interpretation}.
\item
  Analysis of the functional differences between the existing frameworks
  and the different experiment use cases.
\end{itemize}

By 2020: prototype and demonstrator projects for the agreed
architectural concepts and baseline to inform the HL-LHC Computing TDRs
and to demonstrate advances over what is currently deployed. The
following specific items will have to be taken into account:

\begin{itemize}
\item
  These prototypes should be as common as possible between existing
  frameworks, or at least several of them, as a proof-of-concept of
  effort and component sharing between frameworks for their future
  evolution. Possible migration paths to more common implementations
  will be part of this activity.
\item
  In addition to covering the items mentioned for the review phase, they
  should particularly demonstrate possible approaches for scheduling the
  work across heterogeneous resources and using them efficiently, with a
  particular focus on the efficient use of co-processors, such as GPUs.
\item
  They need to identify data model changes that are required for an
  efficient use of new processor architectures (e.g., vectorisation),
  and for scaling I/O performance in the context of concurrency.
\item
  Prototypes of a more advanced integration with workload management,
  taking advantage in particular of the advanced features available at
  facilities for a finer control of the interactions with storage and
  network, and dealing efficiently with the specificities of HPC
  resources.
\end{itemize}

By 2022: production-quality framework libraries usable by several
experiment frameworks, covering the main areas successfully demonstrated
in the previous phase. During these activities we expect at least one
major paradigm shift to take place on this 5-year time scale. It will be
important to continue discussing their impact within the community,
which will be ensured through appropriate cross-experiment workshops
dedicated to data processing frameworks.

\hypertarget{conditions-data}{%
\subsection{Conditions Data}\label{conditions-data}}

\subsubsection*{Scope and Challenges}

Conditions data is defined as the non-event data required by
data-processing software to correctly simulate, digitise or reconstruct
the raw detector event data. The non-event data discussed here consists
mainly of detector calibration and alignment information, with some
additional data describing the detector configuration, the machine
parameters, as well as information from the detector control system.

Conditions data is different from event data in many respects, but one
of the important differences is that its volume scales with time rather
than with the luminosity. As a consequence, its growth is limited, as
compared to event data: conditions data volume is expected to be at the
terabyte scale and the update rate is modest (typically O(1)Hz).
However, conditions data is used by event processing applications
running on a very large distributed computing infrastructure, resulting
in tens of thousands of jobs that may try to access the conditions data
at the same time, and leading to a very significant rate of reading
(typically O(10) kHz).

To successfully serve such rates, some form of caching is needed, either
by using services such as web proxies (CMS and ATLAS use Frontier) or by
delivering the conditions data as files distributed to the jobs. For the
latter approach, CVMFS is an attractive solution due to its embedded
caching, and its advanced snapshotting and branching features. ALICE
have made some promising tests, and started to use this approach in Run
2; Belle II already took the same approach~\cite{WoodACAT2017}, and NA62
has also decided to adopt this solution. However, one particular
challenge to be overcome with the filesystem approach is to design an
efficient mapping of conditions data and metadata to files in order to
use the CVMFS caching layers efficiently.

Efficient caching is especially important in order to support the
high-reading rates that will be necessary for ATLAS and CMS experiments
starting with Run 4. For these experiments, a subset of the conditions
data is linked to the luminosity, leading to an interval of granularity
down to the order of a minute. Insufficient or inefficient caching may
impact the efficiency of the reconstruction processing.

Another important challenge is ensuring the long-term maintainability of
the conditions data storage infrastructure. Shortcomings in the initial
approach used in LHC Run 1 and Run 2, leading to complex
implementations, helped to identify the key requirements for an
efficient and sustainable condition data handling infrastructure. There
is now a consensus among experiments on these requirements~\cite{Laycock2017}:
ATLAS and CMS are working on a common next-generation
conditions database~\cite{1742-6596-898-4-042047}.
The Belle II experiment, which is about to start its data taking, has already
developed a solution based on the same concepts and architecture. One
key point in this new design is to have a server mostly agnostic to the
data content with most of the intelligence on the client side. This new
approach should make it easier to rely on well-established open-source
products (e.g., Boost) or software components developed for the
processing of event data (e.g., CVMFS). With such an approach, it should
be possible to leverage technologies such as REST interfaces to simplify
insertion and read operations, and make them very efficient to reach the
rate levels foreseen. Also, to provide a resilient service to jobs that
depend on it, the client will be able to use multiple proxies or servers
to access the data.

One conditions data challenge may be linked to the use of an event
service, as ATLAS is doing currently, to use efficiently HPC facilities
for event simulation or processing. The event service allows better use
of resources that may be volatile by allocating and bookkeeping the work
done, not at the job granularity, but at the event granularity. This
reduces the possibility for optimising access to the conditions data at
the job level, and may lead to an increased pressure on the conditions
data infrastructure. This approach is still at an early stage, and more
experience is needed to better appreciate the exact impact on the
conditions data.

\subsubsection*{Current Practices}

The data model for conditions data management is an area where the
experiments have converged on something like a best common practice. The
time information for the validity of the Payloads is specified with a
parameter called an \emph{Interval Of Validity} (IOV), which can be
represented by a Run number, the ID of a luminosity section or a
universal timestamp. A fully qualified set of conditions data consists
of a set of payloads and their associate IOVs covering the time span
required by the workload. A label called a \emph{Tag} identifies the
version of the set and the global tag is the top-level configuration of
all conditions data. For a given detector subsystem and a given IOV, a
global tag will resolve to one, and only one, conditions data payload.
The global tag resolves to a particular system tag via the global tag
map table. A system tag consists of many intervals of validity or
entries in the IOV table. Finally, each entry in the IOV table maps to a
payload via its unique hash key.

A relational database is a good choice for implementing this design. One
advantage of this approach is that a payload has a unique identifier,
its hash key, and this identifier is the only way to access it. All
other information, such as tags and IOV, is metadata used to select a
particular payload. This allows a clear separation of the payload data
from the metadata, and may allow use of a different backend technology
to store the data and the metadata. This has potentially several
advantages:

\begin{itemize}
\item
  Payload objects can be cached independently of their metadata, using
  the appropriate technology, without the constraints linked to metadata
  queries.
\item
  Conditions data metadata are typically small compared to the
  conditions data themselves, which makes it easy to export them as a
  single file using technologies such as SQLite. This may help for
  long-term data preservation.
\item
  IOVs, being independent of the payload, can also be cached on their
  own.
\end{itemize}

A recent trend is the move to full reconstruction online, where the
calibrations and alignment are computed and applied in the High Level
Trigger (HLT). This is currently being tested by ALICE and LHCb, who
will adopt it for use in Run 3. This will offer an opportunity to
separate the distribution of conditions data to reconstruction jobs and
analysis jobs, as they will not run on the same infrastructure. However,
running reconstruction in the context of the HLT will put an increased
pressure on the access efficiency to the conditions data, due to the HLT
time budget constraints.

\subsubsection*{Research and Development Programme}

R\&D actions related to Conditions databases are already in progress,
and all the activities described below should be completed by 2020. This
will provide valuable input for the future HL-LHC TDRs, and allow these
services to be deployed during Run 3 to overcome the limitations seen in
today's solutions.

\begin{itemize}
\item
  File-system view of conditions data for analysis jobs: study how to
  leverage advanced snapshotting/branching features of CVMFS for
  efficiently distributing conditions data as well as ways to optimise
  data/metadata layout in order to benefit from CVMFS caching. Prototype
  production of the file-system view from the conditions database.
\item
  Identify and evaluate industry technologies that could replace
  HEP-specific components.
\item
  ATLAS: migrate current implementations based on COOL to the proposed
  REST-based approach; study how to avoid moving too much complexity on
  the client side, in particular for easier adoption by subsystems, e.g.,
  possibility of common modules/libraries. ALICE is also planning to
  explore this approach for the future, as an alternative or to
  complement the current CVMFS-based implementation.
\end{itemize}

\hypertarget{visualisation}{%
\subsection{Visualisation}\label{visualisation}}

\subsubsection*{Scope and Challenges}

In modern High Energy Physics (HEP) experiments, visualisation of data
has a key role in many activities and tasks across the whole data
processing chain: detector development, monitoring, event generation,
reconstruction, detector simulation, data analysis, as well as outreach
and education.

\emph{Event displays} are the main tool to explore experimental data at
the event level and to visualise the detector itself. There are two main
types of application: firstly, those integrated in the experiments'
frameworks, which are able to access and visualise all the experiments'
data, but at a cost in terms of complexity; secondly,
those designed as cross-platform applications, lightweight and fast,
delivering only a simplified version or a subset of the event data. In
the first case, access to data is tied intimately to an experiment's
data model (for both event and geometry data) and this inhibits
portability; in the second, processing the experiment data into a
generic format usually loses some detail and is an extra processing
step. In addition, there are various graphical backends that can be used
to visualise the final product, either standalone or within a browser,
and these can have a substantial impact on the types of devices
supported.

Beyond event displays, HEP also uses \emph{visualisation of statistical
information}, typically histograms, which allow the analyst to quickly
characterise the data. Unlike event displays, these visualisations are
not strongly linked to the detector geometry, and often aggregate data
from multiple events. Other types of visualisation are used to
\emph{display non-spatial data}, such as graphs for describing the
logical structure of the detector or for illustrating dependencies
between the data products of different reconstruction algorithms.

The main challenges in this domain are in the sustainability of the many
experi\-ment specific visualisation tools when common projects could
reduce duplication and increase quality and long-term maintenance. The
ingestion of events and other data could be eased by common formats,
which would need to be defined and satisfy all users. Changes to support
a client-server architecture would help broaden the ability to support
new devices, such as mobile phones. Making a good choice for the
libraries used to render 3D shapes is also key, impacting on the range
of output devices that can be supported and the level of interaction
with the user. Reacting to a fast-changing technology landscape is very
important -- HEP's effort is limited and generic solutions can often be
used with modest effort. This applies strongly to non-event
visualisation, where many open source and industry standard tools can be
exploited.

\subsubsection*{Current Practices}

Three key features characterise almost all HEP event displays:

\begin{itemize}
\item
  \textbf{Event-based workflow}: applications access experimental data
  on an event-by-event basis, visualising the data collections belonging
  to a particular event. Data can be related to the actual physics
  events (e.g., physics objects such as jets or tracks) or to the
  experimental conditions (e.g., detector descriptions, calibrations).
\item
  \textbf{Geometry visualisation}: The application can display the
  geometry of the detector, as retrieved from the experiments' software
  frameworks, or a simplified description, usually for the sake of speed
  or portability.
\item
  \textbf{Interactivity}: applications offer different interfaces and
  tools to users, in order to interact with the visualisation itself,
  select event data, and set cuts on objects' properties.
\end{itemize}

Experiments have often developed multiple event displays that either
take the full integration approach explained above or are standalone and
rely on extracted and simplified data.

The visualisation of data can be achieved through the low level OpenGL
API, by the use of higher-level OpenGL-based libraries, or within a web
browser using WebGL. Using OpenGL directly is robust and avoids other dependencies, but
implies a significant effort. Instead of using the API directly, a
library layer on top of OpenGL (e.g., Coin3D) can more closely match the
underlying data, such as geometry, and offers a higher level API that
simplifies development. However, this carries the risk that if the
library itself becomes deprecated, as has happened with Coin3D, the
experiment needs to migrate to a different solution or to take on the
maintenance burden itself. Standalone applications often use WebGL
technology to render 3D objects inside a web browser. This is a very
convenient way of rendering 3D graphics, due to the cross-platform
nature of web technologies, and offers many portability advantages (e.g.,
easier support for mobile or virtual reality devices), but at some cost
of not supporting the most complex visualisations requiring heavy
interaction with the experiments' data.

In recent years, video game engines, such as Unity~\cite{unity} or
the Unreal Engine~\cite{Sanders:2016:IUE:3099885},
have become particularly popular in the game and architectural
visualisation industry. They provide very sophisticated graphics engines
and offer a lot of tools for user interaction, such as menu systems or
native handling of VR devices. They are well supported by industry and
tend to have a long lifespan (Unreal Engine is now  20 years old and
is still very popular). However, such engines are meant to be used as
development frameworks and their usage in HEP code is not always
evident. Code should be developed within them, while in HEP
framework-based applications we often want to use graphics libraries
that can be integrated in existing code. A number of HEP collaborations
have started experimenting in building event display tools with such
engines, among them Belle II and ATLAS, but their use is currently
limited to the display of simplified data only.

The new client-server architecture proposed as one of the visualisation
R\&D activities will ease the usage of WebGL technologies and game
engines in HEP.

For statistical data, ROOT has been the tool of choice in HEP for many
years and satisfies most use cases. However, increasing use of generic
tools and data formats means Matplotlib (Python) or JavaScript based
solutions (used, for example, in Jupyter notebooks) have made the
landscape more diverse. For visualising trees or graphs interactively,
there are many generic offerings and experiments have
started to take advantage of them.

\subsubsection*{Research and Development Roadmap}

The main goal of R\&D projects in this area will be to develop
techniques and tools that let visualisation applications and event
displays be less dependent on specific experiments' software frameworks,
leveraging common packages and common data formats. Exporters and
interface packages will be designed as bridges between the experiments'
frameworks, needed to access data at a high level of detail, and the
common packages based on the community standards that this group will
develop.

As part of this development work, demonstrators will be designed to show
the usability of our community solutions and tools. The goal will be to
get a final design of those tools so that the experiments can depend on
them in their future developments.

The working group will also work towards a more convenient access to geometry and
event data, through a client-server interface~\cite{1742-6596-898-7-072015}. In collaboration with the
Data Access and Management Working Group, an API or a service to deliver streamed
event data would be designed.

The work above should be completed by 2020.

Beyond that point, the focus will be on developing the actual
community-driven tools, to be used by the experiments for their
visualisation needs in production, potentially taking advantage of new
data access services.

The workshops that were held as part of the CWP process
(\emph{HSF Visualization Workshop}, see Appendix \ref{appendix-a---list-of-workshops})
were felt to be
extremely useful for exchanging knowledge between developers in
different experiments, fostering collaboration and in bringing in ideas
from outside the community. These will now be held as an annual events and
will facilitate work on the common R\&D plan.

\hypertarget{software-development-deployment-validation-and-verification}{%
\subsection{Software Development, Deployment, Validation and
Verification}\label{software-development-deployment-validation-and-verification}}

\subsubsection*{Scope and Challenges}

Modern HEP experiments are often large distributed collaborations
with several hundred people actively writing software. It is
therefore vital that the processes and tools used for development are
streamlined to ease the process of contributing code and to facilitate
collaboration between geographically separated peers. At the same time,
we must properly manage the whole project, ensuring code quality,
reproducibility, and maintainability with the least effort possible.
Making sure this happens is largely a continuous process and shares a
lot with non-HEP specific software industries.

Work is ongoing to track and promote solutions in the following areas:

\begin{itemize}
\item
  Distributed development of software components, including the tools
  and processes required to do so (code organisation, documentation,
  issue tracking, artefact building), and the best practices in terms of
  code and people management.
\item
  Software quality, including aspects such as modularity and reusability
  of the developed components, architectural and performance best
  practices.
\item
  Software sustainability, including both development and maintenance
  efforts, as well as best practices given long timescales of HEP
  experiments.
\item
  Deployment of software and interaction with operations teams.
\item
  Validation of the software both at small scales (e.g., best practices
  on how to write a unit test) and larger ones (large scale validation
  of data produced by an experiment).
\item
  Software licensing and distribution, including their impact on
  software interoperability.
\item
  Recognition of the significant contribution that software makes to HEP
  as a field (also see Section \ref{training-and-careers} regarding
  career recognition).
\end{itemize}

HEP-specific challenges derive from the fact that HEP is a large,
inhomogeneous community with multiple sources of funding, mostly formed
of people belonging to university groups and HEP-focused laboratories.
Software development effort within an experiment usually encompasses a
huge range of experience and skills, from a few more or less full-time
experts to many physicist programmers with little formal software
training. In addition, the community is split between different
experiments that often diverge in timescales, size, and resources.
Experiment software is usually divided in two separate use cases:
production (being it data acquisition, data reconstruction or
simulation) and user analysis, whose requirements and lifecycles are
completely different. The former is very carefully managed in a
centralised and slow-moving manner, following the schedule of the
experiment itself. The latter is much more dynamic and strongly coupled
with conferences or article publication timelines. Finding solutions
that adapt well to both cases is not always obvious or even possible.

\subsubsection*{Current Practices}

Due to significant variations between experiments at various stages of
their lifecycles, there is a huge variation in practice across the
community. Thus, here we describe \emph{best practice}, with the
understanding that this ideal may be far from the reality for some
developers.

It is important that developers can focus on the design and
implementation of the code and do not have to spend a lot of time on
technical issues. Clear procedures and policies must exist to perform
administrative tasks in an easy and quick way. This starts with the
setup of the development environment. Supporting different platforms not
only allows developers to use their machines directly for
development, it also provides a check of code portability. Clear
guidance and support for good design must be available in advance of
actual coding.

To maximise productivity, it is very beneficial to use development tools
that are not HEP-specific. There are many open source projects that are
of similar scale to large experiment software stacks and standard tools
are usually well documented. For source control HEP has generally chosen
to move to \emph{git}~\cite{git},
which is very welcome, as it also brings an
alignment with many open source projects and commercial organisations.
A major benefit that has come with this technical choice is the use
of social coding sites, such as \emph{GitHub}~\cite{GitHub}
and \emph{GitLab}~\cite{GitLab}, where
code sharing and code review are far superior compared to previous
solutions.
Likewise, \emph{CMake}~\cite{CMake}
is widely used for the builds of software
packages, both within HEP and outside. Packaging many build products
together into a software stack is an area that still requires close
attention with respect to active developments (the HSF has an active
working group here).

Proper testing of changes to code should always be done in advance of a
change request to be accepted. Continuous integration, where `merge' or
`pull' requests are built and tested in advance, is now standard practice
in the open source community and in industry. Continuous integration can
run unit and integration tests, and can also incorporate code quality
checks and policy checks that help improve the consistency and quality
of the code at low human cost. Further validation on different platforms
and at large scales must be as automated as possible, including the
deployment of build artefacts for production.

Training (Section \ref{training-and-careers}) and documentation
are key to efficient use
of developer effort. Documentation must cover best practices and
conventions as well as technical issues. For documentation that has to
be specific, the best solutions have a low barrier of entry for new
contributors, but also allow and encourage review of material.
Consequently, it is very useful to host documentation sources in a
repository with a similar workflow to code, and to use an engine that
translates the sources into modern web pages.

Recognition of software work as a key part of science has resulted in a
number of journals where developers can publish their work~\cite{SSI2017}. Journal publication also disseminates information to the
wider community in a permanent way and is the most established mechanism
for academic recognition. Publication in such journals provides proper
peer review, beyond that provided in conference papers, so it is
valuable for recognition as well as dissemination. However, this
practice is not widespread enough in the community and needs further
encouragement.

\subsubsection*{Research and Development Programme}

HEP must endeavour to be as responsive as possible to developments
outside of our field. In terms of hardware and software tools, there
remains great uncertainty as to what the platforms offering the best
value for money will be on the timescale of a decade. It therefore
behoves us to be as generic as possible in our technology choices,
retaining the necessary agility to adapt to this uncertain future.

Our vision is characterised by HEP being current with technologies and
para\-digms that are dominant in the wider software development community,
especially for open-source software, which we believe to be the right
model for our community. In order to achieve that aim, we propose that
the community establishes a development forum that allows for technology
tracking and discussion of new opportunities. The HSF can play a key
role in marshalling this group and in ensuring its findings are widely
disseminated. In addition, having wider and more accessible training for
developers in the field, that will teach the core skills needed for
effective software development, would be of great benefit.

Given our agile focus, it is better to propose here projects and
objectives to be investigated in the short to medium term, alongside
establishing the means to continually review and refocus the community
on the most promising areas. The main idea is to investigate new tools
as demonstrator projects where clear metrics for success in a reasonable
time should be established to avoid wasting community effort on
initially promising products that fail to live up to expectations.

Ongoing activities and short-term projects, to complete by 2020,
include the following:

\begin{itemize}
\item
  Establish a common forum for the discussion of HEP software problems.
  This should be modeled along the lines of the Concurrency Forum~\cite{ConcurrencyForum}, which was very successful in establishing
  demonstrators and prototypes that were used as experiments started to
  develop parallel data processing frameworks.
\item
  Continue the HSF working group on \emph{Packaging}, with more
  prototype implementations based on the strongest candidates identified
  so far.
\item
  Provide practical advice on how to best set up new software packages,
  developing on the current project template work, and working to
  advertise this within the community.
\item
  Work with HEP experiments and other training projects to provide
  accessible core skills training to the community (see Section \ref{training-and-careers}). This
  training should be experiment-neutral, but could be usefully combined
  with the current experiment specific training. Specifically, this work
  can build on, and collaborate with, recent highly successful
  initiatives such as the LHCb \emph{Starterkit}~\cite{LHCbStarterkit}
  and ALICE \emph{Juniors}~\cite{ALICEJuniors}, and with established
  generic training initiatives such as \emph{Software Carpentry}~\cite{SoftwareCarpentry}.
\item
  Strengthen links with software communities and conferences outside of
  the HEP domain, presenting papers on the HEP experience and problem
  domain. The Scientific Computing with Python (SciPy), the
  Supercomputing Conferences (SCxx), the Conference of
  Research Software Engineers \emph{(RSE)}, and the Workshops on
  Sustainable Software for Science: Practice and Experiences (WSSSPE)
  would all be useful meetings to consider.
\item
  Write a paper that looks at case studies of successful and
  unsuccessful HEP software developments and that draws specific
  conclusions and advice for future projects.
\item
  Strengthen the publication record for important HEP software packages.
  Both peer-reviewed journals~\cite{SSI2017} and citable software version
  records (such as DOIs obtained via Zenodo~\cite{Zenodo}).
\end{itemize}

Longer term projects, to conclude by 2022, include the following:

\begin{itemize}
\item
  Prototype C++ refactoring tools, with specific use cases in migrating
  HEP code.
\item
  Prototyping of portable solutions for exploiting modern vector
  hardware on heterogenous platforms.
\item
  Support the adoption of industry standards and solutions over
  HEP-specific implementations whenever possible.
\item
  Develop tooling and instrumentation to measure software performance
  where tools with sufficient capabilities are not available from
  industry, especially in the domain of concurrency. This should
  primarily aim to further developments of existing tools, such as
  \emph{igprof}~\cite{Eulisse:2005zz},
  rather than to develop new ones.
\item
  Develop a common infrastructure to gather and analyse data about
  experiments' software, including profiling information and code
  metrics, and to ease sharing across different user communities.
\item
  Undertake a feasibility study of a common toolkit for statistical
  analysis that would be of use in regression testing for experiment's
  simulation and reconstruction software.
\end{itemize}

\hypertarget{data-and-software-preservation}{%
\subsection{Data and Software
Preservation}\label{data-and-software-preservation}}

\subsubsection*{Scope and Challenges}

Given the very large investment in particle physics experiments, it is
incumbent upon physicists to preserve the data and the knowledge that
leads to scientific results in a manner such that this investment is not
lost to future generations of scientists. For preserving ``data'', at
whatever stage of production, many of the aspects of the low level
bit-wise preservation have been covered by the Data Preservation for HEP
group~\cite{DPHEP}. ``Knowledge'' preservation encompasses the more
challenging aspects of retaining processing and analysis software,
documentation, and other components necessary for reusing a given
dataset. Preservation of this type can enable new analyses on older
data, as well as a way to revisit the details of a result after
publication. The latter can be especially important in resolving
conflicts between published results, applying new theoretical
assumptions, evaluating different theoretical models, or tuning new
modeling techniques.

Preservation enabling reuse can offer tangible benefits within a given
experiment. The preservation of software and workflows such that they
can be shared enhances collaborative work between analysts and analysis
groups, providing a way of capturing the knowledge behind a given
analysis during the review process. It enables easy transfer of knowledge
to new students or analysis teams, and could establish a manner by which
results can be generated automatically for submission to central
repositories, such as HEPData~\cite{Maguire:2017ypu}.
Preservation within an experiment can
provide ways of reprocessing and reanalysing data that could have been
collected more than a decade earlier. Benefits from preservation are
derived internally whether or not analysis work is approved through the
publication approval process for an experiment. Providing such immediate
benefits makes the adoption of data preservation in experiment workflows
particularly desirable.

A final series of motivations comes from the potential re-use by others
outside of the HEP experimental community. Significant outreach efforts
to bring the excitement of analysis and discovery to younger students
have been enabled by the preservation of experimental data and software
in an accessible format. Many examples also exist of phenomenology
papers reinterpreting the results of a particular analysis in a new
context. This has been extended further with published results based on
the reanalysis of processed data by scientists outside of the
collaborations. Engagement of external communities, such as machine
learning specialists, can be enhanced by providing the capability to
process and understand low-level HEP data in portable and relatively
platform-independent way, as happened with the Kaggle ML
challenges~\cite{pmlr-v42-cowa14}.
This allows external users direct access to the same tools
and data as the experimentalists working in the collaborations.
Connections with industrial partners, such as those fostered by CERN
OpenLab, can be facilitated in a similar manner.

Preserving the knowledge of analysis, given the extremely wide scope of
how analysts do their work and experiments manage their workflows, is
far from easy. The level of reuse that is applicable needs to be
identified, and so a variety of preservation systems will probably be
appropriate given the different preservation needs between large central
experiment workflows and the work of an individual analyst. The larger
question is to what extent common low-level tools can be provided that
address similar needs across a wide scale of preservation problems.
These would range from capture tools, that preserve the details of an
analysis and its requirements, to ensuring that software and services
needed for a workflow would continue to function as required.

The above-mentioned steps are consistent with the
FAIR data principles that are increasingly being mandated by funding
agencies~\cite{FAIRdata}.

\subsubsection*{Current Practices}

Each of the LHC experiments has adopted a data access and/or data
preservation policy, all of which can be found on the CERN Open Data
Portal~\cite{ODP}. All of the LHC experiments support public access to
some subset of the data in a highly reduced data format for the purposes
of outreach and education. CMS has gone one step further, releasing
substantial datasets in an Analysis Object Data (AOD) format that can be
used for new analyses. The current data release includes simulated data,
virtual machines that can instantiate the added analysis examples, and
extensive documentation~\cite{CMS-OpenData}. ALICE has promised to
release 10\% of their processed data after a five-year embargo and has
released 2010 data at this time~\cite{ALICE-OpenData}.
LHCb is willing to make access to reconstructed data available, but is
unable to commit to a specific timescale due to resource limitations. A
release of ntuple-level data for one high profile analysis, aimed
primarily at educational activities, is currently in preparation. ATLAS
has chosen a different direction for data release: data associated with
journal publications is made available, and ATLAS also strives to make
available additional material that allows reuse and reinterpretations of
the data in the context of new theoretical models~\cite{ATL-CB-PUB-2015-001}.
ATLAS is  exploring how
to provide the capability for reinterpretation of searches in the future
via a service such as
RECAST~\cite{Cranmer:1299950}, in which the original internal analysis
code (including full detector simulation and reconstruction) is
preserved, as opposed to the re-coding approach with object-efficiency
calibrations used by external reinterpretation toolkits. All experiments
frequently provide detailed supplemental data along with publications to
allow for more detailed comparisons between results, or even
reinterpretation.

The LHC experiments have not yet set a formal policy addressing the new
capabilities of the CERN Analysis Preservation Portal (CAP)~\cite{CAP}
and whether or not some use of it will be required or merely encouraged.
All of them support some mechanisms for internal preservation of the
knowledge surrounding a physics publication~\cite{Shiers2017}.

\subsubsection*{Research and Development Programme}

There is a significant programme of work already happening in the data
preservation area. The feasibility and cost of common base services have
been studied for bit preservation, the preservation of executable
software environments, and the structured capturing of analysis metadata~\cite{Berghaus2016}.

The goals presented here should be orchestrated in conjunction with
projects conducted by the R\&D programmes of other working groups, since
the questions addressed are common. Goals to address on the timescale of
2020 are:

\begin{itemize}
\item
  Include embedded elements for the capture of preservation information
  and metadata and tools for the archiving of this information in
  developing a prototype analysis ecosystem(s). This should include an
  early demonstration of the CAP analysis preservation portal with a
  working UI.
\item
  Demonstrate the capability to provision and execute production
  workflows for experiments that are composed of multiple independent
  containers.
\item
  Collection of analysis use cases and elements that are necessary to
  preserve in order to enable re-use and to ensure these analyses can be
  captured in developing systems. This should track analysis evolution
  towards possible \emph{Big Data} environments and determine any elements
  that are difficult to capture, spawning further R\&D.
\item
  Evaluate, in the preservation area, the full potential and limitations
  of sandbox and ``freezing'' technologies, possibly coupled with
  version and history control software distribution systems.
\item
  Develop prototypes for the preservation and validation of large-scale
  production executables and workflows.
\item
  Integrate preservation capabilities into newly developed computing
  tools and workflows.
\item
  Extension and standardisation of the \emph{final data and analysis}
  preservation sche\-me via HEPData, Rivet and/or other reinterpretation
  tools. This could be used to preserve a sufficiently detailed
  re-usable record of many LHC Run 2 research outputs.
\end{itemize}

This would then lead naturally to deployed solutions that support data
preservation in the 2020-2022 time frame for the HEP experimental
programmes, in particular an analysis ecosystem that enables reuse for
any analysis that can be conducted in the ecosystem, and a system for
the preservation and validation of large-scale production workflows.

\hypertarget{security}{%
\subsection{Security}\label{security}}

\subsubsection*{Scope and Challenges}

Security is a cross-cutting area that impacts our projects,
collaborative work, users, and software infrastructure fundamentally. It
crucially shapes our reputation, our collaboration, the trust between
participants, and the users' perception of the quality and ease of use
of our services.

There are three key areas:

\begin{itemize}
\item
  Trust and policies; this includes trust models, policies, compliance,
  data protection issues.
\item
  Operational security; this includes threat intelligence, security
  operations, incident response.
\item
  Authentication and Authorisation; this includes identity management,
  identity federation, access control.
\end{itemize}

The evolution in the security domain requires the HEP community to work in
collaboration with the various national security organisations and
policy groups, building on many relationships that are already
established.

\paragraph{Trust and Policies}

Data Protection defines the boundaries that enable HEP work to be
conducted, in particular regarding data sharing aspects, for example
between the EU and the US. It is essential to establish a trusted
personal data exchange framework, minimising the amount of personal data
to be processed and ensuring legal compliance.

Beyond legal compliance and best practice, offering open access to
scientific resources and achieving shared goals requires prioritising
the protection of people and science, including the mitigation of the
effects of surveillance programs on scientific collaborations.

On the technical side, it is necessary to adapt the current, aging trust
model and security architecture relying solely on X.509 (which is no longer
the direction industry is taking), in order to include modern data
exchange design, for example involving commercial providers or hybrid
clouds. The future of our infrastructure involves increasingly diverse
resource providers connected through cloud gateways. For example,
HEPCloud~\cite{HEPCloud} at FNAL aims to connect Amazon,
Google Clouds, and HPC centres with our traditional grid computing
resources. The HNSciCloud European Project~\cite{HNSciCloud} aims to
support the enhancement of commercial cloud providers in order to be
leveraged by the scientific community. These are just two out of a
number of endeavours. As part of this modernisation, a transition is
needed from a model in which all participating organisations are bound
by custom HEP security policies to a more flexible approach where some
partners are not in a position to adopt such policies.

\paragraph{Operational Security and Threat Intelligence}

As attacks have become extremely sophisticated and costly to defend
against, the only cost-effective strategy is to address security threats
together, as a community. This involves constantly striving to liaise
with external organisations, including security vendors and law
enforcement entities, to enable the sharing of indicators of compromise
and threat intelligence between all actors. For organisations from all
sectors, including private companies, governments, and academia, threat
intelligence has become the main means by which to detect and manage
security breaches.

In addition, a global forum for HEP and the larger Research and Education
(R\&E) community needs to be built, where security experts feel
confident enough to share threat intelligence and security expertise. A
key to success is to ensure a closer collaboration between HEP security
contacts and campus security. The current gap at many HEP organisations
is both undermining the community's security posture and reducing the
effectiveness of the HEP security strategy.

There are several very active trust groups in the HEP community where
HEP participants share threat intelligence and organise coordinated
incident response \cite{EGI-CSIRT,REN-ISAC,XSEDE}.
There is unfortunately still no global Research and Education forum for
incident response, operational security, and threat intelligence
sharing. With its mature security operations and dense, global network
of HEP organisations, both of which are quite unique in the research
sector, the HEP community is ideally positioned to contribute to such a
forum and to benefit from the resulting threat intelligence, as it has
exposure, sufficient expertise, and connections to lead such an
initiative. It may play a key role in protecting multiple scientific
domains at a very limited cost.

There will be many technology evolutions as we start to take a serious
look at the next generation internet. For example, IPv6 is one upcoming
change that has yet to be fully understood from the security
perspective. Another high impact area is the internet of things (IoT),
connected devices on our networks that create new vectors of attack.

It will become necessary to evaluate and maintain operational security
in connected environments spanning public, private, and hybrid clouds.
The trust relationship between our community and such providers has yet
to be determined, including the allocation of responsibility for
coordinating and performing vulnerability management and incident
response. Incompatibilities between the e-Infrastructure approach to
community-based incident response and the ``pay-for-what-you-break''
model of certain commercial companies may come to light and must be
resolved.

\paragraph{Authentication and Authorisation Infrastructure}

It is now largely acknowledged that end-user certificates are
challenging to manage and create a certain entrance barrier to our
infrastructure for early career researchers. Integrating our access
control management system with new, user-friendly technologies and
removing our dependency on X.509 certificates is a key area of interest
for the HEP Community.

An initial step is to identify other technologies that can satisfy
traceability, isolation, privilege management and other requirements
necessary for HEP workflows. The chosen solution should prioritise
limiting the amount of change required to our services and follow
accepted standards to ease integration with external entities, such as
commercial clouds and HPC centres.

Trust federations and inter-federations, such as the R\&E standard
eduGAIN \cite{eduGAIN}, provide a needed functionality for
Authentication. They can remove the burden of identity provisioning from
our community and allow users to leverage their home organisation
credentials to access distributed computing resources. Although certain
web-based services have enabled authentication via such federations,
uptake is not yet widespread. The challenge remains to have the
necessary attributes published by each federation to provide robust
authentication.

The existing technologies leveraged by identity federations, e.g., the
Security Assertion Markup Language (SAML), have not supported non-web
applications historically. There is momentum within the wider community
to develop next-generation identity federations~\cite{OICFederation}
that natively support a
wider range of clients. In the meantime there are several viable interim
solutions that are able to provision users with the token required to
access a service (such as X.509) transparently, translated from their
home organisation identity.

Although non-X509 federated identity provides a potential solution for our
challenges in Authentication, Authorisation should continue to be
tightly controlled by the HEP community. Enabling Virtual Organisation
(VO) membership for federated credentials and integrating such a
workflow with existing identity vetting processes is a major topic
currently being worked on, in particular within the WLCG community.
Commercial clouds and HPC centres have fundamentally different access
control models and technologies from our grid environment. We shall need
to enhance our access control model to ensure compatibility and
translate our grid-based identity attributes into those consumable by
such services.

\subsubsection*{Current Activities}

Multiple groups are working on policies and establishing a common trust
framework, including the EGI Security Policy Group
\cite{EGISecurityPolicyGroup} and the Security for Collaboration among
Infrastructures working group~\cite{SCI_WG}.

Operational security for the HEP community is being followed up in the
WLCG Working Group on Security Operations Centres~\cite{WLCG-SOC-WG}. The
HEP Community is actively involved in multiple operational security
groups and trust groups, facilitating the exchange of threat
intelligence and incident response communication. WISE~\cite{WISE}
provides a forum for e-Infrastructures to share and develop security
best practices and offers the opportunity to build relationships between
security representatives at multiple e-infrastructures of interest to
the HEP community.

The evolution of Authentication and Authorisation is being evaluated in
the recently created WLCG Working Group on Authorisation.
In parallel, HEP is contributing to a wider effort
to document requirements for multiple Research Communities through the
work of FIM4R~\cite{FIM4R}. Participation of CERN and a few other major
WLCG sites in the European
Authentication and Authorisation for Research and Collaboration (AARC)
project~\cite{AARC} provides the opportunity to ensure
that any directions chosen are consistent with those taken by the wider
community of research collaborations. The flow of attributes between
federated entities continues to be problematic, disrupting the
authentication flow. Trust between service providers and identity
providers is still evolving, and efforts within the R\&E Federations
Group (REFEDS)~\cite{REFEDS} and the AARC project aim to address the
visibility of both the level of assurance of identities and the security
capability of federation participants (through Sirtfi~\cite{Sirtfi}).

\subsubsection*{Research and Development Programme}

Over the next decade, it is expected that considerable changes will be
made to address security in the domains highlighted above. The
individual groups, in particular those mentioned above, working in the
areas of trust and policies, operational security, authentication and
authorisation, and technology evolutions, are driving the R\&D
activities. These groups are generally much broader than just the HEP
community. The list below summarises the most important actions:

\paragraph{Trust and Policies}
\begin{itemize}
\item
By 2020:
\begin{itemize}
    \item
      Define and adopt policies in line with new EU Data Protection
      requirements.
    \item
      Develop frameworks to ensure trustworthy interoperability of
      infrastructures and communities.
\end{itemize}

\item
By 2022:
\begin{itemize}
    \item
      Create and promote community driven incident response policies and
      procedures.
\end{itemize}
\end{itemize}

\paragraph{Operational Security and threat intelligence}
\begin{itemize}
\item
By 2020:
\begin{itemize}
\item
  Offer a reference implementation, or at least specific guidance, for a
  Security Operation Centre deployment at HEP sites, enabling them to
  take action based on threat intelligence shared within the HEP
  community.
\end{itemize}
\item
By 2022:
\begin{itemize}
\item
  Participate in the founding of a global Research and Education Forum
  for incident response, since responding as a global community is the only
  effective solution against global security threats.
\item
  Build the capabilities to accommodate more participating organisations
  and streamline communication workflows, within and outside HEP,
  including maintaining a list of security contacts, secure communications
  channels, and security incident response mechanisms.
\item
  Reinforce the integration of HEP security capabilities with their
  respective home organisation, to ensure adequate integration of HEP
  security teams and site security teams.
\end{itemize}
\item
By 2025:
\begin{itemize}
\item
  Prepare adequately as a community, in order to enable HEP
  organisations to operate defendable services against more
  sophisticated threats, stemming both from global cyber-criminal gangs
  targeting HEP resources (finance systems, intellectual property,
  ransomware), as well as from state actors targeting the energy and
  research sectors with advanced malware.
\end{itemize}
\end{itemize}

\paragraph{Authentication and Authorisation}

\begin{itemize}
\item
By 2020:
\begin{itemize}
\item
  Ensure that ongoing efforts in trust frameworks are sufficient to
  raise the level of confidence in non-X509 federated identities to the
  equivalent of X.509, at which stage they could be a viable alternative
  to both grid certificates and CERN accounts.
\item
  Participate in setting directions for the future of identity
  federations, through the FIM4R~\cite{FIM4R} community.
\end{itemize}
\item
By 2022:
\begin{itemize}
\item
  Overhaul the current Authentication and Authorisation infrastructure,
  including Token Translation, integration with Community IdP-SP
  Proxies, and Membership Management tools. Enhancements in this area
  are needed to support a wider range of user identities for WLCG
  services.
\end{itemize}
\end{itemize}

\hypertarget{training-and-careers}{%
\section{Training and Careers}\label{training-and-careers}}

For HEP computing to be as successful as possible, the careers and
skills of the individuals who participate must be considered. Ensuring
that software developers can acquire the necessary skills and obtain
successful careers is considered an essential goal of the HSF, which has
the following specific objectives in its mission:

\begin{itemize}
\item
  To provide training opportunities for developers; this should include
  the support to the software schools for young scientists and computer
  engineers, and of a permanent training infrastructure for accomplished
  developers;
\item
  To provide career support for developers, for instance by listing job
  opportunities and by helping to shape well-defined career paths that
  provide advancement opportunities on a par with those in, for example,
  detector construction;
\item
  To increase the visibility of the value of software developers in HEP,
  recognising that it has scientific research value on an equal footing
  with other activities, and acknowledging and advocating for researchers
  who choose this as their speciality.
\end{itemize}

\subsection{Training Challenges}

HEP is facing major challenges with its software and computing that
require innovative solutions based on the proper adoption of new
technologies. More and more technologies are emerging as scientific
communities and industry face similar challenges and produce solutions
relevant to us. Integrating such technologies in our software and
computing infrastructure requires specialists, but it is also important that a
large fraction of the community is able to use these new tools and
paradigms. Specific solutions and optimisations must be implemented by
the HEP community itself, since many advanced requirements are unique to
our field.

There is a very close collaboration, even overlap,
in HEP between users of software and developers. This has given experiments
an agility that was often essential for success in the past.
Many details of experiment data
cannot be known before data taking has started, and each change in
detector technology or machine performance improvement can have
important consequences for the software and computing infrastructure. In
the case of detectors, engineers and physicists are required to have a
good understanding of each other's field of expertise. In the same way,
it is necessary that physicists understand some of the complexities of
writing software, and that software experts are able to fathom the
requirements of physics problems.

Training must address an audience with very diverse computing skills,
ranging from novice programmers to advanced developers and users.
It must be used to spread best software engineering practices and
software technologies to a very large number of people, including the
physicists involved across the whole spectrum of data processing tasks,
from triggering to analysis. It must be done by people who have a sound
knowledge of the scientific and technical details, who prepare training
material despite the many calls on their time. Training thus needs
proper recognition to ensure that it happens and is carried out well.

HEP is seen as an interesting, innovative, and challenging field.
This is a great advantage in attracting talented young people looking
for experience in a challenging and diverse environment in which
they can acquire skills that will be valuable, even in other fields.
As discussed in Software Development (Section
\ref{software-development-deployment-validation-and-verification}),
using industry standard tools across different experiments,
and training people in how to use them properly,
helps with people's later career prospects
and makes our field even more attractive.
At the same time, experiments have a scientific programme to accomplish
and also to focus on the specific training required to
accomplish their specific goals. The right balance must be found between
these two requirements. It is necessary to find the right incentives to
favour training activities that bring more benefits in the medium to
long term, for the experiment, the community, and the careers of the
trainees.

\subsection{Possible Directions for Training}

To increase training activities in the community, whilst taking into
account the constraints of both the attendees and the trainers, we should
explore new approaches to training. The current ``school''
model is well established, as exemplified by three well-known successful
schools, the CERN School of Computing~\cite{CSC},
the Bertinoro School of Computing~\cite{ESC2017}
and the GridKa School of Computing~\cite{GridKASchool}.
They require a significant amount of
dedicated time of all the participants, at the same time and location,
and therefore are difficult to scale to meet the needs of a large number
of students. In view of this, we should identify opportunities to work
with HEP experiments and other training projects to provide accessible
core skills training to the community by basing them at laboratories
where students can easily travel. A number of highly successful
experiment-specific examples exist, such as the LHCb StarterKit~\cite{LHCbStarterkit}
and ALICE Juniors~\cite{ALICEJuniors}, as well as
established generic training initiatives, such as
Software Carpentry~\cite{SoftwareCarpentry}. As with hands-on tutorials organised during
conferences and workshops, the resulting networking is an important and
distinctive benefit of these events, where people build relationships
with other colleagues and experts.

In recent years, several R\&D projects, such as DIANA-HEP~\cite{DIANA-HEP} and
AMVA4\-New\-Physics~\cite{AMVA4NewPhysics}, have had training as one
of their core activities. This
has provided an incentive to organise training events and has resulted
in the spread of expertise on advanced topics. We believe that training
should become an integral part of future major R\&D projects.

New pedagogical methods, such as active training and peer training, that
are complementary to schools or topical tutorials, also deserve more
attention. Online material can be shared by a student and a teacher to
provide the exchange of real examples and practical exercises. For
example, notebook technologies, such as Jupyter, support embedding of
runnable code and comments into the same document. The initial material
can be easily enriched by allowing other students and experts to add
comments and more examples in a collaborative way. The HSF started to
experiment with this approach with WikiToLearn~\cite{WikiToLearn}, a
platform developed in Italy outside HEP that promotes this kind of
training and collaborative enrichment of the training material. Projects
such as ROOT have also started to provide some
training material based
on notebooks.

A lot of initiatives have been undertaken by the software community that
HEP can benefit from, and materials have been made available in the form
of online tutorials, active training, and Massive Open Online Courses
(MOOCs). Some effort needs to be invested to evaluate existing courses
and build a repository of selected ones that are appropriate to HEP
needs. This is not a negligible task and would require some dedicated
effort to reach the appropriate level of support. It should help to
increase training efficiency by making it easier to identify appropriate
courses or initiatives.

A model that emerged in recent years as a very valuable means of sharing
expertise is to use Question and Answer (Q\&A) systems, such as Stack
Overflow~\cite{Stackoverflow}. A few such systems are run by experiments for their own needs,
but this is not necessarily optimal, as the value of these services is
increased by a large number of contributors with diverse backgrounds.
Running a cross-experiment Q\&A system has been discussed, but it has
not yet been possible to converge on a viable approach, both technically
and because of the effort required to run and support such a service.

\subsection{Career Support and Recognition}

Computer specialists in HEP are often physicists who have chosen to
specialise in computing. This has always been the case and needs to
continue. Nevertheless, for young people in particular, this leads to a
career recognition problem, as software and computing activities are not
well-recognised roles in various institutions supporting HEP research
and recruiting people working in the field. The exact situation is
highly dependent on policies and boundary conditions of the organisation
or country, but recognition of physicists tends to be based generally on
participation in data analysis or hardware developments. This is even a
bigger problem if the person is spending time contributing to training
efforts. This negatively impacts the future of these people and reduces
the possibility of HEP engaging them in the training effort of the
community when the community actually needs more people to participate
in this activity. Recognition of training efforts, either by direct
participation in training activities or by providing materials, is an
important issue to address, complementary to the incentives mentioned
above.

There is no easy solution to this problem. Part of the difficulty is
that organisations, and in particular the people inside them in charge
of the candidate selections for new positions and promotions, need to
adapt their expectations to these needs and to the importance of having
computing experts with a strong physics background as permanent members
of the community. Experts writing properly engineered and optimised
software can significantly reduce resource consumption and increase
physics reach, which provides huge financial value to modern HEP
experiments. The actual path for improvements in career recognition, as
the possible incentives for participating in the training efforts,
depends on the local conditions.

\hypertarget{conclusions}{%
\section{Conclusions}\label{conclusions}}

Future challenges for High Energy Physics in the domain of software and
computing are not simply an extrapolation of the challenges faced today.
The needs of the HEP programme in the high luminosity era far exceed those
that can be met by simply making incremental changes to today's code and
scaling up computing facilities within the anticipated budget. At the
same time, the limitation in single core CPU performance is making the
landscape of computing hardware far more diverse and challenging to
exploit, whilst offering huge performance boosts for suitable code.
Exploiting parallelism and other new techniques, such as modern machine
learning, offer great promise, but will require substantial work from
the community to adapt to our problems. If there were any lingering
notion that software or computing could be done cheaply by a few junior
people for modern experimental programmes, it should now be thoroughly
dispelled.

We believe HEP Software and Computing requires a step change in its profile and
effort to match the challenges ahead. We need investment in people who
can understand the problems we face, the solutions employed today, and
have the correct skills to provide innovative solutions for the future.
There needs to be recognition from the whole community for the work done
in this area, with a recognised career path for these experts. In
addition, we will need to invest heavily in training for the whole
software community as the contributions of the bulk of non-expert
physicists are also vital for our success.

We know that in any future scenario development effort will be
constrained, so it is vital that successful R\&D projects provide
sustainable software for the future. It is important to emphasise that the goal
is to support the HEP physics programme in a cost effective manner, so the
deployment consequences of a particular technology choice or direction must be
understood with partners in distributed computing. In many areas it is recognised
that different experiments could have adopted common solutions,
reducing overall development effort and increasing robustness and
functionality. That model of duplicated development is not sustainable. We must
endeavour to achieve better coherence within HEP for future developments
to build advanced, open-source projects that can be shared and supported
in common. The HSF has already established itself as a forum that can
facilitate this. Establishing links outside of HEP, to other academic
disciplines, to industry, and to the computer science community, can
strengthen both the research and production phases of new solutions. We
should ensure that the best products are chosen, from inside and outside
HEP, and that they receive support from all parties, aiming at technical
excellence and economy of scale.

We have presented programmes of work that the community has identified
as being part of the roadmap for the future. While there is always some
scope to reorient current effort in the field, we would highlight the
following work programmes as being of the highest priority for
investment to address the goals that were set in the introduction.

\paragraph{Improvements in software efficiency, scalability and performance}

\begin{quote}
The bulk of CPU cycles consumed by experiments relate to the fundamental
challenges of simulation and reconstruction. Thus, the work programmes
in these areas, together with the frameworks that support them, are of
critical importance. The sheer volumes of data involved make research
into appropriate data formats and event content to reduce storage
requirements vital. Optimisation of our distributed computing systems,
including data and workload management, is paramount.
\end{quote}

\paragraph{Enable new approaches that can radically extend physics reach}

\begin{quote}
New techniques in simulation and reconstruction will be vital here.
Phys\-ics analysis is an area where new ideas can be particularly
fruitful. Exploring the full potential of machine learning is one common
theme that underpins many new approaches and the community should
endeavour to share knowledge widely across subdomains. New data analysis
paradigms coming from the Big Data industry, based on innovative
parallelised data processing on large computing farms, could transform
data analysis.
\end{quote}

\paragraph{Ensure the long-term sustainability of the software}

\begin{quote}
Applying modern software development techniques to our codes has
increased, and will continue to increase, developer productivity and
code quality. There is ample scope for more common tools and common
training to equip the community with the correct skills. Data
Preservation makes sustainability an immediate goal of development and
analysis and helps to reap the benefits of our experiments for decades
to come. Support for common software used across the community needs to
be recognised and accepted as a common task, borne by labs, institutes,
experiments, and funding agencies.
\end{quote}

The R\&D actions proposed in this Roadmap have taken into account the
charges that were laid down. When considering a specific project
proposal addressing our computing challenges, that project's impact, measured
against the charges, should be evaluated. Over the next decade, there
will almost certainly be disruptive changes that cannot be planned for,
and we must remain agile enough to adapt to these.

The HEP community has many natural subdivisions, between different
regional funding agencies, between universities and laboratories, and
between different experiments. It was in an attempt to overcome these
obstacles, and to encourage the community to work together in an
efficient and effective way, that the HEP Software Foundation was
established in 2014. This Community White Paper process has been
possible only because of the success of that effort in bringing the
community together. The need for more common developments in the future,
as underlined here, reinforces the importance of the HSF as a common
point of contact between all the parties involved, strengthening our
community spirit and continuing to help share expertise and identify
priorities. Even though this evolution will also require projects and
experiments to define clear priorities about these common developments,
we believe that the HSF, as a community effort, must be strongly
supported as part of our roadmap to success.

\onecolumn

%% file: appendices.tex
\begin{appendices}

\hypertarget{appendix-a---list-of-workshops}{%
\section{List of Workshops}\label{appendix-a---list-of-workshops}}

\textbf{HEP Software Foundation Workshop}\\
\emph{Date:} 23-26 Jan, 2017\\
\emph{Location:} UCSD/SDSC (La Jolla, CA, USA)\\
\emph{URL:}
\href{http://indico.cern.ch/event/570249/}{{http://indico.cern.ch/event/570249/}}\\
\emph{Description:} This HSF workshop at SDSC/UCSD was the first
workshop supporting the CWP process. There were plenary sessions
covering topics of general interest as well as parallel sessions for the
many topical working groups in progress for the CWP.\\

\noindent
\textbf{Software Triggers and Event Reconstruction WG meeting}\\
\emph{Date:} 9 Mar, 2017\\
\emph{Location:} LAL-Orsay (Orsay, France)\\
\emph{URL:}
\href{https://indico.cern.ch/event/614111/}{{https://indico.cern.ch/event/614111/}}\\
\emph{Description:} This was a meeting of the Software Triggers and Event
Reconstruction CWP working group. It was held as a parallel session at
the ``Connecting the Dots'' workshop, which focuses on forward-looking
pattern recognition and machine learning algorithms for use in HEP.\\

\noindent
\textbf{IML Topical Machine Learning Workshop}\\
\emph{Date:} 20-22 Mar, 2017\\
\emph{Location:} CERN (Geneva, Switzerland)\\
\emph{URL:}
\href{https://indico.cern.ch/event/595059}{{https://indico.cern.ch/event/595059}}\\
\emph{Description:} This was a meeting of the Machine Learning CWP
working group. It was held as a parallel session at the
``Inter-experimental Machine Learning (IML)'' workshop, an organisation
formed in 2016 to facilitate communication regarding R\&D on ML
applications in the LHC experiments.\\

\noindent
\textbf{Community White Paper Follow-up at FNAL}\\
\emph{Date:} 23 Mar, 2017\\
\emph{Location:} FNAL (Batavia, IL, USA)\\
\emph{URL:}
\href{https://indico.fnal.gov/conferenceDisplay.py?confId=14032}{{https://indico.fnal.gov/conferenceDisplay.py?confId=14032}}\\
\emph{Description:} This one-day workshop was organised to engage with
the experimental HEP community involved in computing and software for
Intensity Frontier experiments at FNAL. Plans for the CWP were
described, with discussion about commonalities between the HL-LHC
challenges and the challenges of the FNAL neutrino and muon
experiments\\

\noindent
\textbf{CWP Visualisation Workshop}\\
\emph{Date:} 28-30 Mar, 2017\\
\emph{Location:} CERN (Geneva, Switzerland)\\
\emph{URL:}
\href{https://indico.cern.ch/event/617054/}{{https://indico.cern.ch/event/617054/}}\\
\emph{Description:} This workshop was organised by the Visualisation CWP
working group. It explored the current landscape of HEP visualisation
tools as well as visions for how these could evolve. There was
participation both from HEP developers and industry.\\

\noindent
\textbf{DS@HEP 2017 (Data Science in High Energy Physics)}\\
\emph{Date:} 8-12 May, 2017\\
\emph{Location:} FNAL (Batava, IL, USA)\\
\emph{URL:}
\href{https://indico.fnal.gov/conferenceDisplay.py?confId=13497}{{https://indico.fnal.gov/conferenceDisplay.py?confId=13497}}\\
\emph{Description:} This was a meeting of the Machine Learning CWP
working group. It was held as a parallel session at the ``Data Science
in High Energy Physics (DS@HEP)'' workshop, a workshop series begun in
2015 to facilitate communication regarding R\&D on ML applications in
HEP.\\

\noindent
\textbf{HEP Analysis Ecosystem Retreat}\\
\emph{Date:} 22-24 May, 2017\\
\emph{Location:} Amsterdam, the Netherlands\\
\emph{URL:}
\href{http://indico.cern.ch/event/613842/}{{http://indico.cern.ch/event/613842/}}\\
\emph{Summary report:}
\href{http://hepsoftwarefoundation.org/assets/AnalysisEcosystemReport20170804.pdf}{{http://cern.ch/go/mT8w}}\\
\emph{Description:} This was a general workshop, organised about the
HSF, about the ecosystem of analysis tools used in HEP and the ROOT
software framework. The workshop focused both on the current status and
the 5-10 year time scale covered by the CWP.\\

\noindent
\textbf{CWP Event Processing Frameworks Workshop}\\
\emph{Date:} 5-6 Jun, 2017\\
\emph{Location:} FNAL (Batavia, IL, USA)\\
\emph{URL:}
\href{https://indico.fnal.gov/conferenceDisplay.py?confId=14186}{{https://indico.fnal.gov/conferenceDisplay.py?confId=14186}}\\
\emph{Description:} This was a workshop held by the Event Processing Frameworks
CWP working group focused on writing an initial draft of the framework
white paper. Representatives from most of the current practice
frameworks participated.\\

\noindent
\textbf{HEP Software Foundation Workshop}\\
\emph{Date:} 26-30 Jun, 2017\\
\emph{Location:} LAPP (Annecy, France)\\
\emph{URL:}
\href{https://indico.cern.ch/event/613093/}{{https://indico.cern.ch/event/613093/}}\\
\emph{Description:} This was the final general workshop for the CWP
process. The CWP working groups came together to present their status
and plans, and develop consensus on the organisation and context for the
community roadmap. Plans were also made for the CWP writing phase that
followed in the few months following this last workshop.

\newpage
\hypertarget{appendix-b-glossary}{%
\section{Glossary}\label{appendix-b-glossary}}

\begin{description}

\item[AOD] Analysis Object Data is a summary of the reconstructed event and
contains sufficient information for common physics analyses.

\item[ALPGEN] An event generator designed for the generation of Standard Model
processes in hadronic collisions, with emphasis on final states with
large jet multiplicities. It is based on the exact LO evaluation of
partonic matrix elements, as well as top quark and gauge boson decays
with helicity correlations.

\item[BSM] Physics beyond the Standard Model (BSM) refers to the theoretical
developments needed to explain the deficiencies of the Standard Model
(SM), such as the
\href{https://en.wikipedia.org/wiki/Origin_of_mass}{origin of mass}, the
\href{https://en.wikipedia.org/wiki/Strong_CP_problem}{strong CP
problem},
\href{https://en.wikipedia.org/wiki/Neutrino_oscillation}{neutrino
oscillations},
\href{https://en.wikipedia.org/wiki/Baryon_asymmetry}{matter--antimatter
asymmetry}, and the nature of
\href{https://en.wikipedia.org/wiki/Dark_matter}{dark matter} and
\href{https://en.wikipedia.org/wiki/Dark_energy}{dark energy}.

\item[Coin3D] A C++ object oriented retained mode 3D graphics API
used to provide a higher layer of programming for OpenGL.

\item[COOL] LHC Conditions Database Project, a subproject of the POOL persistency framework.

\item[Concurrency Forum] Software engineering is moving towards a paradigm shift in order to accommodate new CPU
architectures with many cores, in which concurrency will play a more fundamental role in programming languages
and libraries. The forum on concurrent programming models and frameworks aims to share knowledge among interested
parties that work together to develop 'demonstrators' and agree on technology so that they can share code and compare results.

\item[CRSG] Computing Resources Scrutiny Group, a WLCG committee in charge of scrutinizing and assessing LHC experiment yearly resource requests to prepare funding agency decisions.

\item[CSIRT] Computer Security Incident Response Team. A CSIRT provides a reliable and trusted single point of contact for reporting computer security incidents and taking the appropriate measures in response tothem.

\item[CVMFS] The CERN Virtual Machine File System is a network file system
based on HTTP and optimised to deliver experiment software in a fast,
scalable, and reliable way through sophisticated caching strategies.

\item[CWP] The Community White Paper (this document) is the result of an
organised effort to describe the community strategy and a roadmap for
software and computing R\&D in HEP for the 2020s. This activity is
organised under the umbrella of the HSF.

\item[Deep Learning (DL)] one class of Machine Learning algorithms, based on a
high number of neural network layers.

\item[DNN] Deep Neural Network, class of neural networks with typically a
large number of hidden layers through which data is processed.

\item[DPHEP] The Data Preservation in HEP project is a collaboration for data
preservation and long term analysis.

\item[EGI] European Grid Initiative. A European organisation in charge of
delivering advanced computing services to support scientists,
multinational projects and research infrastructures, partially funded by
the European Union. It is operating both a grid infrastructure (many
WLCG sites in Europe are also EGI sites) and a federated cloud
infrastructure. It is also responsible for security incident response
for these infrastructures (CSIRT).

\item[FAIR] The Facility for Antiproton and Ion Research (FAIR) is located at
GSI Darmstadt. It is an international accelerator facility for research
with antiprotons and ions.

\item[FAIR] An abbreviation for a set of desirable data properties: Findable,
Accessible, Interoperable, and Re-usable.

\item[FCC] Future Circular Collider, a proposed new accelerator complex for
CERN, presently under study.

\item[FCC-hh] A 100 TeV proton-proton collider version of the FCC (the ``h''
stands for ``hadron'').

\item[GAN] Generative Adversarial Networks are a class of
\href{https://en.wikipedia.org/wiki/Artificial_intelligence}{artificial
intelligence} algorithms used in
\href{https://en.wikipedia.org/wiki/Unsupervised_machine_learning}{unsupervised
machine learning}, implemented by a system of two
\href{https://en.wikipedia.org/wiki/Neural_network}{neural networks}
contesting with each other in a
\href{https://en.wikipedia.org/wiki/Zero-sum_game}{zero-sum game}
framework.

\item[Geant4] A toolkit for the simulation of the passage of
particles through matter.

\item[GeantV] An R\&D project that aims to fully exploit the
parallelism, which is increasingly offered by the new generations of CPUs, in the field of detector simulation.

\item[GPGPU] General-Purpose computing on Graphics Processing Units is the use of a Graphics Processing Unit (GPU), which typically handles computation only for computer graphics, to perform computation in applications traditionally handled by the Central Processing Unit (CPU). Programming for GPUs is typically more challenging, but can offer significant gains in arithmetic throughput.

\item[HEPData] The Durham High Energy Physics Database is an open access
repository for scattering data from experimental particle physics.

\item[HERWIG] This is an event generator containing a wide range of Standard
Model, Higgs and supersymmetric processes. It uses the parton-shower approach
for initial- and final-state QCD radiation, including colour coherence effects and
azimuthal correlations both within and between jets.

\item[HL-LHC] The High Luminosity Large Hadron Collider is a proposed upgrade to
the Large Hadron Collider to be made in 2026. The upgrade aims at increasing the
luminosity of the machine by a factor of 10, up to
$10^{35}\mathrm{cm}^{-2}\mathrm{s}^{-1}$,
providing a better chance to see rare processes and improving
statistically marginal measurements.

\item[HLT] High Level Trigger. The computing resources, generally a large farm, close to the detector which process the events in real-time and select those who must be stored for further analysis.

\item[HPC] High Performance Computing.

\item[HS06] HEP-wide benchmark for measuring CPU performance based on the SPEC2006 benchmark
(\href{https://www.spec.org}{{https://www.spec.org}}).

\item[HSF] The HEP Software Foundation facilitates coordination and common
efforts in high energy physics (HEP) software and computing
internationally.

\item[IML] The Inter-experimental LHC Machine Learning (IML) Working Group is
focused on the development of modern state-of-the art machine learning
methods, techniques and practices for high-energy physics problems.

\item[IOV] Interval Of Validity, the period of time for which a specific piece
of conditions data is valid.

\item[JavaScript] A high-level, dynamic, weakly typed,
prototype-based, multi-paradigm, and interpreted programming language.
Alongside HTML and CSS, JavaScript is one of the three core technologies
of World Wide Web content production.

\item[Jupyter Notebook] This is a server-client application that allows
editing and running notebook documents via a web browser. Notebooks are
documents produced by the Jupyter Notebook App, which contain both
computer code (e.g., python) and rich text elements (paragraph,
equations, figures, links, etc...). Notebook documents are both
human-readable documents containing the analysis description and the
results (figures, tables, etc..) as well as executable documents which
can be run to perform data analysis.

\item[LHC] Large Hadron Collider, the main particle accelerator at CERN.

\item[LHCONE] A set of network circuits, managed worldwide by the National
Research and Education Networks, to provide dedicated transfer paths for
LHC T1/T2/T3 sites on the standard academic and research physical
network infrastructure.

\item[LHCOPN] LHC Optical Private Network. It is the private physical and IP
network that connects the Tier0 and the Tier1 sites of the WLCG.

\item[MADEVENT] This is a multi-purpose tree-level event generator. It is
powered by the matrix element event generator MADGRAPH, which generates
the amplitudes for all relevant sub-processes and produces the mappings
for the integration over the phase space.

\item[Matplotlib] This is a Python 2D plotting library that provides
publication quality figures in a variety of hardcopy formats and
interactive environments across platforms.

\item[ML] Machine learning is a field of computer science that gives computers
the ability to learn without being explicitly programmed. It focuses on
prediction making through the use of computers and emcompasses a lot of
algorithm classes (boosted decision trees, neural networks\ldots{}).

\item[MONARC] A model of large scale distributed computing based on many
regional centers, with a focus on LHC experiments at CERN. As part of
the MONARC project, a simulation framework was developed that provides a
design and optimisation tool. The MONARC model has been the initial
reference for building the WLCG infrastructure and to organise the data
transfers around it.

\item[OpenGL] Open Graphics Library is a cross-language, cross-platform
application programming interface(API) for rendering 2D and 3D vector
graphics. The API is typically used to interact with a graphics
processing unit(GPU), to achieve hardware-accelerated rendering.

\item[Openlab] CERN openlab is a public-private partnership that accelerates
the development of cutting-edge solutions for the worldwide LHC
community and wider scientific research.

\item[P5] The Particle Physics Project Prioritization Panel is a scientific
advisory panel tasked with recommending plans for U.S. investment in
particle physics research over the next ten years.

\item[PRNG] A PseudoRandom Number Generator is an algorithm for generating a
sequence of numbers whose properties approximate the properties of
sequences of random numbers.

\item[PyROOT] A \href{http://www.python.org/}{Python} extension module that
allows the user to interact with any ROOT class from the Python
interpreter.

\item[PYTHIA] A program for the generation of high-energy physics events, i.e.,
for the description of collisions at high energies between elementary
particles such as e+, e-, p and pbar in various combinations. It
contains theory and models for a number of physics aspects, including
hard and soft interactions, parton distributions, initial- and
final-state parton showers, multiparton interactions, fragmentation and
decay.

\item[QCD] Quantum Chromodynamics, the theory describing the strong
interaction between quarks and gluons.

\item[REST] Representational State Transfer
\href{https://en.wikipedia.org/wiki/Web_service}{web services} are a way
of providing interoperability between computer systems on the Internet.
One of its main features is stateless interactions between clients and
servers (every interaction is totally independent of the others),
allowing for very efficient caching.

\item[ROOT] A modular scientific software framework widely used in HEP data
processing applications.

\item[SAML] Security Assertion Markup Language. It is an open, XML-based,
standard for exchanging authentication and authorisation data between
parties, in particular, between an identity provider and a service
provider.

\item[SDN] Software-defined networking is an umbrella term encompassing several kinds of network technology aimed at making the network as agile
and flexible as the virtualised server and storage infrastructure of the modern data center.

\item[SHERPA] Sherpa is a Monte Carlo event generator for the Simulation of
High-Energy Reactions of PArticles in lepton-lepton, lepton-photon,
photon-photon, lepton-hadron and hadron-hadron collisions.

\item[SIMD] Single instruction, multiple data (\textbf{SIMD}), describes
computers with multiple processing elements that perform the same
operation on multiple data points simultaneously.

\item[SM] The Standard Model is the name given in the 1970s to a theory of
fundamental particles and how they interact. It is the currently
dominant theory explaining the elementary particles and their dynamics.

\item[SWAN] Service for Web based ANalysis is a platform for interactive data
mining in the CERN cloud using the Jupyter notebook interface.

\item[TBB] Intel Threading Building Blocks is a widely used C++ template
library for task parallelism. It lets you easily write parallel C++
programs that take full advantage of multicore performance.

\item[TMVA] The Toolkit for Multivariate Data Analysis with ROOT is a
standalone project that provides a ROOT-integrated machine learning
environment for the processing and parallel evaluation of sophisticated
multivariate classification techniques.

\item[VecGeom] The vectorised geometry library for particle-detector simulation.

\item[VO] Virtual Organisation. A group of users sharing a common interest
(for example, each LHC experiment is a VO), centrally managed, and used
in particular as the basis for authorisations in the WLCG
infrastructure.

\item[WebGL] The Web Graphics Library is a JavaScript API for rendering
interactive 2D and 3D graphics within any compatible web browser without
the use of plug-ins.

\item[WLCG] The Worldwide LHC Computing Grid project is a global collaboration
of more than 170 computing centres in 42 countries, linking up national
and international grid infrastructures. The mission of the WLCG project
is to provide global computing resources to store, distribute and
analyse data generated by the Large Hadron Collider (LHC) at CERN.

\item[X.509] A cryptographic standard which defines how to implement service
security using electronic certificates, based on the use of a private
and public key combination. It is widely used on web servers accessed
using the https protocol and is the main authentication mechanism on the
WLCG infrastructure.

\item[x86\_64] 64-bit version of the x86 instruction set.

\item[XRootD] Software framework that is a fully generic suite for fast, low latency and scalable data access.

\end{description}

\end{appendices}

%% file: authors.tex
Albrecht, Johannes$^{69}$;
Alves Jr, Antonio Augusto$^{81}$;
Amadio, Guilherme$^{5}$;
Andronico, Giuseppe$^{27}$;
Anh-Ky, Nguyen$^{122}$;
Aphecetche, Laurent$^{66}$;
Apostolakis, John$^{5}$;
Asai, Makoto$^{63,p}$;
Atzori, Luca$^{5}$;
Babik, Marian$^{5}$;
% ...
\bigskip
\par {\footnotesize $^{1}$ High Energy Physics Division, Argonne National Laboratory, Argonne, IL, USA}
\par {\footnotesize $^{2}$ Physics Department, Brookhaven National Laboratory, Upton, NY, USA}
\par {\footnotesize $^{3}$ Centro Brasileiro de Pesquisas Físicas, Rio de Janeiro, Brazil}
\par {\footnotesize $^{4}$ Centre de Calcul de l’IN2P3, Villeurbanne, Lyon, France}
\par {\footnotesize $^{5}$ CERN, Geneva, Switzerland}
\par {\footnotesize $^{6}$ Center for High Throughput  Computing, Computer Sciences Department, University of Wisconsin-Madison, Madison, WI, USA}
\par {\footnotesize $^{7}$ Centro de Investigaciones Energéticas Medioambientales y Tecnológicas (CIEMAT), Madrid, Spain}
\par {\footnotesize $^{8}$ Cinvestav, Mexico City, Mexico}
\par {\footnotesize $^{9}$ Centro Nazionale Analisi Fotogrammi (CNAF), INFN, Bologna, Italy}
\par {\footnotesize $^{10}$ Center for High Performance Computing, Cape Town, South Africa}
\par {\footnotesize $^{11}$ California Institute of Technology, Pasadena, California, CA, USA}
\par {\footnotesize $^{12}$ Cornell University, Ithaca, NY, USA}
\par {\footnotesize $^{13}$ Deutsches Elektronen-Synchrotron, Hamburg, Germany}
\par {\footnotesize $^{14}$ Institute of Physics, École Polytechnique Fédérale de Lausanne (EPFL), Lausanne, Switzerland}
\par {\footnotesize $^{15}$ ETH Zürich - Institute for Particle Physics and Astrophysics (IPA), Zürich, Switzerland}
\par {\footnotesize $^{16}$ Fermi National Accelerator Laboratory, Batavia, IL, USA}
\par {\footnotesize $^{17}$ University of Hamburg, Hamburg, Germany}
\par {\footnotesize $^{18}$ Helsinki Institute of Physics, Helsinki, Finland}
\par {\footnotesize $^{19}$ International Center for Theoretical Physics, Trieste, Italy}
\par {\footnotesize $^{20}$ Institut de Física d’Altes Energies and Departament de Física de la Universitat Autònoma de Barcelona and ICREA, Barcelona, Spain}
\par {\footnotesize $^{21}$ Instituto de Física de Cantabria (IFCA), CSIC-Universidad de Cantabria, Santander, Spain}
\par {\footnotesize $^{22}$ Instituto de Física Corpuscular, Centro Mixto Universidad de Valencia - CSIC, Valencia, Spain}
\par {\footnotesize $^{23}$ Institute of High Energy Physics, Chinese Academy of Sciences, Beijing}
\par {\footnotesize $^{24}$ Department of Physics, Indiana University, Bloomington, IN, USA}
\par {\footnotesize $^{25}$ INFN Sezione di Bologna, Università di Bologna, Bologna, Italy}
\par {\footnotesize $^{26}$ INFN Sezione di Bari, Università di Bari, Politecnico di Bari, Bari, Italy}
\par {\footnotesize $^{27}$ INFN Sezione di Catania, Università di Catania, Catania, Italy}
\par {\footnotesize $^{28}$ Università e INFN, Ferrara, Ferrara, Italy}
\par {\footnotesize $^{29}$ INFN Sezione di Genova, Genova, Italy}
\par {\footnotesize $^{30}$ INFN Sezione di Milano-Bicocca, Milano, Italy}
\par {\footnotesize $^{31}$ INFN Sezione di Napoli, Università di Napoli, Napoli, Italy}
\par {\footnotesize $^{32}$ INFN Sezione di Pisa, Università di Pisa, Scuola Normale Superiore di Pisa, Pisa, Italy}
\par {\footnotesize $^{33}$ INFN Sezione di Padova, Università di Padova b, Padova, Italy}
\par {\footnotesize $^{34}$ INFN Sezione di Roma I, Università La Sapienza, Roma, Italy}
\par {\footnotesize $^{35}$ INFN Sezione di Trieste, Università di Trieste, Trieste, Italy}
\par {\footnotesize $^{36}$ INFN Sezione di Torino, Torino, Italy}
\par {\footnotesize $^{37}$ Université de Strasbourg, CNRS, IPHC UMR 7178, F-67000 Strasbourg, France}
\par {\footnotesize $^{38}$ Imperial College London, London, United Kingdom}
\par {\footnotesize $^{39}$ DSM/IRFU (Institut de Recherches sur les Lois Fondamentales de l’Univers), CEA Saclay (Commissariat à l’Énergie Atomique), Gif-sur-Yvette, France}
\par {\footnotesize $^{40}$ Thomas Jefferson National Accelerator Facility, Newport News, Virginia, VA, USA}
\par {\footnotesize $^{41}$ Jožef Stefan Institute, Ljubljana, Slovenia}
\par {\footnotesize $^{42}$ Karlsruhe Institute of Technology, Karlsruhe, Germany}
\par {\footnotesize $^{43}$ AGH University of Science and Technology, Faculty of Physics and Applied Computer Science, Krakow, Poland}
\par {\footnotesize $^{44}$ Kyungpook National University, Daegu, Republic of Korea}
\par {\footnotesize $^{45}$ LAL, Université Paris-Sud and CNRS/IN2P3, Orsay, France}
\par {\footnotesize $^{46}$ Lawrence Berkeley National Laboratory and University of California, Berkeley, CA, USA}
\par {\footnotesize $^{47}$ Laboratório de Instrumentação e Física Experimental de Partículas (LIP), Lisboa, Portugal}
\par {\footnotesize $^{48}$ Laboratoire Leprince-Ringuet, École Polytechnique, CNRS/IN2P3, Université Paris-Saclay, Palaiseau, France}
\par {\footnotesize $^{49}$ Fakultät für Physik, Ludwig-Maximilians-Universität München, München, Germany}
\par {\footnotesize $^{50}$ Laboratoire de Physique Corpusculaire, Clermont Université and Université Blaise Pascal and CNRS/IN2P3, Clermont-Ferrand, France}
\par {\footnotesize $^{51}$ Laboratoire de Physique Nucléaire et de Hautes Energies (LPNHE), Sorbonne Université, Université Paris Diderot, CNRS/IN2P3, Paris, France}
\par {\footnotesize $^{52}$ Laboratoire de Physique Subatomique et de Cosmologie, Université Joseph Fourier and CNRS/IN2P3 and Institut National Polytechnique de Grenoble, Grenoble, France}
\par {\footnotesize $^{53}$ Massachusetts Institute of Technology, Cambridge, MA, USA}
\par {\footnotesize $^{54}$ Max-Planck-Institut für Physik (Werner-Heisenberg-Institut), München, Germany}
\par {\footnotesize $^{55}$ Niels Bohr Institute, University of Copenhagen, Kobenhavn, Denmark}
\par {\footnotesize $^{56}$ Nikhef National Institute for Subatomic Physics, Amsterdam, Netherlands}
\par {\footnotesize $^{57}$ Department of Physics, New York University, New York, NY, USA}
\par {\footnotesize $^{58}$ Center for High Energy Physics, University of Oregon, Eugene, OR, USA}
\par {\footnotesize $^{59}$ Port d’Informació Científica (PIC), Universitat Autònoma de Barcelona (UAB), Barcelona, Spain}
\par {\footnotesize $^{60}$ High Energy Physics (IHEP), Protvino, Russia}
\par {\footnotesize $^{61}$ STFC Rutherford Appleton Laboratory, Didcot, United Kingdom}
\par {\footnotesize $^{62}$ Department of Physics, Royal Holloway University of London, Surrey, United Kingdom}
\par {\footnotesize $^{63}$ SLAC National Accelerator Laboratory, Menlo Park, CA, USA}
\par {\footnotesize $^{64}$ Simons Foundation, New York, NY, USA}
\par {\footnotesize $^{65}$ SoundCloud, Berlin, Germany}
\par {\footnotesize $^{66}$ SUBATECH, IMT Atlantique, Université de Nantes, CNRS-IN2P3, Nantes, France}
\par {\footnotesize $^{67}$ National Taiwan University, Taipei, Taiwan}
\par {\footnotesize $^{68}$ Tata Institute of Fundamental Research, Mumbai, India}
\par {\footnotesize $^{69}$ Technische Universit\"at Dortmund, Dortmund, Germany}
\par {\footnotesize $^{70}$ Department of Physics, University of Alberta, Edmonton, AB, Canada}
\par {\footnotesize $^{71}$ Department of Physics, University of Arizona, Tucson, AZ, USA}
\par {\footnotesize $^{72}$ H.H. Wills Physics Laboratory, University of Bristol, Bristol, United Kingdom}
\par {\footnotesize $^{73}$ Department of Physics, University of British Columbia, Vancouver, BC, Canada}
\par {\footnotesize $^{74}$ Department of Physics and Astronomy, University of California Irvine, Irvine, CA, USA}
\par {\footnotesize $^{75}$ University of California, San Diego, La Jolla, CA, USA}
\par {\footnotesize $^{76}$ Cavendish Laboratory, University of Cambridge, Cambridge, United Kingdom}
\par {\footnotesize $^{77}$ Physics Department, University of Cape Town, Cape Town, South Africa}
\par {\footnotesize $^{78}$ Carleton University, Ottawa, ON, Canada}
\par {\footnotesize $^{79}$ University of Illinois Urbana-Champaign, Champaign, Illinois, IL, USA}
\par {\footnotesize $^{80}$ Enrico Fermi Institute, University of Chicago, Chicago, IL, USA}
\par {\footnotesize $^{81}$ University of Cincinnati, Cincinnati, OH, USA}
\par {\footnotesize $^{82}$ Department of Physics, Duke University, Durham, NC, USA}
\par {\footnotesize $^{83}$ SUPA - School of Physics and Astronomy, University of Edinburgh, Edinburgh, United Kingdom}
\par {\footnotesize $^{84}$ Gangneung-Wonju National University, South Korea}
\par {\footnotesize $^{85}$ Section de Physique, Université de Genève, Geneva, Switzerland}
\par {\footnotesize $^{86}$ SUPA - School of Physics and Astronomy, University of Glasgow, Glasgow, United Kingdom}
\par {\footnotesize $^{87}$ Physikalisches Institut, Ruprecht-Karls-Universitat Heidelberg, Heidelberg, Germany}
\par {\footnotesize $^{88}$ Department of Physics and Astronomy, University of Kentucky, Lexington, KY, USA}
\par {\footnotesize $^{89}$ Physics Department, Lancaster University, Lancaster, United Kingdom}
\par {\footnotesize $^{90}$ Department of Physics, Jožef Stefan Institute and University of Ljubljana, Ljubljana, Slovenia}
\par {\footnotesize $^{91}$ Fysiska institutionen, Lunds Universitet, Lund, Sweden}
\par {\footnotesize $^{92}$ School of Physics and Astronomy, University of Manchester, Manchester, United Kingdom}
\par {\footnotesize $^{93}$ University of Maribor, Ljubljana, Slovenia}
\par {\footnotesize $^{94}$ Department of Physics, University of Massachusetts, Amherst, MA, USA}
\par {\footnotesize $^{95}$ Department of Physics, The University of Michigan, Ann Arbor, MI, USA}
\par {\footnotesize $^{96}$ Departamento de Física, Universidade do Minho, Braga, Portugal}
\par {\footnotesize $^{97}$ University of Nebraska-Lincoln, Lincoln, NE, USA}
\par {\footnotesize $^{98}$ Department of Physics and Astronomy, University of New Mexico, Albuquerque, NM, USA}
\par {\footnotesize $^{99}$ University of Notre Dame, Notre Dame, IN, USA}
\par {\footnotesize $^{100}$ Homer L. Dodge Department of Physics and Astronomy, University of Oklahoma, Norman, OK, USA}
\par {\footnotesize $^{101}$ The Ohio State University, Columbus, OH, USA}
\par {\footnotesize $^{102}$ Department of Physics, University of Oslo, Oslo, Norway}
\par {\footnotesize $^{103}$ Universidad de Oviedo, Oviedo, Spain}
\par {\footnotesize $^{104}$ Department of Physics, Oxford University, Oxford, UK}
\par {\footnotesize $^{105}$ Department of Physics and Astronomy, University of Pittsburgh, Pittsburgh, PA, USA}
\par {\footnotesize $^{106}$ Princeton University, Princeton, NJ, USA}
\par {\footnotesize $^{107}$ Rice University, Houston, TX, USA}
\par {\footnotesize $^{108}$ Universidade Estadual Paulista, São Paulo, Brazil}
\par {\footnotesize $^{109}$ Dipartimento di Fisica, Università La Sapienza, Roma, Italy}
\par {\footnotesize $^{110}$ University of Washington, Seattle, WA, USA}
\par {\footnotesize $^{111}$ School of Physics, Shandong University, Shandong, China}
\par {\footnotesize $^{112}$ Department of Physics and Astronomy, University of Sheffield, Sheffield, United Kingdom}
\par {\footnotesize $^{113}$ University of Sofia, Sofia, Bulgaria}
\par {\footnotesize $^{114}$ Department of Physics, The University of Texas at Arlington, Arlington, TX, USA}
\par {\footnotesize $^{115}$ National Research Tomsk Polytechnic University, Tomsk, Russia}
\par {\footnotesize $^{116}$ Department of Physics, University of Toronto, Toronto, ON, Canada}
\par {\footnotesize $^{117}$ Dipartimento di Chimica, Fisica e Ambiente, Università di Udine, Udine, Italy}
\par {\footnotesize $^{118}$ Department of Physics, University of Warwick, Coventry, United Kingdom}
\par {\footnotesize $^{119}$ University of Wisconsin - Madison, Madison, WI, USA}
\par {\footnotesize $^{120}$ Department of Physics, Yale University, New Haven, CT, USA}
\par {\footnotesize $^{121}$ Physik-Institut, Universität Zürich, Zürich, Switzerland}
\par {\footnotesize $^{122}$ IOP and GUST, Vietnam Academy of Science and Technology (VAST), Hanoi, Vietnam}
\par {\footnotesize $^{123}$ Virginia Tech, Blacksburg, Virginia, VA, USA}
\par {\footnotesize $^{124}$ Vrije Universiteit Amsterdam, Amsterdam, The Netherlands}
\bigskip
\par {\footnotesize $^{a}$ Community White Paper Editorial Board Member}
\par {\footnotesize $^{b}$ Vladimir V. Gligorov acknowledges funding from the European Research Council (ERC) under the European Union's Horizon 2020 research and innovation programme under grant agreement No 724777 “RECEPT”}
\par {\footnotesize $^{c}$ Supported by the US-DOE, DE-AC02-07CH11359}
\par {\footnotesize $^{d}$ Supported by FCT-Portugal, IF/00050/2013/CP1172/CT0002}
\par {\footnotesize $^{e}$ Supported by the US-NSF, ACI-1450319}
\par {\footnotesize $^{f}$ Supported by the US-NSF, PHY-1607578}
\par {\footnotesize $^{g}$ Supported by ES-MINECO, FPA2016-80994-c2-1-R \& MDM-2015-0509}
\par {\footnotesize $^{h}$ Caterina Doglioni acknowledges funding from the European Research Council (ERC) under the European Union's Horizon 2020 research and innovation programme under grant agreement No 679305 “DARKJETS”}
\par {\footnotesize $^{i}$ Supported by the US-DOE, DE-SC0009973}
\par {\footnotesize $^{j}$ Supported by the ES-MINECO FPA2016-80994-C2-2-R}
\par {\footnotesize $^{k}$ Supported by the US-DOE, DE-SC0018098 and US-NSF ACI-1558233}
\par {\footnotesize $^{l}$ Supported by the US-DOE, DE-SC0007859 and US-NSF, 76749/1136652/2}
\par {\footnotesize $^{m}$ Supported by funding from the European Union’s Horizon 2020 research and innovation programme under the Marie Skłodowska-Curie grant agreement No. 752730}
\par {\footnotesize $^{n}$ Supported by Swiss National Science Foundation, Early Postdoc Mobility Fellowship, project number P2ELP2\_168556}
\par {\footnotesize $^{o}$ Supported by CONACYT (Mexico)}
\par {\footnotesize $^{p}$ Supported by the US-DOE, DE-AC02-76SF0051}